\documentclass[10pt,english,aps,prd,nofootinbib,eqsecnum,superscriptaddress]{revtex4-2}
\usepackage[T1]{fontenc}
\usepackage[utf8]{luainputenc}
\usepackage{color}
\usepackage{amsmath}
\usepackage{amssymb}
\usepackage{graphicx}
\usepackage{wasysym}

\makeatletter

\newcommand{\lyxdot}{.}

\usepackage[marginal]{footmisc}
\usepackage{array}
\usepackage{braket}
\allowdisplaybreaks[4]
\usepackage[section]{placeins}

\usepackage{verbatim}
\usepackage{fancyhdr}
\usepackage{tikz}
\usepackage[colorlinks,linkcolor=blue,anchorcolor=blue,citecolor=blue,urlcolor=blue]{hyperref}

\usetikzlibrary{chains,patterns,scopes,positioning,backgrounds,shapes,fit,shadows,calc,arrows.meta,decorations.pathreplacing,calligraphy} 

\DeclareMathSizes{12}{12}{7}{7}
\renewcommand{\theequation}{\arabic{section}.\arabic{equation}}

\def\frontmatter@abstractheading{\centerline{\textbf{Abstract}}}
\g@addto@macro\abstract{\ignorespaces}

\makeatother

\usepackage{babel}
\begin{document}
\title{The linear response theory approach to the sub-GeV dark matter in
the Sun}
\author{Zheng-Liang Liang}
\email{liangzl@mail.buct.edu.cn}

\affiliation{College of Mathematics and Physics, Beijing University of Chemical
Technology~\\
Beijing 100029, China}
\author{Ping Zhang}
\email{pzhang2012@qq.com}

\affiliation{School of Physics and Physical Engineering, Qufu Normal University~\\
 Qufu, 273165, China}
\affiliation{Institute of Applied Physics and Computational Mathematics~\\
Beijing, 100088, China}
\begin{abstract}
In recent years, the importance of the electronic in-medium effect
in the sub-GeV dark matter~(DM) direct detection has been recognized
and a coherent formulation of the DM-electron scattering based on
the linear response theory has been well established in the literature.
In this paper, we apply the formulation to the scattering between
DM particles and solar medium, and it is found that the dynamic structure
factor inherently incorporate the particle-particle scattering and
in-medium effect. Using this tool and taking a benchmark model as
an example, we demonstrate how the in-medium effect affects the scattering
of DM particles in the Sun, in both the heavy and the light mediator
limit. Formulae derived in this work lay the foundation for accurately
calculating the spectra of solar-accelerated DM particles, which is
of particular importance for the detection of DM particles via plasmon
in semiconductor targets.
\end{abstract}
\maketitle

\section{Introduction}

The sub-GeV \textit{dark matter}~(DM) as an alternative candidate
to the \textit{weakly interacting massive particles}~(WIMPs), has
attracted increasing attention for its theoretical motivations and
detection feasibility. The sub-GeV DM particles are expected to reveal
itself via the weak DM-electron interaction in semiconductor targets~(\textit{e.g}.,
SENSEI~\citep{Barak:2020fql}, DAMIC~\citep{Castello-Mor:2020jhd},
SuperCDMS~\citep{Amaral:2020ryn}, CDEX~\citep{CDEX:2022kcd} and
EDELWEISS~\citep{Arnaud:2020svb}). In the theoretical front, since
the introduction of the first-principles \textit{density functional
theory}~(DFT)~\citep{Essig:2015cda} method into the interpretation
of DM signals in semiconductor detectors, similar investigations have
been generalized to a wider range of target materials~\citep{Hochberg:2015pha,Essig:2016crl,Hochberg:2016ntt,Hochberg:2016sqx,Derenzo:2016fse,Hochberg:2017wce,Knapen2018,Griffin:2018bjn,Griffin:2019mvc,Trickle:2019ovy,Kurinsky:2019pgb,Trickle:2019nya,Campbell-Deem:2019hdx,Coskuner:2019odd,Geilhufe:2019ndy,Griffin:2020lgd,Prabhu:2022dtm,Esposito:2022bnu},
and have spurred further discussions on the methodology~\citep{Liang:2018bdb,Griffin:2021znd,Kahn:2021ttr,Trickle:2022fwt,Dreyer:2023ovn}
and interpretations of the DM-electron interactions~\citep{Emken:2019tni,Essig:2019xkx,Trickle:2020oki,Andersson:2020uwc,Su:2020zny,Mitridate:2021ctr,Vahsen:2021gnb,Catena:2021qsr,Chen:2022pyd,Catena:2022fnk,Li:2022acp,Wang:2023xgm}.

Recently, the importance of nontrivial collective behavior of the
electrons in solid detectors has come to be recognized~\citep{Kurinsky:2020dpb,Gelmini_2020,Kozaczuk:2020uzb,Mitridate:2021ctr,Knapen:2021run,Hochberg:2021pkt,Knapen:2021bwg,Liang:2021zkg}.
Phenomena such as screening and the plasmon excitation that cannot
be explained in terms of standard two-body scattering, and non-interacting
single-particle states, can be well described with the dielectric
function. The in-medium effect induced by the DM-electron interaction
has been thoroughly investigated in Refs.~\citep{Knapen:2021run,Hochberg:2021pkt,Liang:2021zkg,Knapen:2021bwg,Chen:2022xzi},
and it proved to have a significant impact on DM signals, either destructively
as screening or constructively as collective resonance (plasmon).

Additionally, the in-medium effect in stellar media is also important
for relevant physical processes. For example, the screening effect
plays a vital role when considering the long-range interaction between
DM particles and electrons in the Sun. The authors of Ref.~\citep{An:2021qdl}
introduced the dielectric function to account for the screening effect
in simulating the DM-electron scattering and DM propagation in the
Sun. Moreover, in Ref.~\citep{DeRocco:2022rze} the authors has established
a systematic approach based on the relativistic thermal field theory
for describing the collective effect in DM scattering in various stellar
environments. 

In the light of these progresses, the key insight in this paper is
that, a well-established tool widely used in condensed matter physics,
\textit{i.e}., the dynamic structure factor in the context of the
linear response theory, a physical quantity that has already been
introduced recently in the DM direct detection community~\citep{Knapen:2021run,Hochberg:2021pkt,Liang:2021zkg,Knapen:2021bwg,Chen:2022xzi},
suffices to give a full description of the in-medium effect in the
Sun, where all particles involved scatter non-relativistically. As
a result, the description of the in-medium effect in the solar source
and the detection end can be unified under a general framework. Instead
of starting with a particle-particle scattering amplitude and then
adding the dielectric function by hand to account for the in-medium
effect, we find that it is natural and convenient to describe the
DM-solar medium interaction using the dynamic structure factor $S\left(\mathbf{Q},\omega\right)$.
By doing so, we extend the argument of the energy transfer $\omega$
from the range $\omega>0$ in direct detection analysis, where DM
particles deposit energy to detectors, to the range $\omega<0$, in
order to further describe the DM particles absorbing and releasing
energy in the solar medium on a general ground. In contrast to the
case of direct detection where only specified final states of target
are of concern, here we only focus on the final states of DM particles,
which is an exclusive process where all the intermediate and final
states of the target are summed over. Therefore, it is advantageous
to use the dynamic structure factor to describe the scattering between
the DM particle and solar medium in the Sun.

As a benchmark study, we discuss a Dirac particle coupling exclusively
to electrons via a massive vector boson, a model widely discussed
in direct detection analyses~(\textit{e.g.}, Refs~\citep{Knapen:2021run,Hochberg:2021pkt,Kahn:2021ttr}).
On the one hand, such a simple model with a heavy mediator corresponds
to a contact DM-electron interaction, and hence is a good prototype
for the study of the solar DM particles and relevant phenomenology~\citep{An:2017ojc,Garani:2017jcj,Liang:2018cjn,Emken_2018,Liang:2021zkg,Gaidau:2021vyr,Emken:2021lgc,Bose:2021cou,Acevedo:2023owd}.
However, even for this simple model the in-medium effect of the solar
environment has not been taken into consideration in the previous
studies~\citep{Liang:2021zkg,An:2017ojc,Emken:2021lgc}. In Ref.~\citep{Liang:2021zkg}
we have discussed the in-medium effect in the detection end, but for
the sake of self-consistency, the in-medium effect in solar environment
should also be taken into consideration.

On the other hand, the collective effect in detectors can also play
a role in probing DM signals. The authors of Ref.~\citep{Kurinsky:2020dpb}
proposed that if the incident DM particle is fast enough~($v_{\chi}\apprge10^{-2}$)
it can excite a collective mode of electrons, namely, the plasmon
in semiconductor detectors, and such resonance behavior will significantly
increase the ionization yield. This fact is particularly important
for scenarios where DM particles interact with electrons via a light
mediator, because in this case the amplitude squared $\sim1/Q^{4}$
(with $Q$ being the transferred momentum) significantly prefers contribution
from the small $Q$ region, where the resonance locates ($Q<5\,\mathrm{keV},$
$\omega\sim15\,\mathrm{eV}$ for silicon and germanium semiconductors).
While the velocity of the halo DM particles is far below the threshold~($v_{\chi}\apprge10^{-2}$)
to trigger a plasmon excitation, the solar-reflected DM particles
are potentially capable of producing such signals, which, also requires
us to first investigate the in-medium effect in the Sun for the light
mediator scenario.

As an interesting application, in this work we also discuss the ionic
response to the electron density primarily triggered by the DM field.
Although the DM-nucleus interaction is absent in our benchmark model
in the first place, the polarized electron density induced by the
DM particles can still introduce an effective DM-ion interaction in
the Sun. While it is presumed that ionic in-medium effect can be neglected
due their heavy masses and slow response to electrons and DM particles,
in the small transferred energy~(or low frequency) regime, there
may be remarkable ionic response in the wake of the slow responding
electrons in the solar medium. So it is necessary to give a quantitative
description of such process.

In short, through this simple model one can learn how to describe
the DM particles in solar medium under the framework of the linear
response theory, and apply it to more complex and more realistic DM
models in the Sun. The rest of this paper is organized as follows.
We begin Sec.~\ref{sec:DM-electron-interaction} by giving the formula
describing the scattering rate of DM particles in the solar electronic
medium with the linear response theory. Based on this discussion,
we then generalize the formalism to include the ionic response in
Sec.~\ref{sec:Ionic Response}. We summarize and conclude in Sec.~\ref{sec:Conclusions}.
A short review on the plasma dispersion function is provided in Appendix~\ref{sec:appendix1}.

Our discussion will proceed in natural units, where $\hbar=c=k_{B}=1$.

\section{\label{sec:DM-electron-interaction} Scattering rate between DM particles
and electronic medium}

\begin{figure}[h]
\begin{centering}
\includegraphics[scale=0.7]{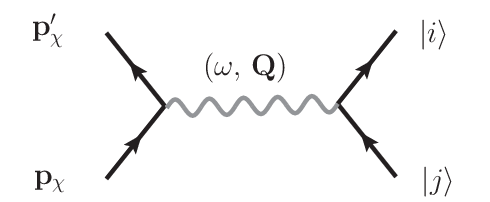}
\par\end{centering}
\caption{\label{fig:Feynman Diagram}Feynman diagram of the scattering between
a DM particle and an electron bound to a fixed target, where the DM
particle transfer momentum $\mathbf{Q}$ and energy $\omega$ to the
electron, and boosts it from the initial state $\ket{j}$ to a final
state $\ket{i}$.}
\end{figure}

The theoretical description for a non-relativistic DM-electron interaction
involving the in-medium effect in solids has been explored in the
literature~\citep{Hochberg:2021pkt,Knapen:2021run,Knapen:2021bwg,Mitridate:2021ctr,Liang:2021zkg,Boyd:2022tcn}.
Refs.~\citep{Knapen:2021run,Hochberg:2021pkt} pointed out that the
many-body in-medium effect can be naturally accounted for by the dielectric
function $\epsilon\left(\mathbf{Q},\omega\right)$, (or the \textit{energy
loss function}~(ELF) $\mathrm{Im}\left(-\epsilon^{-1}\left(\mathbf{Q},\omega\right)\right)$),
a physical quantity that can be drawn directly from experimental measurements,
while Refs.~\citep{Knapen:2021run,Knapen:2021bwg,Liang:2021zkg}
described the electronic excitation process in the context of the
response theory at a finite temperature, and performed computations
based on the DFT for various semiconductor materials. 

In this paper, since we are discussing DM particles shuttling back
and forth within the Sun, and only the movements of DM particles are
of our concern, it is straightforward to extend the positive deposited
energy region $\omega>0$ in the formalism of excitation event rates
in detectors to the negative absorption energy region $\omega<0$,
so as to describe the energy gain and loss for the DM particles in
the solar medium in a consistent manner. 

We consider a concrete freeze-in DM model, where the DM candidate
is a Dirac particle~($\chi$) coupling to standard model particles
via a massive vector boson $A_{\mu}^{'}$, with the interaction 
\begin{eqnarray}
\mathcal{L}_{\mathrm{int}} & = & g_{\chi}\bar{\chi}\gamma^{\mu}\chi A_{\mu}^{'}+g_{e}\bar{e}\gamma^{\mu}eA_{\mu}^{'},\label{eq:leptopihilic_interaction}
\end{eqnarray}
where $g_{\chi}$ and $g_{e}$ represent the strengths of the mediator
coupling to the DM particle and electron, respectively. As has been
pointed out in Refs.~\citep{Knapen:2021run,Hochberg:2021pkt}, for
this generic type of model, the DM component only couples to the electric
density at the leading order. Following the procedure in Ref.~\citep{Mitridate:2021ctr},
we obtain the effective electron Lagrangian by matching onto the \textit{nonrelativistic}~(NR)
\textit{effective field theory}~(EFT) in the beginning, which reads
as 
\begin{eqnarray}
\mathcal{L}_{A'e}^{\mathrm{eff}} & \supset & g_{e}A_{0}^{'}\psi_{e}^{*}\psi_{e}+\frac{ig_{e}}{2m_{e}}\mathbf{A}^{'}\cdot\left(\psi_{e}^{*}\overrightarrow{\nabla}\psi_{e}-\psi_{e}^{*}\overleftarrow{\nabla}\psi_{e}\right)+\cdots,\label{eq:DM-electron interaction}
\end{eqnarray}
where $\psi_{e}$ is the NR electron wavefunction. Since the terms
(\textit{e.g.}, the second term representing the electric current
coupling) beyond the first are subject to electron velocity suppression
$\frac{\nabla}{m_{e}}\sim v_{e}\sim\mathcal{O}\left(10^{-2}\sim10^{-1}\right)$
in the solar medium, only the leading order term representing the
density coupling $g_{e}A_{0}^{'}\psi_{e}^{*}\psi_{e}$ is preserved
in $\mathcal{L}_{A'e}^{\mathrm{eff}}$ in our discussion. Similarly,
only the longitudinal component in the NR effective electron-photon
interaction
\begin{eqnarray}
\mathcal{L}_{Ae}^{\mathrm{eff}} & \supset & -eA_{0}\psi_{e}^{*}\psi_{e}-\frac{ie}{2m_{e}}\mathbf{A}\cdot\left(\psi_{e}^{*}\overrightarrow{\nabla}\psi_{e}-\psi_{e}^{*}\overleftarrow{\nabla}\psi_{e}\right)+\cdots
\end{eqnarray}
$A_{0}$~(or more specifically, the Coulomb interaction) is retained
for the description of electron-electron~(and electron-ion) interaction
in this work.

We start our discussion with the NR calculation of scattering between
an incident DM particle and a fixed target, and then derive the cross
section and the scattering event rates at a finite temperature. The
convention $\ket{\mathbf{p},s}=\hat{a}_{\mathbf{p},s}^{\dagger}\ket{0}$
and the anti-commutation rule $\left\{ \hat{a}_{\mathbf{p},s},\hat{a}_{\mathbf{p'},s'}^{\dagger}\right\} =\left(2\pi\right)^{3}\delta^{\left(3\right)}\left(\mathbf{p}-\mathbf{p}'\right)\delta_{s,s'}$
for a Dirac fermion DM particle are adopted, with $s$~($s'$) labeling
the two spin orientations. Considering an DM particle $\chi$ with
velocity $v_{\chi}$ impinging on a fixed target (see Fig.~\ref{fig:Feynman Diagram}),
if the $T$-matrix is expressed in the form
\begin{eqnarray}
\braket{\mathbf{p}_{\chi}',i|\,iT\,|\mathbf{p}_{\chi},j} & = & i\mathcal{M}_{i\rightarrow j}\,2\pi\delta\left(\omega_{p'p}+\varepsilon_{i}-\varepsilon_{j}\right),
\end{eqnarray}
where $\mathcal{M}_{i\rightarrow j}$ describes the relevant scattering
amplitude, $\omega_{p'p}=\left(p'_{\chi}\right)^{2}/2m_{\chi}-p_{\chi}^{2}/2m_{\chi}$
is the energy gain of the DM particle~(with $m_{\chi}$ being the
DM particle mass), and $\varepsilon_{j}$~($\varepsilon_{i}$) represents
the initial~(final) energy of the fixed target, the effective cross
section for the excitation channel $i\rightarrow j$ can be written
as 
\begin{eqnarray}
\sigma_{i\rightarrow j} & = & \int\frac{\mathrm{d}^{3}p_{\chi}'}{\left(2\pi\right)^{3}}\frac{\left|\mathcal{M}_{i\rightarrow j}\right|^{2}}{v_{\chi}}2\pi\delta\left(\omega_{p'p}+\varepsilon_{i}-\varepsilon_{j}\right).
\end{eqnarray}
 Given the interaction in Eq.~(\ref{eq:DM-electron interaction}),
the $T$-matrix of the scattering process in Fig.~\ref{fig:Feynman Diagram}
is read as 
\begin{eqnarray}
i\mathcal{T} & = & \frac{-i\,g_{\chi}g_{e}}{\left(\varepsilon_{i}-\varepsilon_{j}\right)^{2}-Q^{2}-m_{A'}^{2}}\braket{i|e^{i\mathbf{Q}\cdot\hat{\mathbf{x}}}|j}2\pi\delta\left(\omega_{p'p}+\varepsilon_{i}-\varepsilon_{j}\right)\nonumber \\
 & \simeq & \frac{i\,g_{\chi}g_{e}}{Q^{2}+m_{A'}^{2}}\braket{i|e^{i\mathbf{Q}\cdot\hat{\mathbf{x}}}|j}2\pi\delta\left(\omega_{p'p}+\varepsilon_{i}-\varepsilon_{j}\right),
\end{eqnarray}
where $Q=\left|\mathbf{\mathbf{Q}}\right|=\left|\mathbf{\mathbf{p}}_{\chi}-\mathbf{\mathbf{p}}_{\chi}'\right|$
is the magnitude of the momentum transfer, $m_{A'}$ is the mass of
the vector mediator. In order to discuss the excitation process in
the context of the response theory, we introduce the density operator
$\hat{\rho}_{e}\left(\mathbf{x}\right)\equiv\hat{\psi}_{e}^{\dagger}\left(\mathbf{x}\right)\hat{\psi}_{e}\left(\mathbf{x}\right)$,
and thus the total cross section~(by averaging over the initial states
and summing over the final states) of an electron gas in a volume
$V$ can be formulated as 
\begin{eqnarray}
\sigma\left(\mathbf{v}_{\chi}\right) & = & \sum_{i,j}p_{j}\sigma_{j\rightarrow i}\nonumber \\
 & = & \sum_{i,j}\int\mathrm{d}\omega\,\delta\left(\omega-\varepsilon_{i}+\varepsilon_{j}\right)\int\mathrm{d}^{3}Q\,\delta^{(3)}\left(\mathbf{Q}+\mathbf{\mathbf{p}}_{\chi}'-\mathbf{\mathbf{p}}_{\chi}\right)\int\frac{\mathrm{d}^{3}p_{\chi}'}{\left(2\pi\right)^{3}}\frac{1}{v_{\chi}}\left(\frac{g_{\chi}g_{e}}{Q^{2}+m_{A'}^{2}}\right)^{2}\int_{V}\mathrm{d}^{3}x\,\mathrm{d}^{3}x'\,p_{j}\braket{j|e^{-i\mathbf{\mathbf{\mathbf{Q}}}\cdot\mathbf{x}}\hat{\rho}_{e}\left(\mathbf{x}\right)|i}\nonumber \\
 &  & \times\braket{i|e^{i\mathbf{\mathbf{\mathbf{Q}}}\cdot\mathbf{x}'}\hat{\rho}_{e}\left(\mathbf{x}'\right)|j}\,2\pi\delta\left(\omega_{p'p}+\omega\right)\nonumber \\
\nonumber \\
 & = & \int\mathrm{d}\omega\int\frac{\mathrm{d}^{3}Q}{\left(2\pi\right)^{3}}\frac{1}{v_{\chi}}\left(\frac{g_{\chi}g_{e}}{Q^{2}+m_{A'}^{2}}\right)^{2}\int_{V}\mathrm{d}^{3}x\,\mathrm{d}^{3}x'\,\int_{-\infty}^{+\infty}\mathrm{d}t\braket{\hat{\rho}_{eI}\left(\mathbf{x},0\right)\hat{\rho}_{eI}\left(\mathbf{x}',t\right)}e^{i\omega\left(0-t\right)}e^{-i\mathbf{\mathbf{\mathbf{Q}}}\cdot\left(\mathbf{x}-\mathbf{x}'\right)}\,\delta\left(\omega_{p'p}+\omega\right)\nonumber \\
\nonumber \\
 & = & V\int\mathrm{d}\omega\int\frac{\mathrm{d}^{3}Q}{\left(2\pi\right)^{3}}\frac{1}{v_{\chi}}\frac{\pi\bar{\sigma}_{e}}{\mu_{\chi e}^{2}}\left(\frac{\alpha^{2}m_{e}^{2}+m_{A'}^{2}}{Q^{2}+m_{A'}^{2}}\right)^{2}\frac{\left(-2\right)\,\mathrm{Im}\left[\chi_{\hat{\rho}\hat{\rho}}^{\mathrm{r}}\left(\mathbf{\mathbf{Q}},\,\omega\right)\right]}{1-e^{-\beta\omega}}\,\delta\left(\frac{Q^{2}}{2m_{\chi}}-\mathbf{v}_{\chi}\cdot\mathbf{Q}+\omega\right),\label{eq:cross_section}
\end{eqnarray}
where $p_{j}$ is the thermal distribution of the initial state $\ket{j}$,
the symbol $\left\langle \cdots\right\rangle $ represents the thermal
average, and $\hat{\rho}_{eI}\left(\mathbf{x}',t\right)\equiv e^{i\hat{H}_{0}t}\hat{\rho}_{e}\left(\mathbf{x}'\right)e^{-i\hat{H}_{0}t}$,
with $\hat{H}_{0}$ being the unperturbed Hamiltonian of the electronic
system. In the last line, $\alpha=e^{2}/4\pi$ is the electromagnetic
fine structure constant, and a reference cross section
\begin{eqnarray}
\bar{\sigma}_{e} & \equiv & \frac{\mu_{\chi e}^{2}}{\pi}\left(\frac{g_{\chi}g_{e}}{\alpha^{2}m_{e}^{2}+m_{A'}^{2}}\right)^{2}\label{eq:ReferenceCrossSection}
\end{eqnarray}
defined at a benchmark momentum transfer $\alpha m_{e}$ is introduced
to parameterize the DM-electron coupling strength for a typical scattering
process, with $\mu_{\chi e}$ being the reduced mass of the DM-electron
pair. In the above derivation we invoke the fluctuation-dissipation
theorem
\begin{eqnarray}
S_{\hat{\rho}\hat{\rho}}\left(\mathbf{\mathbf{Q}},\,\omega\right) & = & \frac{1}{V}\int_{V}\mathrm{d}^{3}x\,\mathrm{d}^{3}x'\,\int_{-\infty}^{+\infty}\mathrm{d}t\braket{\hat{\rho}_{eI}\left(\mathbf{x},0\right)\hat{\rho}_{eI}\left(\mathbf{x}',t\right)}e^{i\omega\left(0-t\right)}e^{-i\mathbf{\mathbf{\mathbf{Q}}}\cdot\left(\mathbf{x}-\mathbf{x}'\right)}\nonumber \\
\nonumber \\
 & = & i\frac{\left[\chi_{\hat{\rho}\hat{\rho}}\left(\mathbf{\mathbf{Q}},\,\omega+i0^{+}\right)-\chi_{\hat{\rho}\hat{\rho}}\left(\mathbf{\mathbf{Q}},\,\omega-i0^{+}\right)\right]}{1-e^{-\beta\omega}}\nonumber \\
 & = & \frac{-2\,\mathrm{Im}\left[\chi_{\hat{\rho}\hat{\rho}}^{\mathrm{r}}\left(\mathbf{\mathbf{Q}},\,\omega\right)\right]}{1-e^{-\beta\omega}}\nonumber \\
 & = & \frac{2}{V_{e}\left(Q\right)}\frac{1}{1-e^{-\beta\omega}}\,\mathrm{Im}\left[\frac{-1}{\epsilon\left(\mathbf{\mathbf{Q}},\omega\right)}\right],\label{eq:Shh}
\end{eqnarray}
where $S_{\hat{\rho}\hat{\rho}}\left(\mathbf{\mathbf{Q}},\,\omega\right)$
is the dynamic structure factor associated with the density-density
correlation, $\beta=1/T$ is the inverse temperature, $V_{e}\left(Q\right)=4\pi\alpha/Q^{2}$
is the electron Coulomb interaction in momentum space. In the linear
response theory, the inverse dielectric function in the last line
connects the polarization function $\chi_{\hat{\rho}\hat{\rho}}^{\mathrm{r}}\left(\mathbf{\mathbf{Q}},\,\omega\right)$
through the following relation,
\begin{eqnarray}
\frac{1}{\epsilon\left(\mathbf{Q},\omega\right)} & = & 1+V_{e}\left(Q\right)\chi_{\hat{\rho}\hat{\rho}}^{\mathrm{r}}\left(\mathbf{Q},\,\omega\right),\label{eq:inverse_epsilon0}
\end{eqnarray}
where the unity $1$ faithfully reflects the bare external field,
while the second term describes the Coulomb potential sourced by the
polarized electrons, and hence the whole expression describes the
net effect caused by the external field in the momentum space. In
practice~\citep{bruus2004many}, one first evaluates the master function
$\chi_{\hat{\rho}\hat{\rho}}\left(\mathbf{Q},\,z\right)$ using the
Matsubara Green's function within the framework of finite temperature
field theory, and then the retarded polarizability function $\chi_{\hat{\rho}\hat{\rho}}^{\mathrm{r}}\left(\mathbf{\mathbf{Q}},\,\omega\right)$
can be derived by performing the analytic continuation $\chi_{\hat{\rho}\hat{\rho}}^{\mathrm{r}}\left(\mathbf{\mathbf{Q}},\omega\right)=\chi_{\hat{\rho}\hat{\rho}}\left(\mathbf{\mathbf{Q}},z\rightarrow\omega+i0^{+}\right)$,
which is presented as the sum of all possible diagrams that connect
the two density operators as follows,
\tikzset{ pattern size/.store in=\mcSize,  
pattern size = 5pt, 
pattern thickness/.store in=\mcThickness,  
pattern thickness = 0.3pt, 
pattern radius/.store in=\mcRadius,  
pattern radius = 1pt} 
\makeatletter 

\pgfutil@ifundefined{pgf@pattern@name@_2rn8z5b7p}{ 
\pgfdeclarepatternformonly[\mcThickness,\mcSize]{_2rn8z5b7p} {\pgfqpoint{0pt}{0pt}} {\pgfpoint{\mcSize}{\mcSize}} {\pgfpoint{\mcSize}{\mcSize}} { \pgfsetcolor{\tikz@pattern@color} \pgfsetlinewidth{\mcThickness} \pgfpathmoveto{\pgfqpoint{0pt}{\mcSize}} \pgfpathlineto{\pgfpoint{\mcSize+\mcThickness}{-\mcThickness}} \pgfpathmoveto{\pgfqpoint{0pt}{0pt}} \pgfpathlineto{\pgfpoint{\mcSize+\mcThickness}{\mcSize+\mcThickness}} \pgfusepath{stroke} }} 

\pgfutil@ifundefined{pgf@pattern@name@_5mf4knbal}{ 
\pgfdeclarepatternformonly[\mcThickness,\mcSize]{_5mf4knbal} {\pgfqpoint{0pt}{0pt}} {\pgfpoint{\mcSize+\mcThickness}{\mcSize+\mcThickness}} {\pgfpoint{\mcSize}{\mcSize}} { \pgfsetcolor{\tikz@pattern@color} \pgfsetlinewidth{\mcThickness} \pgfpathmoveto{\pgfqpoint{0pt}{0pt}} \pgfpathlineto{\pgfpoint{\mcSize+\mcThickness}{\mcSize+\mcThickness}} \pgfusepath{stroke} }}

\pgfutil@ifundefined{pgf@pattern@name@_ls0b50db1}{ 
\pgfdeclarepatternformonly[\mcThickness,\mcSize]{_ls0b50db1} {\pgfqpoint{0pt}{0pt}} {\pgfpoint{\mcSize+\mcThickness}{\mcSize+\mcThickness}} {\pgfpoint{\mcSize}{\mcSize}} { \pgfsetcolor{\tikz@pattern@color} \pgfsetlinewidth{\mcThickness} \pgfpathmoveto{\pgfqpoint{0pt}{0pt}} \pgfpathlineto{\pgfpoint{\mcSize+\mcThickness}{\mcSize+\mcThickness}} \pgfusepath{stroke} }}

\pgfutil@ifundefined{pgf@pattern@name@_4lbrki03p}{ 
\pgfdeclarepatternformonly[\mcThickness,\mcSize]{_4lbrki03p} {\pgfqpoint{0pt}{0pt}} {\pgfpoint{\mcSize+\mcThickness}{\mcSize+\mcThickness}} {\pgfpoint{\mcSize}{\mcSize}} { \pgfsetcolor{\tikz@pattern@color} \pgfsetlinewidth{\mcThickness} \pgfpathmoveto{\pgfqpoint{0pt}{0pt}} \pgfpathlineto{\pgfpoint{\mcSize+\mcThickness}{\mcSize+\mcThickness}} \pgfusepath{stroke} }}

\pgfutil@ifundefined{pgf@pattern@name@_fmz8jlkvv}{ 
\pgfdeclarepatternformonly[\mcThickness,\mcSize]{_fmz8jlkvv} {\pgfqpoint{0pt}{0pt}} {\pgfpoint{\mcSize+\mcThickness}{\mcSize+\mcThickness}} {\pgfpoint{\mcSize}{\mcSize}} { \pgfsetcolor{\tikz@pattern@color} \pgfsetlinewidth{\mcThickness} \pgfpathmoveto{\pgfqpoint{0pt}{0pt}} \pgfpathlineto{\pgfpoint{\mcSize+\mcThickness}{\mcSize+\mcThickness}} \pgfusepath{stroke} }}

\pgfutil@ifundefined{pgf@pattern@name@_j5kdpgcqs}{ 
\pgfdeclarepatternformonly[\mcThickness,\mcSize]{_j5kdpgcqs} {\pgfqpoint{0pt}{0pt}} {\pgfpoint{\mcSize+\mcThickness}{\mcSize+\mcThickness}} {\pgfpoint{\mcSize}{\mcSize}} { \pgfsetcolor{\tikz@pattern@color} \pgfsetlinewidth{\mcThickness} \pgfpathmoveto{\pgfqpoint{0pt}{0pt}} \pgfpathlineto{\pgfpoint{\mcSize+\mcThickness}{\mcSize+\mcThickness}} \pgfusepath{stroke} }}

\pgfutil@ifundefined{pgf@pattern@name@_bnpcfob31}{ 
\pgfdeclarepatternformonly[\mcThickness,\mcSize]{_bnpcfob31} {\pgfqpoint{0pt}{0pt}} {\pgfpoint{\mcSize+\mcThickness}{\mcSize+\mcThickness}} {\pgfpoint{\mcSize}{\mcSize}} { \pgfsetcolor{\tikz@pattern@color} \pgfsetlinewidth{\mcThickness} \pgfpathmoveto{\pgfqpoint{0pt}{0pt}} \pgfpathlineto{\pgfpoint{\mcSize+\mcThickness}{\mcSize+\mcThickness}} \pgfusepath{stroke} }}

\pgfutil@ifundefined{pgf@pattern@name@_vmr81ywv3}{ 
\pgfdeclarepatternformonly[\mcThickness,\mcSize]{_vmr81ywv3} {\pgfqpoint{0pt}{0pt}} {\pgfpoint{\mcSize+\mcThickness}{\mcSize+\mcThickness}} {\pgfpoint{\mcSize}{\mcSize}} { \pgfsetcolor{\tikz@pattern@color} \pgfsetlinewidth{\mcThickness} \pgfpathmoveto{\pgfqpoint{0pt}{0pt}} \pgfpathlineto{\pgfpoint{\mcSize+\mcThickness}{\mcSize+\mcThickness}} \pgfusepath{stroke} }}

\pgfutil@ifundefined{pgf@pattern@name@_edo9a8k50}{ 
\pgfdeclarepatternformonly[\mcThickness,\mcSize]{_edo9a8k50} {\pgfqpoint{0pt}{0pt}} {\pgfpoint{\mcSize+\mcThickness}{\mcSize+\mcThickness}} {\pgfpoint{\mcSize}{\mcSize}} { \pgfsetcolor{\tikz@pattern@color} \pgfsetlinewidth{\mcThickness} \pgfpathmoveto{\pgfqpoint{0pt}{0pt}} \pgfpathlineto{\pgfpoint{\mcSize+\mcThickness}{\mcSize+\mcThickness}} \pgfusepath{stroke} }}

\makeatother \tikzset{every picture/.style={line width=0.75pt}} 

\begin{eqnarray}    
\label{sum_electron_diagrams} \chi_{\hat{\rho}\hat{\rho}}^{\mathrm{r}}~&=&\vcenter{\hbox{
\begin{tikzpicture}[x=0.45pt,y=0.45pt,yscale=-1,xscale=1] 
\draw  [pattern=_2rn8z5b7p,pattern size=5pt,pattern thickness=0.75pt,pattern radius=0pt, pattern color={rgb, 255:red, 0; green, 0; blue, 0}] (239.77,78.37) -- (333.61,78.37) -- (333.61,137) -- (239.77,137) -- cycle ; 
\draw   (239.77,78.37) -- (333.61,78.37) .. controls (345.8,78.37) and (355.68,91.5) .. (355.68,107.69) .. controls (355.68,123.88) and (345.8,137) .. (333.61,137) -- (239.77,137) .. controls (227.58,137) and (217.69,123.88) .. (217.69,107.69) .. controls (217.69,91.5) and (227.58,78.37) .. (239.77,78.37) -- cycle ; 
\draw  [fill={rgb, 255:red, 128; green, 128; blue, 128 }  ,fill opacity=1 ] (215.94,109.19) .. controls (215.94,110.02) and (216.73,110.69) .. (217.69,110.69) .. controls (218.66,110.69) and (219.44,110.02) .. (219.44,109.19) .. controls (219.44,108.36) and (218.66,107.69) .. (217.69,107.69) .. controls (216.73,107.69) and (215.94,108.36) .. (215.94,109.19) -- cycle ; 
\draw  [fill={rgb, 255:red, 128; green, 128; blue, 128 }  ,fill opacity=1 ] (353.93,106.19) .. controls (353.93,107.02) and (354.72,107.69) .. (355.68,107.69) .. controls (356.65,107.69) and (357.43,107.02) .. (357.43,106.19) .. controls (357.43,105.36) and (356.65,104.69) .. (355.68,104.69) .. controls (354.72,104.69) and (353.93,105.36) .. (353.93,106.19) -- cycle ;
\end{tikzpicture} 
}} \nonumber    \\ 
\nonumber    \\
~&=&\vcenter{\hbox{
\begin{tikzpicture}[x=0.40pt,y=0.40pt,yscale=-1,xscale=1] 
\draw  [pattern=_5mf4knbal,pattern size=4pt,pattern thickness=0.75pt,pattern radius=0pt, pattern color={rgb, 255:red, 0; green, 0; blue, 0}][line width=0.75]  (242.27,121) .. controls (242.27,103.88) and (256.54,90) .. (274.14,90) .. controls (291.73,90) and (306,103.88) .. (306,121) .. controls (306,138.12) and (291.73,152) .. (274.14,152) .. controls (256.54,152) and (242.27,138.12) .. (242.27,121) -- cycle ; 
\draw  [fill={rgb, 255:red, 128; green, 128; blue, 128 }  ,fill opacity=1 ] (306,121) .. controls (306,122.1) and (306.71,123) .. (307.6,123) .. controls (308.48,123) and (309.19,122.1) .. (309.19,121) .. controls (309.19,119.9) and (308.48,119) .. (307.6,119) .. controls (306.71,119) and (306,119.9) .. (306,121) -- cycle ; 
\draw  [fill={rgb, 255:red, 128; green, 128; blue, 128 }  ,fill opacity=1 ] (239.08,121) .. controls (239.08,122.1) and (239.79,123) .. (240.67,123) .. controls (241.56,123) and (242.27,122.1) .. (242.27,121) .. controls (242.27,119.9) and (241.56,119) .. (240.67,119) .. controls (239.79,119) and (239.08,119.9) .. (239.08,121) -- cycle ;
\end{tikzpicture}
}}+\vcenter{\hbox{
\begin{tikzpicture}[x=0.40pt,y=0.40pt,yscale=-1,xscale=1] 
\draw  [pattern=_ls0b50db1,pattern size=4pt,pattern thickness=0.75pt,pattern radius=0pt, pattern color={rgb, 255:red, 0; green, 0; blue, 0}][line width=0.75]  (243.27,132) .. controls (243.27,114.88) and (257.54,101) .. (275.14,101) .. controls (292.73,101) and (307,114.88) .. (307,132) .. controls (307,149.12) and (292.73,163) .. (275.14,163) .. controls (257.54,163) and (243.27,149.12) .. (243.27,132) -- cycle ; 
\draw  [fill={rgb, 255:red, 128; green, 128; blue, 128 }  ,fill opacity=1 ] (307,132) .. controls (307,133.1) and (307.71,134) .. (308.6,134) .. controls (309.48,134) and (310.19,133.1) .. (310.19,132) .. controls (310.19,130.9) and (309.48,130) .. (308.6,130) .. controls (307.71,130) and (307,130.9) .. (307,132) -- cycle ; 
\draw  [fill={rgb, 255:red, 128; green, 128; blue, 128 }  ,fill opacity=1 ] (240.08,132) .. controls (240.08,133.1) and (240.79,134) .. (241.67,134) .. controls (242.56,134) and (243.27,133.1) .. (243.27,132) .. controls (243.27,130.9) and (242.56,130) .. (241.67,130) .. controls (240.79,130) and (240.08,130.9) .. (240.08,132) -- cycle ;
\draw    (308.08,133) .. controls (309.75,131.33) and (311.41,131.33) .. (313.08,133) .. controls (314.75,134.67) and (316.41,134.67) .. (318.08,133) .. controls (319.75,131.33) and (321.41,131.33) .. (323.08,133) .. controls (324.75,134.67) and (326.41,134.67) .. (328.08,133) .. controls (329.75,131.33) and (331.41,131.33) .. (333.08,133) .. controls (334.75,134.67) and (336.41,134.67) .. (338.08,133) .. controls (339.75,131.33) and (341.41,131.33) .. (343.08,133) .. controls (344.75,134.67) and (346.41,134.67) .. (348.08,133) -- (349.08,133) -- (349.08,133) ; 
\draw  [pattern=_4lbrki03p,pattern size=4pt,pattern thickness=0.75pt,pattern radius=0pt, pattern color={rgb, 255:red, 0; green, 0; blue, 0}][line width=0.75]  (350.27,133) .. controls (350.27,115.88) and (364.54,102) .. (382.14,102) .. controls (399.73,102) and (414,115.88) .. (414,133) .. controls (414,150.12) and (399.73,164) .. (382.14,164) .. controls (364.54,164) and (350.27,150.12) .. (350.27,133) -- cycle ; 
\draw  [fill={rgb, 255:red, 128; green, 128; blue, 128 }  ,fill opacity=1 ] (414,133) .. controls (414,134.1) and (414.71,135) .. (415.6,135) .. controls (416.48,135) and (417.19,134.1) .. (417.19,133) .. controls (417.19,131.9) and (416.48,131) .. (415.6,131) .. controls (414.71,131) and (414,131.9) .. (414,133) -- cycle ; 
\draw  [fill={rgb, 255:red, 128; green, 128; blue, 128 }  ,fill opacity=1 ] (347.08,133) .. controls (347.08,134.1) and (347.79,135) .. (348.67,135) .. controls (349.56,135) and (350.27,134.1) .. (350.27,133) .. controls (350.27,131.9) and (349.56,131) .. (348.67,131) .. controls (347.79,131) and (347.08,131.9) .. (347.08,133) -- cycle ;
\end{tikzpicture}  
}}+\vcenter{\hbox{
\begin{tikzpicture}[x=0.40pt,y=0.40pt,yscale=-1,xscale=1] 
\draw  [pattern=_fmz8jlkvv,pattern size=4pt,pattern thickness=0.75pt,pattern radius=0pt, pattern color={rgb, 255:red, 0; green, 0; blue, 0}][line width=0.75]  (185.27,132) .. controls (185.27,114.88) and (199.54,101) .. (217.14,101) .. controls (234.73,101) and (249,114.88) .. (249,132) .. controls (249,149.12) and (234.73,163) .. (217.14,163) .. controls (199.54,163) and (185.27,149.12) .. (185.27,132) -- cycle ; 
\draw  [fill={rgb, 255:red, 128; green, 128; blue, 128 }  ,fill opacity=1 ] (249,132) .. controls (249,133.1) and (249.71,134) .. (250.6,134) .. controls (251.48,134) and (252.19,133.1) .. (252.19,132) .. controls (252.19,130.9) and (251.48,130) .. (250.6,130) .. controls (249.71,130) and (249,130.9) .. (249,132) -- cycle ; 
\draw  [fill={rgb, 255:red, 128; green, 128; blue, 128 }  ,fill opacity=1 ] (182.08,132) .. controls (182.08,133.1) and (182.79,134) .. (183.67,134) .. controls (184.56,134) and (185.27,133.1) .. (185.27,132) .. controls (185.27,130.9) and (184.56,130) .. (183.67,130) .. controls (182.79,130) and (182.08,130.9) .. (182.08,132) -- cycle ;
\draw    (249.08,132) .. controls (250.75,130.33) and (252.41,130.33) .. (254.08,132) .. controls (255.75,133.67) and (257.41,133.67) .. (259.08,132) .. controls (260.75,130.33) and (262.41,130.33) .. (264.08,132) .. controls (265.75,133.67) and (267.41,133.67) .. (269.08,132) .. controls (270.75,130.33) and (272.41,130.33) .. (274.08,132) .. controls (275.75,133.67) and (277.41,133.67) .. (279.08,132) .. controls (280.75,130.33) and (282.41,130.33) .. (284.08,132) .. controls (285.75,133.67) and (287.41,133.67) .. (289.08,132) -- (290.08,132) -- (290.08,132) ; 
\draw  [pattern=_j5kdpgcqs,pattern size=4pt,pattern thickness=0.75pt,pattern radius=0pt, pattern color={rgb, 255:red, 0; green, 0; blue, 0}][line width=0.75]  (293.27,132) .. controls (293.27,114.88) and (307.54,101) .. (325.14,101) .. controls (342.73,101) and (357,114.88) .. (357,132) .. controls (357,149.12) and (342.73,163) .. (325.14,163) .. controls (307.54,163) and (293.27,149.12) .. (293.27,132) -- cycle ; 
\draw  [fill={rgb, 255:red, 128; green, 128; blue, 128 }  ,fill opacity=1 ] (357,132) .. controls (357,133.1) and (357.71,134) .. (358.6,134) .. controls (359.48,134) and (360.19,133.1) .. (360.19,132) .. controls (360.19,130.9) and (359.48,130) .. (358.6,130) .. controls (357.71,130) and (357,130.9) .. (357,132) -- cycle ; 
\draw  [fill={rgb, 255:red, 128; green, 128; blue, 128 }  ,fill opacity=1 ] (290.08,132) .. controls (290.08,133.1) and (290.79,134) .. (291.67,134) .. controls (292.56,134) and (293.27,133.1) .. (293.27,132) .. controls (293.27,130.9) and (292.56,130) .. (291.67,130) .. controls (290.79,130) and (290.08,130.9) .. (290.08,132) -- cycle ;
\draw    (360.19,132) .. controls (361.86,130.33) and (363.52,130.33) .. (365.19,132) .. controls (366.86,133.67) and (368.52,133.67) .. (370.19,132) .. controls (371.86,130.33) and (373.52,130.33) .. (375.19,132) .. controls (376.86,133.67) and (378.52,133.67) .. (380.19,132) .. controls (381.86,130.33) and (383.52,130.33) .. (385.19,132) .. controls (386.86,133.67) and (388.52,133.67) .. (390.19,132) .. controls (391.86,130.33) and (393.52,130.33) .. (395.19,132) .. controls (396.86,133.67) and (398.52,133.67) .. (400.19,132) -- (401.19,132) -- (401.19,132) ; 
\draw  [pattern=_bnpcfob31,pattern size=4pt,pattern thickness=0.75pt,pattern radius=0pt, pattern color={rgb, 255:red, 0; green, 0; blue, 0}][line width=0.75]  (403.27,132) .. controls (403.27,114.88) and (417.54,101) .. (435.14,101) .. controls (452.73,101) and (467,114.88) .. (467,132) .. controls (467,149.12) and (452.73,163) .. (435.14,163) .. controls (417.54,163) and (403.27,149.12) .. (403.27,132) -- cycle ; 
\draw  [fill={rgb, 255:red, 128; green, 128; blue, 128 }  ,fill opacity=1 ] (467,132) .. controls (467,133.1) and (467.71,134) .. (468.6,134) .. controls (469.48,134) and (470.19,133.1) .. (470.19,132) .. controls (470.19,130.9) and (469.48,130) .. (468.6,130) .. controls (467.71,130) and (467,130.9) .. (467,132) -- cycle ; 
\draw  [fill={rgb, 255:red, 128; green, 128; blue, 128 }  ,fill opacity=1 ] (400.08,132) .. controls (400.08,133.1) and (400.79,134) .. (401.67,134) .. controls (402.56,134) and (403.27,133.1) .. (403.27,132) .. controls (403.27,130.9) and (402.56,130) .. (401.67,130) .. controls (400.79,130) and (400.08,130.9) .. (400.08,132) -- cycle ;
\end{tikzpicture} 
}}+\cdots\nonumber    \\ 
\nonumber    \\ 
~&=&\frac{\vcenter{\hbox{
\begin{tikzpicture}[x=0.40pt,y=0.40pt,yscale=-1,xscale=1] 
\draw  [pattern=_edo9a8k50,pattern size=4pt,pattern thickness=0.75pt,pattern radius=0pt, pattern color={rgb, 255:red, 0; green, 0; blue, 0}][line width=0.75]  (303.27,131) .. controls (303.27,113.88) and (317.54,100) .. (335.14,100) .. controls (352.73,100) and (367,113.88) .. (367,131) .. controls (367,148.12) and (352.73,162) .. (335.14,162) .. controls (317.54,162) and (303.27,148.12) .. (303.27,131) -- cycle ; 
\draw  [fill={rgb, 255:red, 128; green, 128; blue, 128 }  ,fill opacity=1 ] (367,131) .. controls (367,132.1) and (367.71,133) .. (368.6,133) .. controls (369.48,133) and (370.19,132.1) .. (370.19,131) .. controls (370.19,129.9) and (369.48,129) .. (368.6,129) .. controls (367.71,129) and (367,129.9) .. (367,131) -- cycle ; 
\draw  [fill={rgb, 255:red, 128; green, 128; blue, 128 }  ,fill opacity=1 ] (300.08,131) .. controls (300.08,132.1) and (300.79,133) .. (301.67,133) .. controls (302.56,133) and (303.27,132.1) .. (303.27,131) .. controls (303.27,129.9) and (302.56,129) .. (301.67,129) .. controls (300.79,129) and (300.08,129.9) .. (300.08,131) -- cycle ;
\end{tikzpicture}
}}}{1-\vcenter{\hbox{
\begin{tikzpicture}[x=0.40pt,y=0.40pt,yscale=-1,xscale=1] 
\draw    (260.19,131) .. controls (261.86,129.33) and (263.52,129.33) .. (265.19,131) .. controls (266.86,132.67) and (268.52,132.67) .. (270.19,131) .. controls (271.86,129.33) and (273.52,129.33) .. (275.19,131) .. controls (276.86,132.67) and (278.52,132.67) .. (280.19,131) .. controls (281.86,129.33) and (283.52,129.33) .. (285.19,131) .. controls (286.86,132.67) and (288.52,132.67) .. (290.19,131) .. controls (291.86,129.33) and (293.52,129.33) .. (295.19,131) .. controls (296.86,132.67) and (298.52,132.67) .. (300.19,131) -- (301.19,131) -- (301.19,131) ; 
\draw  [pattern=_vmr81ywv3,pattern size=4pt,pattern thickness=0.75pt,pattern radius=0pt, pattern color={rgb, 255:red, 0; green, 0; blue, 0}][line width=0.75]  (303.27,131) .. controls (303.27,113.88) and (317.54,100) .. (335.14,100) .. controls (352.73,100) and (367,113.88) .. (367,131) .. controls (367,148.12) and (352.73,162) .. (335.14,162) .. controls (317.54,162) and (303.27,148.12) .. (303.27,131) -- cycle ; 
\draw  [fill={rgb, 255:red, 128; green, 128; blue, 128 }  ,fill opacity=1 ] (367,131) .. controls (367,132.1) and (367.71,133) .. (368.6,133) .. controls (369.48,133) and (370.19,132.1) .. (370.19,131) .. controls (370.19,129.9) and (369.48,129) .. (368.6,129) .. controls (367.71,129) and (367,129.9) .. (367,131) -- cycle ; 
\draw  [fill={rgb, 255:red, 128; green, 128; blue, 128 }  ,fill opacity=1 ] (300.08,131) .. controls (300.08,132.1) and (300.79,133) .. (301.67,133) .. controls (302.56,133) and (303.27,132.1) .. (303.27,131) .. controls (303.27,129.9) and (302.56,129) .. (301.67,129) .. controls (300.79,129) and (300.08,129.9) .. (300.08,131) -- cycle ;
\end{tikzpicture}
}}}\nonumber    \\
~&=&\frac{\Pi_{e}}{1-V_{e}\,\Pi_{e}}, 
\end{eqnarray}where \begin{tikzpicture}[x=0.40pt,y=0.40pt,yscale=-1,xscale=1] 
\draw    (262,99) .. controls (263.69,97.37) and (265.36,97.4) .. (267,99.09) .. controls (268.64,100.78) and (270.31,100.81) .. (272,99.18) .. controls (273.69,97.55) and (275.36,97.58) .. (277,99.27) .. controls (278.64,100.96) and (280.31,100.99) .. (282,99.36) .. controls (283.69,97.73) and (285.36,97.76) .. (287,99.45) .. controls (288.63,101.15) and (290.3,101.18) .. (292,99.55) .. controls (293.69,97.92) and (295.36,97.95) .. (296.99,99.64) .. controls (298.63,101.33) and (300.3,101.36) .. (301.99,99.73) .. controls (303.68,98.1) and (305.35,98.13) .. (306.99,99.82) .. controls (308.63,101.51) and (310.3,101.54) .. (311.99,99.91) .. controls (313.68,98.28) and (315.35,98.31) .. (316.99,100) -- (317,100) -- (317,100) ;
\end{tikzpicture} represents the electron Coulomb interaction $V_{e}\left(Q\right)$,
and the one-particle-irreducible~(1PI)~(hatched circle) is denoted
as a function $\Pi_{e}\left(\mathbf{Q},\,\omega\right)$, which can
be approximated as the electron pair-bubble diagram in the \textit{random
phase approximation}~(RPA), 
\tikzset{ pattern size/.store in=\mcSize,  pattern size = 5pt, pattern thickness/.store in=\mcThickness,  pattern thickness = 0.3pt, pattern radius/.store in=\mcRadius,  pattern radius = 1pt} \makeatletter \pgfutil@ifundefined{pgf@pattern@name@_wmea26d41}{ 
\pgfdeclarepatternformonly[\mcThickness,\mcSize]{_wmea26d41} {\pgfqpoint{0pt}{0pt}} {\pgfpoint{\mcSize+\mcThickness}{\mcSize+\mcThickness}} {\pgfpoint{\mcSize}{\mcSize}} { \pgfsetcolor{\tikz@pattern@color} \pgfsetlinewidth{\mcThickness} \pgfpathmoveto{\pgfqpoint{0pt}{0pt}} \pgfpathlineto{\pgfpoint{\mcSize+\mcThickness}{\mcSize+\mcThickness}} \pgfusepath{stroke} }} 
\makeatother 
\tikzset{every picture/.style={line width=0.40pt}} 

\begin{eqnarray}    
\label{electron_bubble} 
\vcenter{\hbox{
\begin{tikzpicture}[x=0.40pt,y=0.40pt,yscale=-1,xscale=1] 
\draw  [pattern=_wmea26d41,pattern size=4pt,pattern thickness=0.75pt,pattern radius=0pt, pattern color={rgb, 255:red, 0; green, 0; blue, 0}][line width=0.75]  (286.27,114) .. controls (286.27,96.88) and (300.54,83) .. (318.14,83) .. controls (335.73,83) and (350,96.88) .. (350,114) .. controls (350,131.12) and (335.73,145) .. (318.14,145) .. controls (300.54,145) and (286.27,131.12) .. (286.27,114) -- cycle ; 
\draw  [fill={rgb, 255:red, 128; green, 128; blue, 128 }  ,fill opacity=1 ] (350,114) .. controls (350,115.1) and (350.71,116) .. (351.6,116) .. controls (352.48,116) and (353.19,115.1) .. (353.19,114) .. controls (353.19,112.9) and (352.48,112) .. (351.6,112) .. controls (350.71,112) and (350,112.9) .. (350,114) -- cycle ; 
\draw  [fill={rgb, 255:red, 128; green, 128; blue, 128 }  ,fill opacity=1 ] (283.08,114) .. controls (283.08,115.1) and (283.79,116) .. (284.67,116) .. controls (285.56,116) and (286.27,115.1) .. (286.27,114) .. controls (286.27,112.9) and (285.56,112) .. (284.67,112) .. controls (283.79,112) and (283.08,112.9) .. (283.08,114) -- cycle ;
\end{tikzpicture}
}}~&\simeq&\vcenter{\hbox{
\begin{tikzpicture}[x=0.40pt,y=0.40pt,yscale=-1,xscale=1] 
\draw  [draw opacity=0] (329.61,138.27) .. controls (327.48,138.75) and (325.27,139) .. (323,139) .. controls (306.43,139) and (293,125.57) .. (293,109) .. controls (293,93.47) and (304.79,80.7) .. (319.91,79.16) -- (323,109) -- cycle ; 
\draw[line width=1.1]    (326.62,138.78) .. controls (325.43,138.93) and (324.23,139) .. (323,139) .. controls (306.43,139) and (293,125.57) .. (293,109) .. controls (293,93.13) and (305.33,80.13) .. (320.93,79.07) ; 
\draw [shift={(318.11,79.4)}, rotate = 354.16] [fill={rgb, 255:red, 0; green, 0; blue, 0 }  ][line width=0.08]  [draw opacity=0] (10.72,-5.15) -- (0,0) -- (10.72,5.15) -- (7.12,0) -- cycle    ; 
\draw [shift={(329.61,138.27)}, rotate = 167.33] [fill={rgb, 255:red, 0; green, 0; blue, 0 }  ][line width=0.08]  [draw opacity=0] (10.72,-5.15) -- (0,0) -- (10.72,5.15) -- (7.12,0) -- cycle    ; 
\draw  [draw opacity=0] (324.87,79.06) .. controls (340.57,80.02) and (353,93.06) .. (353,109) .. controls (353,125.26) and (340.06,138.5) .. (323.91,138.99) -- (323,109) -- cycle ; 
\draw[line width=1.1]   (324.87,79.06) .. controls (340.57,80.02) and (353,93.06) .. (353,109) .. controls (353,125.26) and (340.06,138.5) .. (323.91,138.99) ;   
\draw  [fill={rgb, 255:red, 128; green, 128; blue, 128 }  ,fill opacity=1 ] (290.4,110) .. controls (290.4,111.1) and (291.12,112) .. (292,112) .. controls (292.88,112) and (293.6,111.1) .. (293.6,110) .. controls (293.6,108.9) and (292.88,108) .. (292,108) .. controls (291.12,108) and (290.4,108.9) .. (290.4,110) -- cycle ; 
\draw  [fill={rgb, 255:red, 128; green, 128; blue, 128 }  ,fill opacity=1 ] (353.4,110) .. controls (353.4,111.1) and (354.12,112) .. (355,112) .. controls (355.88,112) and (356.6,111.1) .. (356.6,110) .. controls (356.6,108.9) and (355.88,108) .. (355,108) .. controls (354.12,108) and (353.4,108.9) .. (353.4,110) -- cycle ;
\end{tikzpicture} 
}}\nonumber    \\ 
\nonumber    \\ 
~&=&\frac{1}{V}\sum_{i,j}\frac{\left|\braket{i|e^{i\mathbf{\mathbf{Q}}\cdot\hat{\mathbf{x}}}|j}\right|^{2}}{\varepsilon_{i}-\varepsilon_{j}-\omega-i0^{+}}\left(f_{i}-f_{j}\right),
\end{eqnarray}
      where $f_{i}$~($f_{j}$) denotes the occupation number of the state
$\ket{i}$~($\ket{j}$), and $\left(\omega\,,\mathbf{Q}\right)$
is the 4-momentum going through the 1PI. Thus after a little algebra
the dielectric function can be expressed in terms of the function
$\Pi_{e}\left(\mathbf{Q},\,\omega\right)$:
\begin{eqnarray}
\epsilon\left(\mathbf{Q},\,\omega\right) & = & 1-V_{e}\left(Q\right)\Pi_{e}\left(\mathbf{Q},\,\omega\right)\nonumber \\
 & \simeq & 1-\frac{V_{e}\left(Q\right)}{V}\sum_{i,j}\frac{\left|\braket{i|e^{i\mathbf{\mathbf{Q}}\cdot\hat{\mathbf{x}}}|j}\right|^{2}}{\varepsilon_{i}-\varepsilon_{j}-\omega-i0^{+}}\left(f_{i}-f_{j}\right).\label{eq:RPA-dielectric-fucnition0}
\end{eqnarray}
In the Sun, the Fermi-Dirac distribution of the electron gas can be
approximated as the classical Boltzmann distribution, and hence $\Pi_{e}$
can be semi-analytically written as~\citep{fetter2012quantum} 
\begin{eqnarray}
\Pi_{e}\left(Q,\,\omega\right) & = & -\frac{n_{e}}{Q}\sqrt{\frac{m_{e}}{2\,T}}\left\{ \Phi\left[\sqrt{\frac{m_{e}}{2\,T}}\left(\frac{\omega}{Q}+\frac{Q}{2\,m_{e}}\right)\right]-\Phi\left[\sqrt{\frac{m_{e}}{2\,T}}\left(\frac{\omega}{Q}-\frac{Q}{2\,m_{e}}\right)\right]\right\} \nonumber \\
\nonumber \\
 &  & -i\,n_{e}\sqrt{\frac{2\pi}{m_{e}T}}\left(\frac{m_{e}}{Q}\right)\exp\left[-\left(\frac{m_{e}^{2}\,\omega^{2}}{Q^{2}}+\frac{Q^{2}}{4}\right)\frac{1}{2\,m_{e}T}\right]\sinh\left[\frac{\omega}{2\,T}\right],\label{eq:polarizability}
\end{eqnarray}
where $n_{e}$ is the number density of the electron gas, and this
expression is dubbed as the plasma dispersion function~(see Appendix~\ref{sec:appendix1}
for the definition of function $\Phi\left(x\right)$ and a detailed
derivation). It is evident that the dielectric function $\epsilon\left(\mathbf{\mathbf{Q}},\omega\right)$
is isotropic~(so one has $\epsilon\left(\mathbf{\mathbf{Q}},\omega\right)=\epsilon\left(Q,\omega\right)$
and $\chi_{\hat{\rho}\hat{\rho}}^{\mathrm{r}}\left(\mathbf{Q},\,\omega\right)=\chi_{\hat{\rho}\hat{\rho}}^{\mathrm{r}}\left(Q,\,\omega\right)$),
and thus the scattering rate in electron gas, namely, the probability
per unit time for a DM particle with velocity $\mathbf{v}_{\chi}$
undergoing a scattering, is expressed as 
\begin{eqnarray}
\lambda\left(v_{\chi}\right) & = & \frac{v_{\chi}\sigma\left(\mathbf{v}_{\chi}\right)}{V}\nonumber \\
 & = & \int\mathrm{d}\omega\int\frac{\mathrm{d}^{3}Q}{\left(2\pi\right)^{3}}\frac{\pi\bar{\sigma}_{e}}{\mu_{\chi e}^{2}}\left(\frac{\alpha^{2}m_{e}^{2}+m_{A'}^{2}}{Q^{2}+m_{A'}^{2}}\right)^{2}\frac{\left(-2\right)\,\mathrm{Im}\left[\chi_{\hat{\rho}\hat{\rho}}^{\mathrm{r}}\left(Q,\,\omega\right)\right]}{1-e^{-\beta\omega}}\,\delta\left(\frac{Q^{2}}{2\,m_{\chi}}-\mathbf{v}_{\chi}\cdot\mathbf{Q}+\omega\right)\nonumber \\
\nonumber \\
 & = & \int\mathrm{d}\omega\int\frac{Q\mathrm{d}Q}{\left(2\pi\right)^{2}}\frac{\pi\bar{\sigma}_{e}}{\mu_{\chi e}^{2}v_{\chi}}\left(\frac{\alpha^{2}m_{e}^{2}+m_{A'}^{2}}{Q^{2}+m_{A'}^{2}}\right)^{2}\frac{\left(-2\right)\,\mathrm{Im}\left[\chi_{\hat{\rho}\hat{\rho}}^{\mathrm{r}}\left(Q,\,\omega\right)\right]}{1-e^{-\beta\omega}}\,\Theta\left[v_{\chi}-v_{\mathrm{min}}\left(Q,\,\omega\right)\right]\,\Theta\left[v_{\chi}+v_{\mathrm{min}}\left(Q,\,\omega\right)\right],\nonumber \\
\label{eq:rate_NR}
\end{eqnarray}
or an equivalent form written in terms of the dielectric function,
\begin{eqnarray}
\lambda\left(v_{\chi}\right) & = & \int\mathrm{d}\omega\int\frac{Q^{3}\,\mathrm{d}Q}{\left(2\pi\right)^{2}}\frac{\bar{\sigma}_{e}}{2\,\alpha\,\mu_{\chi e}^{2}v_{\chi}}\left(\frac{\alpha^{2}m_{e}^{2}+m_{A'}^{2}}{Q^{2}+m_{A'}^{2}}\right)^{2}\,\frac{\mathrm{Im}\left[-\epsilon^{-1}\left(Q,\omega\right)\right]}{1-e^{-\omega/T}}\,\Theta\left[v_{\chi}-v_{\mathrm{min}}\left(Q,\,\omega\right)\right]\,\Theta\left[v_{\chi}+v_{\mathrm{min}}\left(Q,\,\omega\right)\right],\nonumber \\
\label{eq:rate_NR_epsilon}
\end{eqnarray}
where the two Heaviside step functions $\Theta$ represent the kinematical
requirement in the scattering, with 
\begin{eqnarray}
v_{\mathrm{min}}\left(Q,\,\omega\right) & = & \frac{Q}{2\,m_{\chi}}+\frac{\omega}{Q}.
\end{eqnarray}
In the heavy~(light) mediator limit, \textit{i.e.}, $m_{A'}\gg Q$~$\left(m_{A'}\ll Q\right)$,
the factor $\left(\alpha^{2}m_{e}^{2}+m_{A'}^{2}\right)/\left(Q^{2}+m_{A'}^{2}\right)$
in Eq.~(\ref{eq:rate_NR}) reduces to $1$~($\alpha^{2}m_{e}^{2}/Q^{2}$). 

It should be noted that the above discussion is based on the linear
response theory, where the in-medium effect is encoded in the ELF
$\mathrm{Im}\left[-\epsilon^{-1}\left(Q,\omega\right)\right]$, which
however, was omitted in early literatures~\citep{An:2017ojc,Emken:2021lgc,Liang:2021zkg}.
In order to demonstrate the in-medium effect due to the solar electrons,
we need to compare the scattering rates obtained from the linear response
theory and from the scattering theory. The latter (in the heavy mediator
scenario) is formulated as~\citep{Liang:2021zkg}
\begin{eqnarray}
\lambda\left(v_{\chi}\right) & = & n_{e}\left\langle \bar{\sigma}_{e}\cdot\left|\mathbf{v}_{\chi}-\mathbf{u}_{e}\right|\right\rangle \nonumber \\
 & = & n_{e}\,\bar{\sigma}_{e}\left[\frac{u_{0}}{\sqrt{\pi}}\,\exp\left(-v_{\chi}^{2}/u_{0}^{2}\right)+\left(v_{\chi}+\frac{u_{0}^{2}}{2\,v_{\chi}}\right)\mathrm{erf}\left(\frac{v_{\chi}}{u_{0}}\right)\right],\label{eq:lambda2}
\end{eqnarray}
which is independent of the DM mass, where $\left\langle \cdots\right\rangle $
denotes the average over the relative velocity $\mathbf{v}_{\chi}-\mathbf{u}_{e}$
between the DM particle and surrounding electrons that follow the
Maxwellian distribution 
\begin{eqnarray}
f_{e}\left(\mathbf{u}_{e}\right) & = & \left(\sqrt{\pi}u_{0}\right)^{-3}\exp\left(-\frac{u_{e}^{2}}{u_{0}^{2}}\right),
\end{eqnarray}
with $u_{0}=$ $\sqrt{2\,T/m_{e}}$\textcolor{black}{{} the electron
mean squared velocity.}
\begin{figure}
\begin{centering}
\includegraphics[scale=0.72]{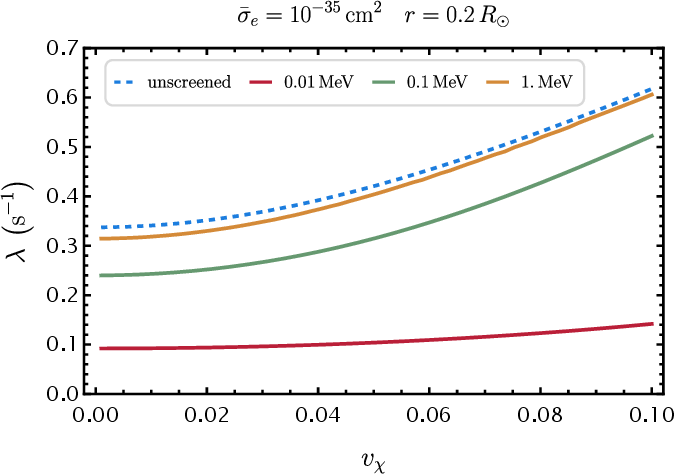}\hspace{1cc}\includegraphics[scale=0.72]{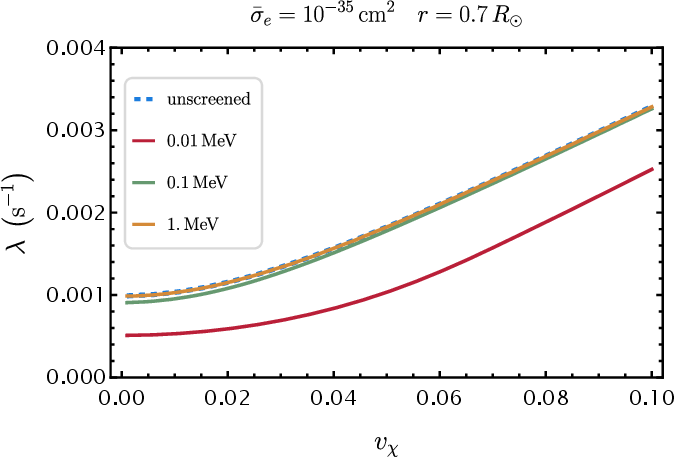}
\par\end{centering}
\caption{\label{fig:lamda_plot} Comparison between the scattering rates calculated
from with~(\textbf{\textit{solid}}) and without~(\textbf{\textit{dashed}})
the in-medium effect for velocities ranging from $10^{-3}$ to $1\times10^{-1}$
at $r=0.2\,R_{\odot}$ (\textbf{\textsl{left}}) and $r=0.7\,R_{\odot}$
(\textbf{\textit{right}}), respectively, with benchmark cross section
$\bar{\sigma}_{e}=10^{-35}\,\mathrm{cm}^{2}$ and DM masses $m_{\chi}=0.01\,\mathrm{MeV}$~(\textbf{\textit{red}}),
$m_{\chi}=0.1\,\mathrm{MeV}$~(\textbf{\textit{green}}), and $m_{\chi}=1\,\mathrm{MeV}$~(\textbf{\textit{orange}}).}
\end{figure}

To get a sense of the relevant physics, \textcolor{black}{in Fig.~\ref{fig:lamda_plot}
we compare the scattering rates calculated with and without the in-medium
effect for a DM particle with masses }$m_{\chi}=0.01\,\mathrm{MeV}$
$0.1\,\mathrm{MeV}$ and $1\,\mathrm{MeV}$,\textcolor{black}{{} traveling}
through and scattering with the solar electron gas with a benchmark
coupling strength $\bar{\sigma}_{e}=10^{-35}\,\mathrm{cm}^{2}$ in
the heavy mediator limi\textcolor{black}{t}, \textcolor{black}{at
distances} $r=0.2\,R_{\odot}$ and $0.7\,R_{\odot}$ from the solar
center~\textcolor{black}{($R_{\odot}$ being the radius of the Sun)},
respectively. The number density of the ionized electrons $n_{e}$
is determined by the condition of charge neutrality~\citep{Emken:2021lgc}
with the Standard Sun Model~AGSS09~\citep{Serenelli:2009yc}. In
practice, we take only hydrogen and helium atoms in our discussion.
Particularly, the electron number density and temperature at $0.2\,R_{\odot}$~($0.7\,R_{\odot}$)
is $1.8\times10^{25}\,\mathrm{cm}^{-3}$~($1.1\times10^{23}\,\mathrm{cm}^{-3}$)
and $9.3\times10^{6}\,\mathrm{K}$~($2.3\times10^{6}\,\mathrm{K}$),
respectively. In turns out that the in-medium effect brings an $\mathcal{O}\left(1\right)$~($\mathcal{O}\left(0.1\right)$)
suppression to the scattering rate in the hotter core of the Sun for
a DM mass $m_{\chi}=0.01\,\mathrm{MeV}$~($0.1\,\mathrm{MeV}$ and
$1\,\mathrm{MeV}$), while in the cooler outer region, the in-medium
effect turns less significant. 

\begin{figure}
\begin{centering}
\includegraphics[scale=0.75]{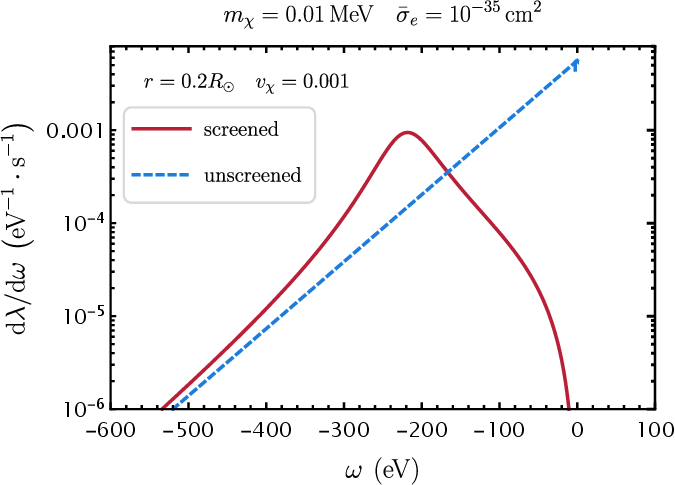}\hspace{1cc}\includegraphics[scale=0.75]{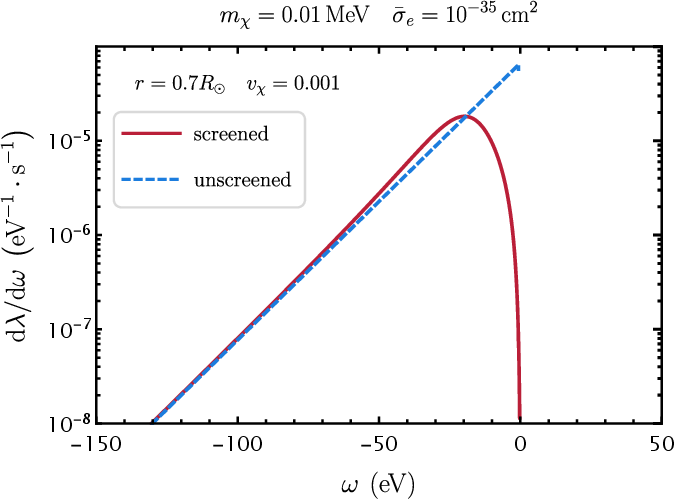}\vspace{0.5cm}
\includegraphics[scale=0.75]{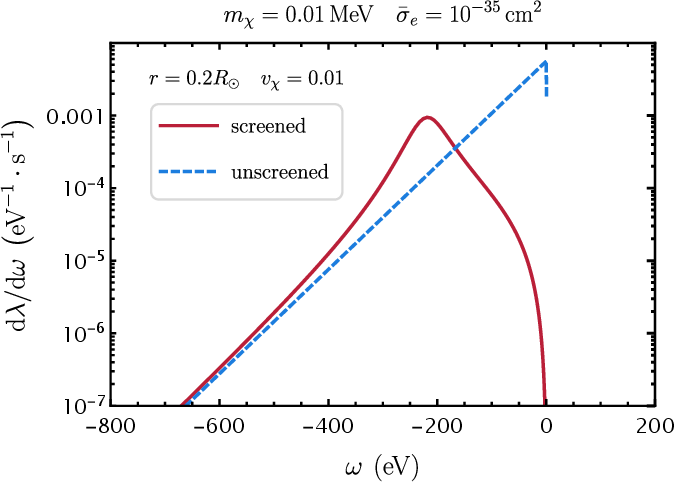}\hspace{1cc}\includegraphics[scale=0.75]{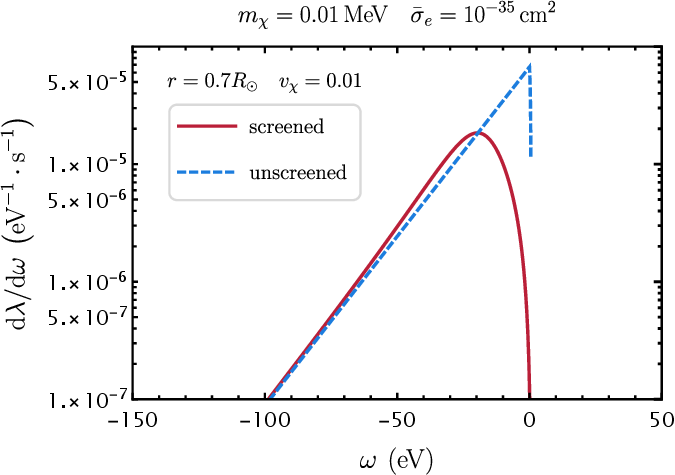}\vspace{0.5cm}
\par\end{centering}
\begin{centering}
\includegraphics[scale=0.75]{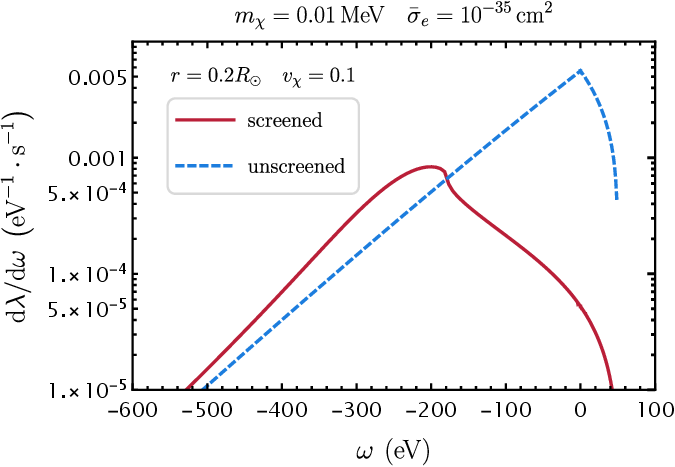}\hspace{1cc}\includegraphics[scale=0.75]{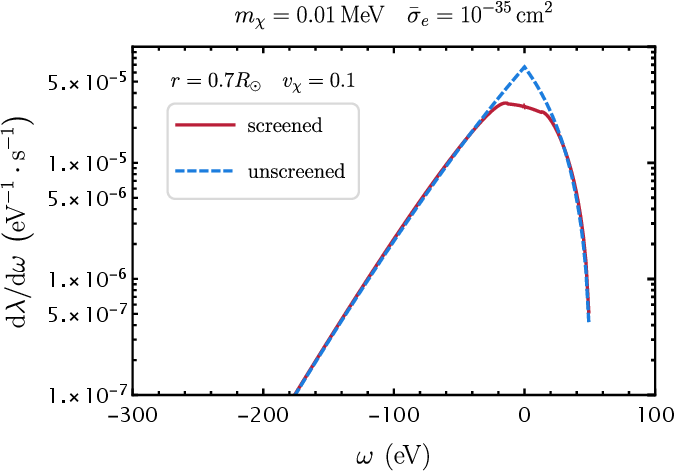}
\par\end{centering}
\textcolor{black}{\caption{\label{fig:differentialScatteringRate-2}The differential scattering
rates for a $0.01\,\mathrm{MeV}$ DM particle with a benchmark cross
section $\bar{\sigma}_{e}=10^{-35}\,\mathrm{cm}^{2}$ in the heavy
mediator limit for various velocities $v_{\chi}=1\times10^{-3}$~(\textbf{\textit{top}}),
$1\times10^{-2}$~(\textbf{\textit{middle}}) $1\times10^{-1}$~(\textbf{\textit{bottom}}),
respectively, at the location $r=0.2\,R_{\odot}$ ~(\textbf{\textit{left}})
and $r=0.7\,R_{\odot}$ ~(\textbf{\textit{right}}) inside the Sun,
with~(\textbf{\textit{red solid}}) and without~(\textbf{\textit{blue
dashed}}) the in-medium effect.}
}
\end{figure}

\begin{figure}
\begin{centering}
\includegraphics[scale=0.75]{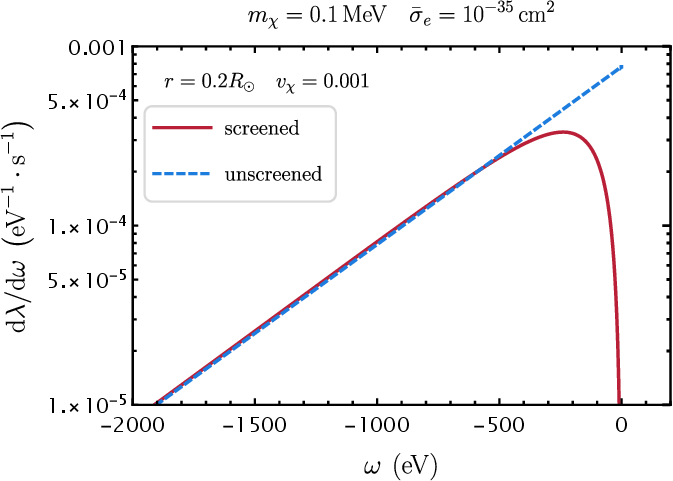}\hspace{1cc}\includegraphics[scale=0.75]{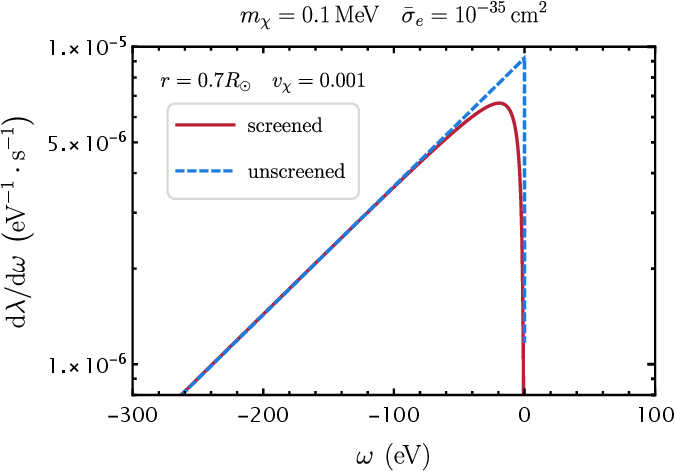}\vspace{0.5cm}
\includegraphics[scale=0.75]{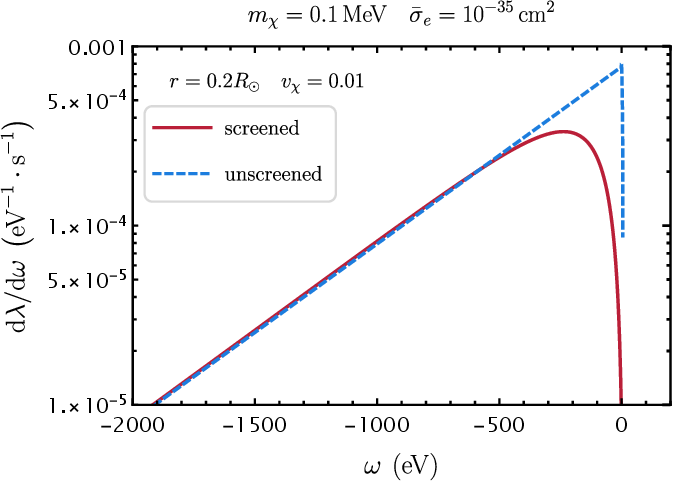}\hspace{1cc}\includegraphics[scale=0.75]{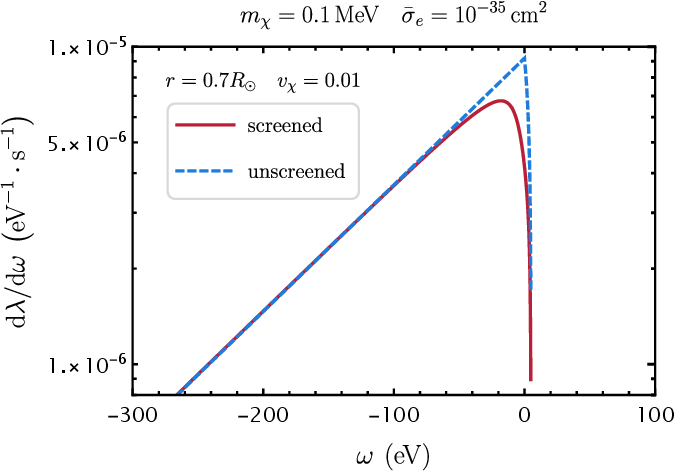}\vspace{0.5cm}
\par\end{centering}
\begin{centering}
\includegraphics[scale=0.75]{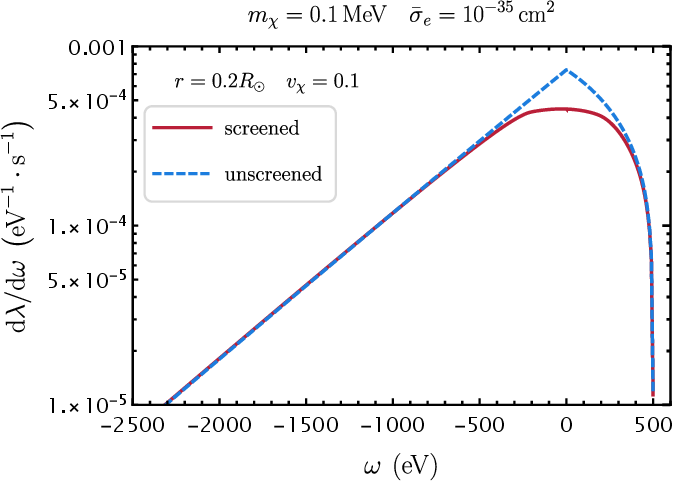}\hspace{1cc}\includegraphics[scale=0.75]{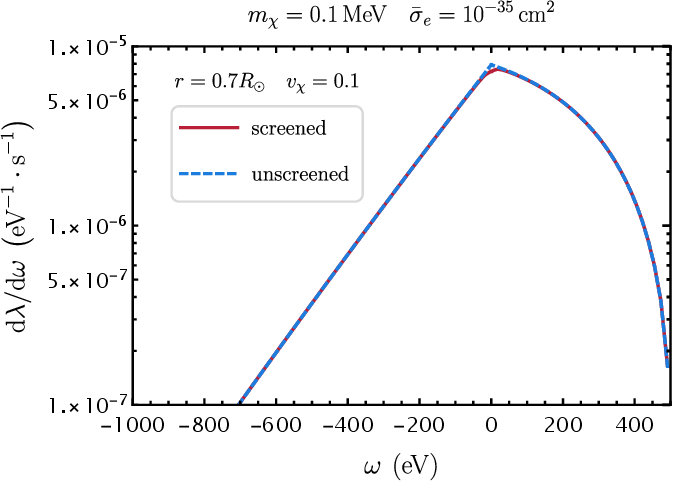}
\par\end{centering}
\textcolor{black}{\caption{\label{fig:differentialScatteringRate-1}The differential scattering
rates for a $0.1\,\mathrm{MeV}$ DM particle with a benchmark cross
section $\bar{\sigma}_{e}=10^{-35}\,\mathrm{cm}^{2}$ in the heavy
mediator limit for velocities $v_{\chi}=1\times10^{-3}$~(\textbf{\textit{top}}),
$1\times10^{-2}$~(\textbf{\textit{middle}}) $1\times10^{-1}$~(\textbf{\textit{bottom}}),
respectively, at the location $r=0.2\,R_{\odot}$ ~(\textbf{\textit{left}})
and $r=0.7\,R_{\odot}$ ~(\textbf{\textit{right}}) inside the Sun,
with~(\textbf{\textit{red solid}}) and without~(\textbf{\textit{blue
dashed}}) the in-medium effect.}
}
\end{figure}

\begin{figure}
\begin{centering}
\includegraphics[scale=0.75]{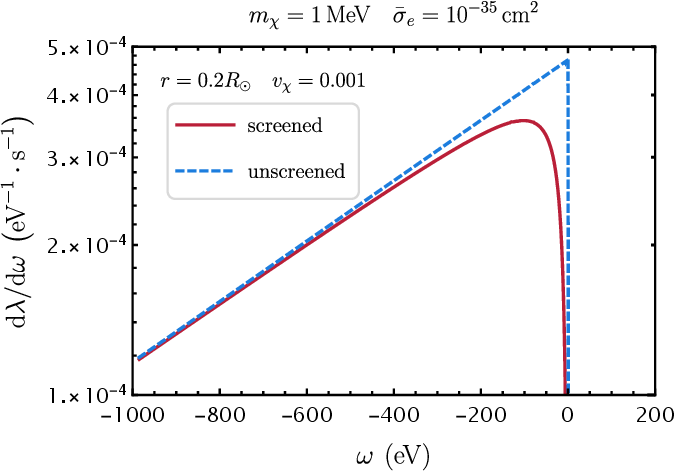}\hspace{1cc}\includegraphics[scale=0.75]{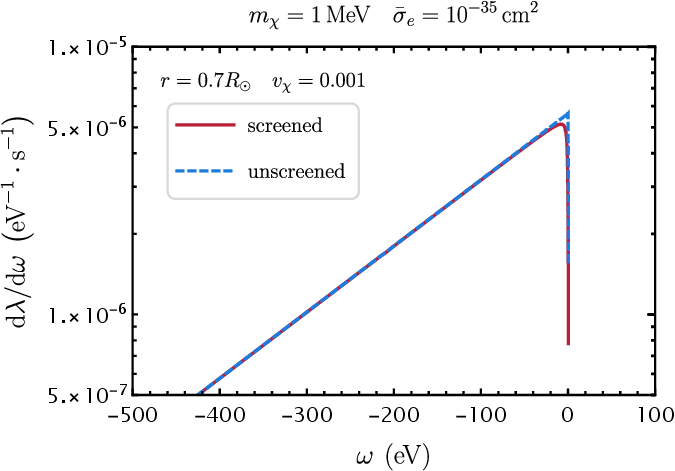}\vspace{0.5cm}
\par\end{centering}
\begin{centering}
\includegraphics[scale=0.75]{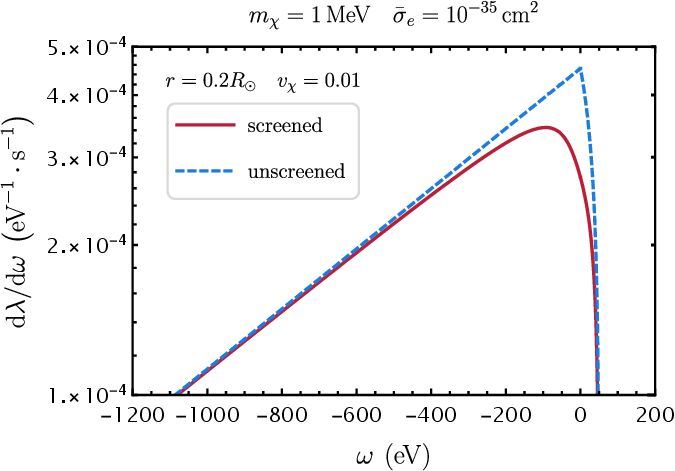}\hspace{1cc}\includegraphics[scale=0.75]{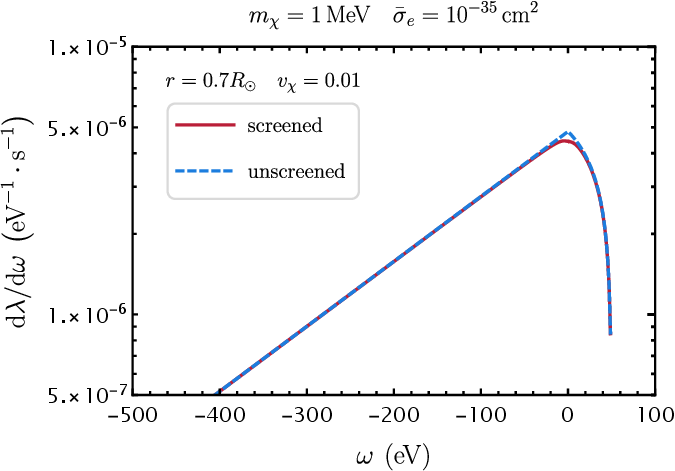}\vspace{0.5cm}
\par\end{centering}
\begin{centering}
\includegraphics[scale=0.75]{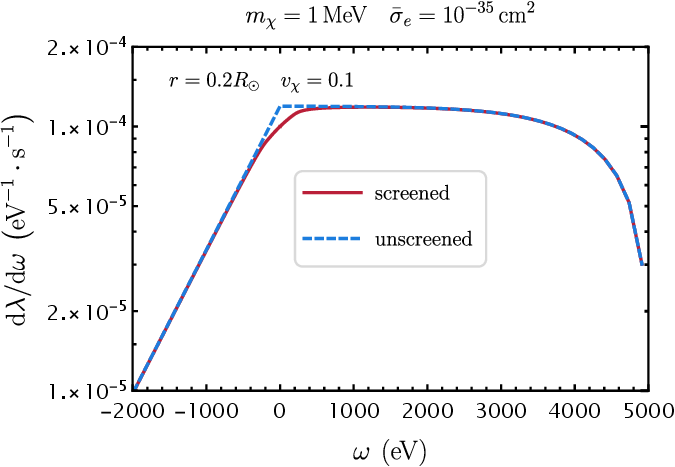}\hspace{1cc}\includegraphics[scale=0.75]{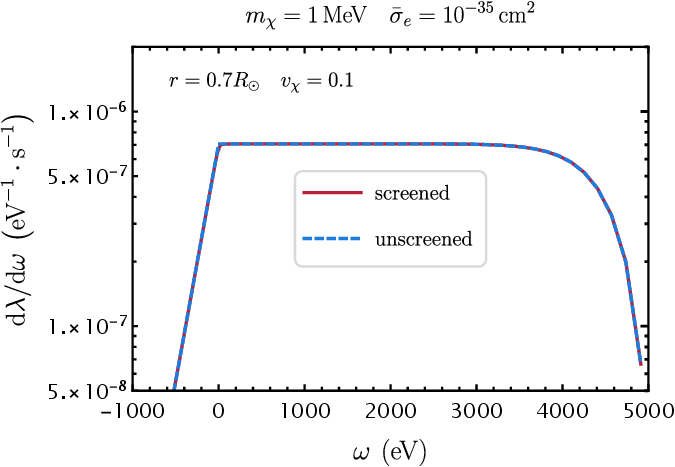}
\par\end{centering}
\caption{\label{fig:differentialScatteringRate}The differential scattering
rates for a $1\,\mathrm{MeV}$ DM particle with a benchmark cross
section $\bar{\sigma}_{e}=10^{-35}\,\mathrm{cm}^{2}$ in the heavy
mediator limit for velocities $v_{\chi}=1\times10^{-3}$~(\textbf{\textit{top}}),
$1\times10^{-2}$~(\textbf{\textit{middle}}) $1\times10^{-1}$~(\textbf{\textit{bottom}}),
respectively, at the location $r=0.2\,R_{\odot}$ ~(\textbf{\textit{left}})
and $r=0.7\,R_{\odot}$ ~(\textbf{\textit{right}}) inside the Sun,
with~(\textbf{\textit{red solid}}) and without~(\textbf{\textit{blue
dashed}}) the in-medium effect.}
\end{figure}

To provide further details, in Figs.~\ref{fig:differentialScatteringRate-2},
\ref{fig:differentialScatteringRate-1}, and \ref{fig:differentialScatteringRate}
we compare the differential scattering rates with and without the
in-medium effect for the same set of parameters. In the left~(right)
panel shown are the differential rates for DM velocities\textcolor{black}{{}
}$v_{\chi}=1\times10^{-3}$, $1\times10^{-2}$, and $1\times10^{-1}$
\textcolor{black}{at} $0.2\,R_{\odot}$~($0.7\,R_{\odot}$), respectively.
The spectra without screening are obtained by approximating $\chi_{\hat{\rho}\hat{\rho}}^{\mathrm{r}}\left(Q,\,\omega\right)$
as $\Pi_{e}\left(Q,\,\omega\right)$ in Eq.~(\ref{eq:rate_NR}),
which corresponds to abandoning all the terms associated with the
electronic Coulomb interaction induced by the polarized electrons
in Eq.~(\ref{sum_electron_diagrams}), or equivalently, substituting
$\mathrm{Im}\left[-\epsilon^{-1}\left(Q,\omega\right)\right]$ with
$\mathrm{Im}\left[\epsilon\left(Q,\omega\right)\right]$ in Eq.~(\ref{eq:rate_NR_epsilon}).
Considering the scattering rate is proportional to the factor (see
Eq.~(\ref{sum_electron_diagrams}) and Eq.~(\ref{eq:rate_NR}))
\begin{eqnarray}
\mathrm{Im}\left(\chi_{\hat{\rho}\hat{\rho}}^{\mathrm{r}}\right) & = & \frac{\mathrm{Im}\left(\Pi_{e}\right)}{\left|\epsilon\left(Q,\omega\right)\right|^{2}}\label{eq:imaginary1}
\end{eqnarray}
the in-medium effect is dictated by the behavior of the denominator
in the kinematically relevant region in the $Q$-$\omega$ space:
$\left|\epsilon\left(Q,\omega\right)\right|^{2}>1$ corresponds to
the screening, while $\left|\epsilon\left(Q,\omega\right)\right|^{2}\ll1$
results in a resonance enhancement of the scattering rate. 

For a lighter DM particle~($m_{\chi}=0.01\,\mathrm{MeV}$ in Fig.~\ref{fig:differentialScatteringRate-2}),
in contrast to the screening that suppresses the scattering rates,
an absorption peak is clearly seen around $\omega=200\,\mathrm{eV}$,
which is due to the presence of a local resonance enclosed in the
contour on the $Q$-$\omega$ plane determined by the kinematical
requirement in Eq.~(\ref{eq:rate_NR}) and Eq.~(\ref{eq:rate_NR_epsilon}).
From Fig.\ref{fig:differentialScatteringRate-1} and Fig.\ref{fig:differentialScatteringRate},
it is observed that for heavier DM particles~($m_{\chi}\geq0.1\,\mathrm{MeV}$),
the calculated scattering rates are overestimated by a factor of a
few if  the in-medium effect is neglected. The differences lie in
the transferred energy region around several hundreds of $\mathrm{eV}$s,
where the electrons in medium are able to respond to the external
DM fields swiftly, and thus a screening can be efficiently established. 

The physics in Figs.~\ref{fig:differentialScatteringRate-2}-\ref{fig:differentialScatteringRate}
is evident. For instance, when a $1\,\mathrm{MeV}$ DM particle moves
slow~($v_{\chi}=1\times10^{-3}$, $1\times10^{-2}$) compared to
\textcolor{black}{electron mean velocity} $u_{0}\approx0.06$ at $r=0.2\,R_{\odot}$,
it can be considered stationary relative to the thermal electrons,
and the scattering rate converges to a saturated value. In this case,
DM particles predominantly absorb energy from the electron medium.
On the other hand, as the velocity increases, DM particles begin to
deposit energy to the electron gas, and at $v_{\chi}=1\times10^{-1}$,
DM particle predominantly loose energy to the electron medium. At
larger radius $r=0.7\,R_{\odot}$, where electron thermal velocity
decreases to $u_{0}\approx0.03$, DM particles are more likely to
loose energy to electron medium.

Lastly, it should be noted that if we combine the term $\left|\epsilon\left(Q,\omega\right)\right|^{-2}$
in Eq.~(\ref{eq:imaginary1}) into the amplitude squared in Eq.~(\ref{eq:rate_NR}),
which corresponds to packing all the in-medium effect into the dielectric
function and approximating $\chi_{\hat{\rho}\hat{\rho}}^{\mathrm{r}}\left(Q,\,\omega\right)\simeq\Pi_{e}\left(Q,\,\omega\right)$
at the same time, Eq.~(\ref{eq:rate_NR}) can also be understood
as the scattering rate of an effective DM-electron interaction~(with
a multiplicative correction of the form $1/\epsilon\left(Q,\omega\right)$)
from a particle-particle scattering perspective. That is to say, Eq.~(\ref{eq:rate_NR})
is equivalent to thermally averaging the DM-electron scattering rate
over the electron velocity distribution, with the in-medium effect
encoded in the dielectric function, a strategy adopted in Ref.~\citep{An:2021qdl}.\textcolor{black}{}
\textcolor{black}{}
\FloatBarrier

\section{\label{sec:Ionic Response}In-medium effect including the ionic response}

In this work, we focus on the models where DM particles only interact
with electrons in the Sun. However, as discussed in the preceding
section, the electrons responding to the external DM field can also
induce further electronic responses via the Coulomb interaction, which
generates screening and collective movement~(plasmon) out of the
electron medium. By the same token, the ions in the solar plasma can
also response to the primarily influenced electrons through the Coulomb
potential, and give rise to additional effective interaction with
DM particles.

The inclusion of the ionic response effect is straightforward under
the framework of the linear response theory. In this picture, the
perturbation exerted on the solar medium remains the same, \textit{i.e.},
still the DM-electron interaction, but ions are brought into consideration
of the solar medium. As a result, the diagram in Eq.~(\ref{sum_electron_diagrams})
should be modified as the following, 

\tikzset{ pattern size/.store in=\mcSize,  
pattern size = 5pt, 
pattern thickness/.store in=\mcThickness,  
pattern thickness = 0.3pt, 
pattern radius/.store in=\mcRadius,  
pattern radius = 1pt} 
\makeatletter 

\pgfutil@ifundefined{pgf@pattern@name@_2rn8z5b7p}{ 
\pgfdeclarepatternformonly[\mcThickness,\mcSize]{_2rn8z5b7p} {\pgfqpoint{0pt}{0pt}} {\pgfpoint{\mcSize}{\mcSize}} {\pgfpoint{\mcSize}{\mcSize}} { \pgfsetcolor{\tikz@pattern@color} \pgfsetlinewidth{\mcThickness} \pgfpathmoveto{\pgfqpoint{0pt}{\mcSize}} \pgfpathlineto{\pgfpoint{\mcSize+\mcThickness}{-\mcThickness}} \pgfpathmoveto{\pgfqpoint{0pt}{0pt}} \pgfpathlineto{\pgfpoint{\mcSize+\mcThickness}{\mcSize+\mcThickness}} \pgfusepath{stroke} }} 

\pgfutil@ifundefined{pgf@pattern@name@_rvi2n1a7u}{ 
\pgfdeclarepatternformonly[\mcThickness,\mcSize]{_rvi2n1a7u} {\pgfqpoint{0pt}{0pt}} {\pgfpoint{\mcSize+\mcThickness}{\mcSize+\mcThickness}} {\pgfpoint{\mcSize}{\mcSize}} { \pgfsetcolor{\tikz@pattern@color} \pgfsetlinewidth{\mcThickness} \pgfpathmoveto{\pgfqpoint{0pt}{0pt}} \pgfpathlineto{\pgfpoint{\mcSize+\mcThickness}{\mcSize+\mcThickness}} \pgfusepath{stroke} }}

\pgfutil@ifundefined{pgf@pattern@name@_mqbru5vbl}{ 
\pgfdeclarepatternformonly[\mcThickness,\mcSize]{_mqbru5vbl} {\pgfqpoint{0pt}{0pt}} {\pgfpoint{\mcSize+\mcThickness}{\mcSize+\mcThickness}} {\pgfpoint{\mcSize}{\mcSize}} { \pgfsetcolor{\tikz@pattern@color} \pgfsetlinewidth{\mcThickness} \pgfpathmoveto{\pgfqpoint{0pt}{0pt}} \pgfpathlineto{\pgfpoint{\mcSize+\mcThickness}{\mcSize+\mcThickness}} \pgfusepath{stroke} }}

\pgfutil@ifundefined{pgf@pattern@name@_dbkyxaspk}{ 
\pgfdeclarepatternformonly[\mcThickness,\mcSize]{_dbkyxaspk} {\pgfqpoint{0pt}{0pt}} {\pgfpoint{\mcSize+\mcThickness}{\mcSize+\mcThickness}} {\pgfpoint{\mcSize}{\mcSize}} { \pgfsetcolor{\tikz@pattern@color} \pgfsetlinewidth{\mcThickness} \pgfpathmoveto{\pgfqpoint{0pt}{0pt}} \pgfpathlineto{\pgfpoint{\mcSize+\mcThickness}{\mcSize+\mcThickness}} \pgfusepath{stroke} }}

\pgfutil@ifundefined{pgf@pattern@name@_21n9tetgz}{ 
\pgfdeclarepatternformonly[\mcThickness,\mcSize]{_21n9tetgz} {\pgfqpoint{0pt}{0pt}} {\pgfpoint{\mcSize+\mcThickness}{\mcSize+\mcThickness}} {\pgfpoint{\mcSize}{\mcSize}} { \pgfsetcolor{\tikz@pattern@color} \pgfsetlinewidth{\mcThickness} \pgfpathmoveto{\pgfqpoint{0pt}{0pt}} \pgfpathlineto{\pgfpoint{\mcSize+\mcThickness}{\mcSize+\mcThickness}} \pgfusepath{stroke} }}

\pgfutil@ifundefined{pgf@pattern@name@_tven52570}{ 
\pgfdeclarepatternformonly[\mcThickness,\mcSize]{_tven52570} {\pgfqpoint{0pt}{0pt}} {\pgfpoint{\mcSize+\mcThickness}{\mcSize+\mcThickness}} {\pgfpoint{\mcSize}{\mcSize}} { \pgfsetcolor{\tikz@pattern@color} \pgfsetlinewidth{\mcThickness} \pgfpathmoveto{\pgfqpoint{0pt}{0pt}} \pgfpathlineto{\pgfpoint{\mcSize+\mcThickness}{\mcSize+\mcThickness}} \pgfusepath{stroke} }}

\pgfutil@ifundefined{pgf@pattern@name@_t3ps3bcqg}{ 
\pgfdeclarepatternformonly[\mcThickness,\mcSize]{_t3ps3bcqg} {\pgfqpoint{0pt}{0pt}} {\pgfpoint{\mcSize+\mcThickness}{\mcSize+\mcThickness}} {\pgfpoint{\mcSize}{\mcSize}} { \pgfsetcolor{\tikz@pattern@color} \pgfsetlinewidth{\mcThickness} \pgfpathmoveto{\pgfqpoint{0pt}{0pt}} \pgfpathlineto{\pgfpoint{\mcSize+\mcThickness}{\mcSize+\mcThickness}} \pgfusepath{stroke} }}

\pgfutil@ifundefined{pgf@pattern@name@_7odshzyid}{ 
\pgfdeclarepatternformonly[\mcThickness,\mcSize]{_7odshzyid} {\pgfqpoint{0pt}{0pt}} {\pgfpoint{\mcSize+\mcThickness}{\mcSize+\mcThickness}} {\pgfpoint{\mcSize}{\mcSize}} { \pgfsetcolor{\tikz@pattern@color} \pgfsetlinewidth{\mcThickness} \pgfpathmoveto{\pgfqpoint{0pt}{0pt}} \pgfpathlineto{\pgfpoint{\mcSize+\mcThickness}{\mcSize+\mcThickness}} \pgfusepath{stroke} }}

\pgfutil@ifundefined{pgf@pattern@name@_izfgnoiaw}{ 
\pgfdeclarepatternformonly[\mcThickness,\mcSize]{_izfgnoiaw} {\pgfqpoint{0pt}{0pt}} {\pgfpoint{\mcSize+\mcThickness}{\mcSize+\mcThickness}} {\pgfpoint{\mcSize}{\mcSize}} { \pgfsetcolor{\tikz@pattern@color} \pgfsetlinewidth{\mcThickness} \pgfpathmoveto{\pgfqpoint{0pt}{0pt}} \pgfpathlineto{\pgfpoint{\mcSize+\mcThickness}{\mcSize+\mcThickness}} \pgfusepath{stroke} }}

\makeatother \tikzset{every picture/.style={line width=0.75pt}} 

\begin{eqnarray}    
\label{sum_iron_diagrams} \chi_{\hat{\rho}\hat{\rho}}^{\mathrm{r}}~&=&\vcenter{\hbox{
\begin{tikzpicture}[x=0.45pt,y=0.45pt,yscale=-1,xscale=1] 
\draw  [pattern=_2rn8z5b7p,pattern size=5pt,pattern thickness=0.75pt,pattern radius=0pt, pattern color={rgb, 255:red, 0; green, 0; blue, 0}] (239.77,78.37) -- (333.61,78.37) -- (333.61,137) -- (239.77,137) -- cycle ; 
\draw   (239.77,78.37) -- (333.61,78.37) .. controls (345.8,78.37) and (355.68,91.5) .. (355.68,107.69) .. controls (355.68,123.88) and (345.8,137) .. (333.61,137) -- (239.77,137) .. controls (227.58,137) and (217.69,123.88) .. (217.69,107.69) .. controls (217.69,91.5) and (227.58,78.37) .. (239.77,78.37) -- cycle ; 
\draw  [fill={rgb, 255:red, 128; green, 128; blue, 128 }  ,fill opacity=1 ] (215.94,109.19) .. controls (215.94,110.02) and (216.73,110.69) .. (217.69,110.69) .. controls (218.66,110.69) and (219.44,110.02) .. (219.44,109.19) .. controls (219.44,108.36) and (218.66,107.69) .. (217.69,107.69) .. controls (216.73,107.69) and (215.94,108.36) .. (215.94,109.19) -- cycle ; 
\draw  [fill={rgb, 255:red, 128; green, 128; blue, 128 }  ,fill opacity=1 ] (353.93,106.19) .. controls (353.93,107.02) and (354.72,107.69) .. (355.68,107.69) .. controls (356.65,107.69) and (357.43,107.02) .. (357.43,106.19) .. controls (357.43,105.36) and (356.65,104.69) .. (355.68,104.69) .. controls (354.72,104.69) and (353.93,105.36) .. (353.93,106.19) -- cycle ;
\end{tikzpicture} 
}} \nonumber    \\ 
\nonumber    \\
~&=&\vcenter{\hbox{
\begin{tikzpicture}[x=0.60pt,y=0.60pt,yscale=-1,xscale=1] 
\draw  [pattern=_rvi2n1a7u,pattern size=4pt,pattern thickness=0.75pt,pattern radius=0pt, pattern color={rgb, 255:red, 0; green, 0; blue, 0}] (286,131) .. controls (286,117.19) and (297.19,106) .. (311,106) .. controls (324.81,106) and (336,117.19) .. (336,131) .. controls (336,144.81) and (324.81,156) .. (311,156) .. controls (297.19,156) and (286,144.81) .. (286,131) -- cycle ; 
\draw  [fill={rgb, 255:red, 128; green, 128; blue, 128 }  ,fill opacity=1 ] (284.25,131) .. controls (284.25,131.83) and (285.03,132.5) .. (286,132.5) .. controls (286.97,132.5) and (287.75,131.83) .. (287.75,131) .. controls (287.75,130.17) and (286.97,129.5) .. (286,129.5) .. controls (285.03,129.5) and (284.25,130.17) .. (284.25,131) -- cycle ; 
\draw  [fill={rgb, 255:red, 128; green, 128; blue, 128 }  ,fill opacity=1 ] (334.25,131) .. controls (334.25,131.83) and (335.03,132.5) .. (336,132.5) .. controls (336.97,132.5) and (337.75,131.83) .. (337.75,131) .. controls (337.75,130.17) and (336.97,129.5) .. (336,129.5) .. controls (335.03,129.5) and (334.25,130.17) .. (334.25,131) -- cycle ;
\end{tikzpicture} 
}}+\vcenter{\hbox{
\begin{tikzpicture}[x=0.40pt,y=0.40pt,yscale=-1,xscale=1] 
\draw  [fill={rgb, 255:red, 128; green, 128; blue, 128 }  ,fill opacity=1 ] (193.32,121.07) .. controls (193.32,122.51) and (194.63,123.69) .. (196.25,123.69) .. controls (197.87,123.69) and (199.18,122.51) .. (199.18,121.07) .. controls (199.18,119.62) and (197.87,118.44) .. (196.25,118.44) .. controls (194.63,118.44) and (193.32,119.62) .. (193.32,121.07) -- cycle ; 
\draw  [fill={rgb, 255:red, 128; green, 128; blue, 128 }  ,fill opacity=1 ] (270.69,121.07) .. controls (270.69,122.51) and (272,123.69) .. (273.62,123.69) .. controls (275.24,123.69) and (276.55,122.51) .. (276.55,121.07) .. controls (276.55,119.62) and (275.24,118.44) .. (273.62,118.44) .. controls (272,118.44) and (270.69,119.62) .. (270.69,121.07) -- cycle ; 
\draw  [pattern=_mqbru5vbl,pattern size=4pt,pattern thickness=0.75pt,pattern radius=0pt, pattern color={rgb, 255:red, 0; green, 0; blue, 0}] (196.25,121.07) .. controls (196.25,100.07) and (213.57,83.04) .. (234.93,83.04) .. controls (256.3,83.04) and (273.62,100.07) .. (273.62,121.07) .. controls (273.62,142.06) and (256.3,159.09) .. (234.93,159.09) .. controls (213.57,159.09) and (196.25,142.06) .. (196.25,121.07) -- cycle ;

\draw  [pattern=_dbkyxaspk,pattern size=4pt,pattern thickness=0.75pt,pattern radius=0pt, pattern color={rgb, 255:red, 0; green, 0; blue, 0}] (347.92,121.02) .. controls (347.92,100.02) and (365.24,83) .. (386.61,83) .. controls (407.97,83) and (425.29,100.02) .. (425.29,121.02) .. controls (425.29,142.02) and (407.97,159.05) .. (386.61,159.05) .. controls (365.24,159.05) and (347.92,142.02) .. (347.92,121.02) -- cycle ; 
\draw  [fill={rgb, 255:red, 128; green, 128; blue, 128 }  ,fill opacity=1 ] (345.21,121.02) .. controls (345.21,122.28) and (346.43,123.3) .. (347.92,123.3) .. controls (349.42,123.3) and (350.63,122.28) .. (350.63,121.02) .. controls (350.63,119.76) and (349.42,118.74) .. (347.92,118.74) .. controls (346.43,118.74) and (345.21,119.76) .. (345.21,121.02) -- cycle ; 
\draw  [fill={rgb, 255:red, 128; green, 128; blue, 128 }  ,fill opacity=1 ] (422.58,121.02) .. controls (422.58,122.28) and (423.8,123.3) .. (425.29,123.3) .. controls (426.79,123.3) and (428,122.28) .. (428,121.02) .. controls (428,119.76) and (426.79,118.74) .. (425.29,118.74) .. controls (423.8,118.74) and (422.58,119.76) .. (422.58,121.02) -- cycle ; 
\draw    (277.53,119.38) .. controls (279.2,117.71) and (280.86,117.72) .. (282.53,119.39) .. controls (284.2,121.06) and (285.86,121.06) .. (287.53,119.4) .. controls (289.2,117.74) and (290.86,117.74) .. (292.53,119.41) .. controls (294.2,121.08) and (295.86,121.08) .. (297.53,119.42) .. controls (299.2,117.76) and (300.86,117.76) .. (302.53,119.43) .. controls (304.2,121.1) and (305.86,121.1) .. (307.53,119.44) .. controls (309.2,117.78) and (310.86,117.78) .. (312.53,119.45) .. controls (314.2,121.12) and (315.86,121.12) .. (317.53,119.46) .. controls (319.2,117.8) and (320.86,117.8) .. (322.53,119.47) .. controls (324.2,121.14) and (325.86,121.14) .. (327.53,119.48) .. controls (329.2,117.82) and (330.86,117.82) .. (332.53,119.49) .. controls (334.2,121.16) and (335.86,121.16) .. (337.53,119.5) .. controls (339.2,117.84) and (340.86,117.84) .. (342.53,119.51) .. controls (344.2,121.18) and (345.86,121.18) .. (347.53,119.52) -- (347.61,119.52) -- (347.61,119.52)(277.52,122.38) .. controls (279.19,120.71) and (280.86,120.72) .. (282.52,122.39) .. controls (284.19,124.06) and (285.85,124.06) .. (287.52,122.4) .. controls (289.19,120.74) and (290.85,120.74) .. (292.52,122.41) .. controls (294.19,124.08) and (295.85,124.08) .. (297.52,122.42) .. controls (299.19,120.76) and (300.85,120.76) .. (302.52,122.43) .. controls (304.19,124.1) and (305.85,124.1) .. (307.52,122.44) .. controls (309.19,120.78) and (310.85,120.78) .. (312.52,122.45) .. controls (314.19,124.12) and (315.85,124.12) .. (317.52,122.46) .. controls (319.19,120.8) and (320.85,120.8) .. (322.52,122.47) .. controls (324.19,124.14) and (325.85,124.14) .. (327.52,122.48) .. controls (329.19,120.82) and (330.85,120.82) .. (332.52,122.49) .. controls (334.19,124.16) and (335.85,124.16) .. (337.52,122.5) .. controls (339.19,120.84) and (340.85,120.84) .. (342.52,122.51) .. controls (344.19,124.18) and (345.85,124.18) .. (347.52,122.52) -- (347.6,122.52) -- (347.6,122.52) ;
\end{tikzpicture}
}}+\vcenter{\hbox{
\begin{tikzpicture}[x=0.40pt,y=0.40pt,yscale=-1,xscale=1] 
\draw  [fill={rgb, 255:red, 128; green, 128; blue, 128 }  ,fill opacity=1 ] (193.32,121.07) .. controls (193.32,122.51) and (194.63,123.69) .. (196.25,123.69) .. controls (197.87,123.69) and (199.18,122.51) .. (199.18,121.07) .. controls (199.18,119.62) and (197.87,118.44) .. (196.25,118.44) .. controls (194.63,118.44) and (193.32,119.62) .. (193.32,121.07) -- cycle ; 
\draw  [fill={rgb, 255:red, 128; green, 128; blue, 128 }  ,fill opacity=1 ] (270.69,121.07) .. controls (270.69,122.51) and (272,123.69) .. (273.62,123.69) .. controls (275.24,123.69) and (276.55,122.51) .. (276.55,121.07) .. controls (276.55,119.62) and (275.24,118.44) .. (273.62,118.44) .. controls (272,118.44) and (270.69,119.62) .. (270.69,121.07) -- cycle ; 
\draw  [pattern=_21n9tetgz,pattern size=4pt,pattern thickness=0.75pt,pattern radius=0pt, pattern color={rgb, 255:red, 0; green, 0; blue, 0}] (196.25,121.07) .. controls (196.25,100.07) and (213.57,83.04) .. (234.93,83.04) .. controls (256.3,83.04) and (273.62,100.07) .. (273.62,121.07) .. controls (273.62,142.06) and (256.3,159.09) .. (234.93,159.09) .. controls (213.57,159.09) and (196.25,142.06) .. (196.25,121.07) -- cycle ;
\draw  [pattern=_tven52570,pattern size=4pt,pattern thickness=0.75pt,pattern radius=0pt, pattern color={rgb, 255:red, 0; green, 0; blue, 0}] (347.92,121.02) .. controls (347.92,100.02) and (365.24,83) .. (386.61,83) .. controls (407.97,83) and (425.29,100.02) .. (425.29,121.02) .. controls (425.29,142.02) and (407.97,159.05) .. (386.61,159.05) .. controls (365.24,159.05) and (347.92,142.02) .. (347.92,121.02) -- cycle ; 
\draw  [fill={rgb, 255:red, 128; green, 128; blue, 128 }  ,fill opacity=1 ] (345.21,121.02) .. controls (345.21,122.28) and (346.43,123.3) .. (347.92,123.3) .. controls (349.42,123.3) and (350.63,122.28) .. (350.63,121.02) .. controls (350.63,119.76) and (349.42,118.74) .. (347.92,118.74) .. controls (346.43,118.74) and (345.21,119.76) .. (345.21,121.02) -- cycle ; 
\draw  [fill={rgb, 255:red, 128; green, 128; blue, 128 }  ,fill opacity=1 ] (422.58,121.02) .. controls (422.58,122.28) and (423.8,123.3) .. (425.29,123.3) .. controls (426.79,123.3) and (428,122.28) .. (428,121.02) .. controls (428,119.76) and (426.79,118.74) .. (425.29,118.74) .. controls (423.8,118.74) and (422.58,119.76) .. (422.58,121.02) -- cycle ; 
\draw    (277.53,119.38) .. controls (279.2,117.71) and (280.86,117.72) .. (282.53,119.39) .. controls (284.2,121.06) and (285.86,121.06) .. (287.53,119.4) .. controls (289.2,117.74) and (290.86,117.74) .. (292.53,119.41) .. controls (294.2,121.08) and (295.86,121.08) .. (297.53,119.42) .. controls (299.2,117.76) and (300.86,117.76) .. (302.53,119.43) .. controls (304.2,121.1) and (305.86,121.1) .. (307.53,119.44) .. controls (309.2,117.78) and (310.86,117.78) .. (312.53,119.45) .. controls (314.2,121.12) and (315.86,121.12) .. (317.53,119.46) .. controls (319.2,117.8) and (320.86,117.8) .. (322.53,119.47) .. controls (324.2,121.14) and (325.86,121.14) .. (327.53,119.48) .. controls (329.2,117.82) and (330.86,117.82) .. (332.53,119.49) .. controls (334.2,121.16) and (335.86,121.16) .. (337.53,119.5) .. controls (339.2,117.84) and (340.86,117.84) .. (342.53,119.51) .. controls (344.2,121.18) and (345.86,121.18) .. (347.53,119.52) -- (347.61,119.52) -- (347.61,119.52)(277.52,122.38) .. controls (279.19,120.71) and (280.86,120.72) .. (282.52,122.39) .. controls (284.19,124.06) and (285.85,124.06) .. (287.52,122.4) .. controls (289.19,120.74) and (290.85,120.74) .. (292.52,122.41) .. controls (294.19,124.08) and (295.85,124.08) .. (297.52,122.42) .. controls (299.19,120.76) and (300.85,120.76) .. (302.52,122.43) .. controls (304.19,124.1) and (305.85,124.1) .. (307.52,122.44) .. controls (309.19,120.78) and (310.85,120.78) .. (312.52,122.45) .. controls (314.19,124.12) and (315.85,124.12) .. (317.52,122.46) .. controls (319.19,120.8) and (320.85,120.8) .. (322.52,122.47) .. controls (324.19,124.14) and (325.85,124.14) .. (327.52,122.48) .. controls (329.19,120.82) and (330.85,120.82) .. (332.52,122.49) .. controls (334.19,124.16) and (335.85,124.16) .. (337.52,122.5) .. controls (339.19,120.84) and (340.85,120.84) .. (342.52,122.51) .. controls (344.19,124.18) and (345.85,124.18) .. (347.52,122.52) -- (347.6,122.52) -- (347.6,122.52) ; 
\draw  [pattern=_t3ps3bcqg,pattern size=4pt,pattern thickness=0.75pt,pattern radius=0pt, pattern color={rgb, 255:red, 0; green, 0; blue, 0}] (496.92,121.02) .. controls (496.92,100.02) and (514.24,83) .. (535.61,83) .. controls (556.97,83) and (574.29,100.02) .. (574.29,121.02) .. controls (574.29,142.02) and (556.97,159.05) .. (535.61,159.05) .. controls (514.24,159.05) and (496.92,142.02) .. (496.92,121.02) -- cycle ; 
\draw  [fill={rgb, 255:red, 128; green, 128; blue, 128 }  ,fill opacity=1 ] (494.21,121.02) .. controls (494.21,122.28) and (495.43,123.3) .. (496.92,123.3) .. controls (498.42,123.3) and (499.63,122.28) .. (499.63,121.02) .. controls (499.63,119.76) and (498.42,118.74) .. (496.92,118.74) .. controls (495.43,118.74) and (494.21,119.76) .. (494.21,121.02) -- cycle ; 
\draw  [fill={rgb, 255:red, 128; green, 128; blue, 128 }  ,fill opacity=1 ] (571.58,121.02) .. controls (571.58,122.28) and (572.8,123.3) .. (574.29,123.3) .. controls (575.79,123.3) and (577,122.28) .. (577,121.02) .. controls (577,119.76) and (575.79,118.74) .. (574.29,118.74) .. controls (572.8,118.74) and (571.58,119.76) .. (571.58,121.02) -- cycle ; 
\draw    (426.53,119.38) .. controls (428.2,117.71) and (429.86,117.72) .. (431.53,119.39) .. controls (433.2,121.06) and (434.86,121.06) .. (436.53,119.4) .. controls (438.2,117.74) and (439.86,117.74) .. (441.53,119.41) .. controls (443.2,121.08) and (444.86,121.08) .. (446.53,119.42) .. controls (448.2,117.76) and (449.86,117.76) .. (451.53,119.43) .. controls (453.2,121.1) and (454.86,121.1) .. (456.53,119.44) .. controls (458.2,117.78) and (459.86,117.78) .. (461.53,119.45) .. controls (463.2,121.12) and (464.86,121.12) .. (466.53,119.46) .. controls (468.2,117.8) and (469.86,117.8) .. (471.53,119.47) .. controls (473.2,121.14) and (474.86,121.14) .. (476.53,119.48) .. controls (478.2,117.82) and (479.86,117.82) .. (481.53,119.49) .. controls (483.2,121.16) and (484.86,121.16) .. (486.53,119.5) .. controls (488.2,117.84) and (489.86,117.84) .. (491.53,119.51) .. controls (493.2,121.18) and (494.86,121.18) .. (496.53,119.52) -- (496.61,119.52) -- (496.61,119.52)(426.52,122.38) .. controls (428.19,120.71) and (429.86,120.72) .. (431.52,122.39) .. controls (433.19,124.06) and (434.85,124.06) .. (436.52,122.4) .. controls (438.19,120.74) and (439.85,120.74) .. (441.52,122.41) .. controls (443.19,124.08) and (444.85,124.08) .. (446.52,122.42) .. controls (448.19,120.76) and (449.85,120.76) .. (451.52,122.43) .. controls (453.19,124.1) and (454.85,124.1) .. (456.52,122.44) .. controls (458.19,120.78) and (459.85,120.78) .. (461.52,122.45) .. controls (463.19,124.12) and (464.85,124.12) .. (466.52,122.46) .. controls (468.19,120.8) and (469.85,120.8) .. (471.52,122.47) .. controls (473.19,124.14) and (474.85,124.14) .. (476.52,122.48) .. controls (478.19,120.82) and (479.85,120.82) .. (481.52,122.49) .. controls (483.19,124.16) and (484.85,124.16) .. (486.52,122.5) .. controls (488.19,120.84) and (489.85,120.84) .. (491.52,122.51) .. controls (493.19,124.18) and (494.85,124.18) .. (496.52,122.52) -- (496.6,122.52) -- (496.6,122.52) ;
\end{tikzpicture} 
}}+\cdots\nonumber    \\ 
\nonumber    \\ 
~&=&\frac{\vcenter{\hbox{
\begin{tikzpicture}[x=0.35pt,y=0.35pt,yscale=-1,xscale=1] 
\draw  [pattern=_izfgnoiaw,pattern size=4pt,pattern thickness=0.75pt,pattern radius=0pt, pattern color={rgb, 255:red, 0; green, 0; blue, 0}] (262.92,145.02) .. controls (262.92,124.02) and (280.24,107) .. (301.61,107) .. controls (322.97,107) and (340.29,124.02) .. (340.29,145.02) .. controls (340.29,166.02) and (322.97,183.05) .. (301.61,183.05) .. controls (280.24,183.05) and (262.92,166.02) .. (262.92,145.02) -- cycle ; 
\draw  [fill={rgb, 255:red, 128; green, 128; blue, 128 }  ,fill opacity=1 ] (260.21,145.02) .. controls (260.21,146.28) and (261.43,147.3) .. (262.92,147.3) .. controls (264.42,147.3) and (265.63,146.28) .. (265.63,145.02) .. controls (265.63,143.76) and (264.42,142.74) .. (262.92,142.74) .. controls (261.43,142.74) and (260.21,143.76) .. (260.21,145.02) -- cycle ; 
\draw  [fill={rgb, 255:red, 128; green, 128; blue, 128 }  ,fill opacity=1 ] (337.58,145.02) .. controls (337.58,146.28) and (338.8,147.3) .. (340.29,147.3) .. controls (341.79,147.3) and (343,146.28) .. (343,145.02) .. controls (343,143.76) and (341.79,142.74) .. (340.29,142.74) .. controls (338.8,142.74) and (337.58,143.76) .. (337.58,145.02) -- cycle ;
\end{tikzpicture} 
}}}{1-\vcenter{\hbox{
\begin{tikzpicture}[x=0.35pt,y=0.35pt,yscale=-1,xscale=1] 
\draw  [fill={rgb, 255:red, 128; green, 128; blue, 128 }  ,fill opacity=1 ] (188.58,145.02) .. controls (188.58,146.28) and (189.8,147.3) .. (191.29,147.3) .. controls (192.79,147.3) and (194,146.28) .. (194,145.02) .. controls (194,143.76) and (192.79,142.74) .. (191.29,142.74) .. controls (189.8,142.74) and (188.58,143.76) .. (188.58,145.02) -- cycle ; 
\draw  [pattern=_7odshzyid,pattern size=4pt,pattern thickness=0.75pt,pattern radius=0pt, pattern color={rgb, 255:red, 0; green, 0; blue, 0}] (262.92,145.02) .. controls (262.92,124.02) and (280.24,107) .. (301.61,107) .. controls (322.97,107) and (340.29,124.02) .. (340.29,145.02) .. controls (340.29,166.02) and (322.97,183.05) .. (301.61,183.05) .. controls (280.24,183.05) and (262.92,166.02) .. (262.92,145.02) -- cycle ; 
\draw  [fill={rgb, 255:red, 128; green, 128; blue, 128 }  ,fill opacity=1 ] (260.21,145.02) .. controls (260.21,146.28) and (261.43,147.3) .. (262.92,147.3) .. controls (264.42,147.3) and (265.63,146.28) .. (265.63,145.02) .. controls (265.63,143.76) and (264.42,142.74) .. (262.92,142.74) .. controls (261.43,142.74) and (260.21,143.76) .. (260.21,145.02) -- cycle ; 
\draw  [fill={rgb, 255:red, 128; green, 128; blue, 128 }  ,fill opacity=1 ] (337.58,145.02) .. controls (337.58,146.28) and (338.8,147.3) .. (340.29,147.3) .. controls (341.79,147.3) and (343,146.28) .. (343,145.02) .. controls (343,143.76) and (341.79,142.74) .. (340.29,142.74) .. controls (338.8,142.74) and (337.58,143.76) .. (337.58,145.02) -- cycle ; 
\draw    (192.53,143.38) .. controls (194.2,141.71) and (195.86,141.72) .. (197.53,143.39) .. controls (199.2,145.06) and (200.86,145.06) .. (202.53,143.4) .. controls (204.2,141.74) and (205.86,141.74) .. (207.53,143.41) .. controls (209.2,145.08) and (210.86,145.08) .. (212.53,143.42) .. controls (214.2,141.76) and (215.86,141.76) .. (217.53,143.43) .. controls (219.2,145.1) and (220.86,145.1) .. (222.53,143.44) .. controls (224.2,141.78) and (225.86,141.78) .. (227.53,143.45) .. controls (229.2,145.12) and (230.86,145.12) .. (232.53,143.46) .. controls (234.2,141.8) and (235.86,141.8) .. (237.53,143.47) .. controls (239.2,145.14) and (240.86,145.14) .. (242.53,143.48) .. controls (244.2,141.82) and (245.86,141.82) .. (247.53,143.49) .. controls (249.2,145.16) and (250.86,145.16) .. (252.53,143.5) .. controls (254.2,141.84) and (255.86,141.84) .. (257.53,143.51) .. controls (259.2,145.18) and (260.86,145.18) .. (262.53,143.52) -- (262.61,143.52) -- (262.61,143.52)(192.52,146.38) .. controls (194.19,144.71) and (195.86,144.72) .. (197.52,146.39) .. controls (199.19,148.06) and (200.85,148.06) .. (202.52,146.4) .. controls (204.19,144.74) and (205.85,144.74) .. (207.52,146.41) .. controls (209.19,148.08) and (210.85,148.08) .. (212.52,146.42) .. controls (214.19,144.76) and (215.85,144.76) .. (217.52,146.43) .. controls (219.19,148.1) and (220.85,148.1) .. (222.52,146.44) .. controls (224.19,144.78) and (225.85,144.78) .. (227.52,146.45) .. controls (229.19,148.12) and (230.85,148.12) .. (232.52,146.46) .. controls (234.19,144.8) and (235.85,144.8) .. (237.52,146.47) .. controls (239.19,148.14) and (240.85,148.14) .. (242.52,146.48) .. controls (244.19,144.82) and (245.85,144.82) .. (247.52,146.49) .. controls (249.19,148.16) and (250.85,148.16) .. (252.52,146.5) .. controls (254.19,144.84) and (255.85,144.84) .. (257.52,146.51) .. controls (259.19,148.18) and (260.85,148.18) .. (262.52,146.52) -- (262.6,146.52) -- (262.6,146.52) ;
\end{tikzpicture} 
}}}, 
\end{eqnarray}where the double wavy line\begin{eqnarray}    
\label{iron_goulomb} 
\begin{tikzpicture}[x=0.75pt,y=0.75pt,yscale=-1,xscale=1] 
\draw    (230,105.5) .. controls (231.67,103.83) and (233.33,103.83) .. (235,105.5) .. controls (236.67,107.17) and (238.33,107.17) .. (240,105.5) .. controls (241.67,103.83) and (243.33,103.83) .. (245,105.5) .. controls (246.67,107.17) and (248.33,107.17) .. (250,105.5) .. controls (251.67,103.83) and (253.33,103.83) .. (255,105.5) .. controls (256.67,107.17) and (258.33,107.17) .. (260,105.5) .. controls (261.67,103.83) and (263.33,103.83) .. (265,105.5) .. controls (266.67,107.17) and (268.33,107.17) .. (270,105.5) .. controls (271.67,103.83) and (273.33,103.83) .. (275,105.5) .. controls (276.67,107.17) and (278.33,107.17) .. (280,105.5) .. controls (281.67,103.83) and (283.33,103.83) .. (285,105.5) -- (287,105.5) -- (287,105.5)(230,108.5) .. controls (231.67,106.83) and (233.33,106.83) .. (235,108.5) .. controls (236.67,110.17) and (238.33,110.17) .. (240,108.5) .. controls (241.67,106.83) and (243.33,106.83) .. (245,108.5) .. controls (246.67,110.17) and (248.33,110.17) .. (250,108.5) .. controls (251.67,106.83) and (253.33,106.83) .. (255,108.5) .. controls (256.67,110.17) and (258.33,110.17) .. (260,108.5) .. controls (261.67,106.83) and (263.33,106.83) .. (265,108.5) .. controls (266.67,110.17) and (268.33,110.17) .. (270,108.5) .. controls (271.67,106.83) and (273.33,106.83) .. (275,108.5) .. controls (276.67,110.17) and (278.33,110.17) .. (280,108.5) .. controls (281.67,106.83) and (283.33,106.83) .. (285,108.5) -- (287,108.5) -- (287,108.5) ;
\end{tikzpicture}~&=&\vcenter{\hbox{
\begin{tikzpicture}[x=0.75pt,y=0.75pt,yscale=-1,xscale=1] 
\draw    (230,107) .. controls (231.67,105.33) and (233.33,105.33) .. (235,107) .. controls (236.67,108.67) and (238.33,108.67) .. (240,107) .. controls (241.67,105.33) and (243.33,105.33) .. (245,107) .. controls (246.67,108.67) and (248.33,108.67) .. (250,107) .. controls (251.67,105.33) and (253.33,105.33) .. (255,107) .. controls (256.67,108.67) and (258.33,108.67) .. (260,107) .. controls (261.67,105.33) and (263.33,105.33) .. (265,107) .. controls (266.67,108.67) and (268.33,108.67) .. (270,107) .. controls (271.67,105.33) and (273.33,105.33) .. (275,107) .. controls (276.67,108.67) and (278.33,108.67) .. (280,107) .. controls (281.67,105.33) and (283.33,105.33) .. (285,107) -- (287,107) -- (287,107) ;
\end{tikzpicture}
}}+\vcenter{\hbox{
\begin{tikzpicture}[x=0.50pt,y=0.50pt,yscale=-1,xscale=1] 
\draw    (225,110) .. controls (226.7,108.37) and (228.37,108.41) .. (230,110.11) .. controls (231.63,111.81) and (233.3,111.84) .. (235,110.21) .. controls (236.7,108.58) and (238.37,108.62) .. (240,110.32) .. controls (241.63,112.02) and (243.3,112.06) .. (245,110.43) .. controls (246.7,108.8) and (248.36,108.84) .. (249.99,110.54) .. controls (251.62,112.24) and (253.29,112.27) .. (254.99,110.64) .. controls (256.69,109.01) and (258.36,109.05) .. (259.99,110.75) -- (260,110.75) -- (260,110.75) ; 
\draw   (264.58,110.5) .. controls (264.58,97.84) and (274.84,87.58) .. (287.5,87.58) .. controls (300.16,87.58) and (310.42,97.84) .. (310.42,110.5) .. controls (310.42,123.16) and (300.16,133.42) .. (287.5,133.42) .. controls (274.84,133.42) and (264.58,123.16) .. (264.58,110.5)(260,110.5) .. controls (260,95.31) and (272.31,83) .. (287.5,83) .. controls (302.69,83) and (315,95.31) .. (315,110.5) .. controls (315,125.69) and (302.69,138) .. (287.5,138) .. controls (272.31,138) and (260,125.69) .. (260,110.5) ; 
\draw    (315,110.5) .. controls (316.7,108.87) and (318.37,108.91) .. (320,110.61) .. controls (321.63,112.31) and (323.3,112.34) .. (325,110.71) .. controls (326.7,109.08) and (328.37,109.12) .. (330,110.82) .. controls (331.63,112.52) and (333.3,112.56) .. (335,110.93) .. controls (336.7,109.3) and (338.36,109.34) .. (339.99,111.04) .. controls (341.62,112.74) and (343.29,112.77) .. (344.99,111.14) .. controls (346.69,109.51) and (348.36,109.55) .. (349.99,111.25) -- (350,111.25) -- (350,111.25) ; 
\draw  [fill={rgb, 255:red, 0; green, 0; blue, 0 }  ,fill opacity=1 ] (282,128) -- (294,134.5) -- (282,141) -- (288,134.5) -- cycle ; 
\draw  [fill={rgb, 255:red, 0; green, 0; blue, 0 }  ,fill opacity=1 ] (292,90) -- (280,83.5) -- (292,77) -- (286,83.5) -- cycle ;
\end{tikzpicture} 
}}+ \vcenter{\hbox{\begin{tikzpicture}[x=0.50pt,y=0.50pt,yscale=-1,xscale=1] 
\draw    (225,110) .. controls (226.7,108.37) and (228.37,108.41) .. (230,110.11) .. controls (231.63,111.81) and (233.3,111.84) .. (235,110.21) .. controls (236.7,108.58) and (238.37,108.62) .. (240,110.32) .. controls (241.63,112.02) and (243.3,112.06) .. (245,110.43) .. controls (246.7,108.8) and (248.36,108.84) .. (249.99,110.54) .. controls (251.62,112.24) and (253.29,112.27) .. (254.99,110.64) .. controls (256.69,109.01) and (258.36,109.05) .. (259.99,110.75) -- (260,110.75) -- (260,110.75) ; 
\draw   (264.58,110.5) .. controls (264.58,97.84) and (274.84,87.58) .. (287.5,87.58) .. controls (300.16,87.58) and (310.42,97.84) .. (310.42,110.5) .. controls (310.42,123.16) and (300.16,133.42) .. (287.5,133.42) .. controls (274.84,133.42) and (264.58,123.16) .. (264.58,110.5)(260,110.5) .. controls (260,95.31) and (272.31,83) .. (287.5,83) .. controls (302.69,83) and (315,95.31) .. (315,110.5) .. controls (315,125.69) and (302.69,138) .. (287.5,138) .. controls (272.31,138) and (260,125.69) .. (260,110.5) ; 
\draw    (315,110.5) .. controls (316.7,108.87) and (318.37,108.91) .. (320,110.61) .. controls (321.63,112.31) and (323.3,112.34) .. (325,110.71) .. controls (326.7,109.08) and (328.37,109.12) .. (330,110.82) .. controls (331.63,112.52) and (333.3,112.56) .. (335,110.93) .. controls (336.7,109.3) and (338.36,109.34) .. (339.99,111.04) .. controls (341.62,112.74) and (343.29,112.77) .. (344.99,111.14) .. controls (346.69,109.51) and (348.36,109.55) .. (349.99,111.25) -- (350,111.25) -- (350,111.25) ; 
\draw  [fill={rgb, 255:red, 0; green, 0; blue, 0 }  ,fill opacity=1 ] (282,128) -- (294,134.5) -- (282,141) -- (288,134.5) -- cycle ; 
\draw  [fill={rgb, 255:red, 0; green, 0; blue, 0 }  ,fill opacity=1 ] (292,90) -- (280,83.5) -- (292,77) -- (286,83.5) -- cycle ; 
\draw   (353.58,111.5) .. controls (353.58,98.84) and (363.84,88.58) .. (376.5,88.58) .. controls (389.16,88.58) and (399.42,98.84) .. (399.42,111.5) .. controls (399.42,124.16) and (389.16,134.42) .. (376.5,134.42) .. controls (363.84,134.42) and (353.58,124.16) .. (353.58,111.5)(349,111.5) .. controls (349,96.31) and (361.31,84) .. (376.5,84) .. controls (391.69,84) and (404,96.31) .. (404,111.5) .. controls (404,126.69) and (391.69,139) .. (376.5,139) .. controls (361.31,139) and (349,126.69) .. (349,111.5) ; 
\draw    (404,111.5) .. controls (405.7,109.87) and (407.37,109.91) .. (409,111.61) .. controls (410.63,113.31) and (412.3,113.34) .. (414,111.71) .. controls (415.7,110.08) and (417.37,110.12) .. (419,111.82) .. controls (420.63,113.52) and (422.3,113.56) .. (424,111.93) .. controls (425.7,110.3) and (427.36,110.34) .. (428.99,112.04) .. controls (430.62,113.74) and (432.29,113.77) .. (433.99,112.14) .. controls (435.69,110.51) and (437.36,110.55) .. (438.99,112.25) -- (439,112.25) -- (439,112.25) ; 
\draw  [fill={rgb, 255:red, 0; green, 0; blue, 0 }  ,fill opacity=1 ] (371,129) -- (383,135.5) -- (371,142) -- (377,135.5) -- cycle ; 
\draw  [fill={rgb, 255:red, 0; green, 0; blue, 0 }  ,fill opacity=1 ] (381,91) -- (369,84.5) -- (381,78) -- (375,84.5) -- cycle ;
\end{tikzpicture}
}}+\cdots\nonumber    \\ 
\nonumber    \\ 
~&=&V_{e}+V_{e}\left(-Z\right)\Pi_{N}V_{e}\left(-Z\right)+V_{e}\left(-Z\right)\Pi_{N}V_{e}\left(-Z\right)^{2}\Pi_{N}V_{e}\left(-Z\right)+\cdots \nonumber    \\ 
\nonumber    \\ 
~&=&\frac{V_{e}}{1-V_{e}\left(-Z\right)^{2}\Pi_{N}}
\end{eqnarray}represents the electron Coulomb interaction screened by the ions,
with $Z$ being the ion charge number, and $\Pi_{N}$ corresponding
to the ion pair-bubble diagram~$\vcenter{\hbox{
\begin{tikzpicture}[x=0.23pt,y=0.23pt,yscale=-1,xscale=1] 
\draw   (284.58,123.5) .. controls (284.58,110.84) and (294.84,100.58) .. (307.5,100.58) .. controls (320.16,100.58) and (330.42,110.84) .. (330.42,123.5) .. controls (330.42,136.16) and (320.16,146.42) .. (307.5,146.42) .. controls (294.84,146.42) and (284.58,136.16) .. (284.58,123.5)(280,123.5) .. controls (280,108.31) and (292.31,96) .. (307.5,96) .. controls (322.69,96) and (335,108.31) .. (335,123.5) .. controls (335,138.69) and (322.69,151) .. (307.5,151) .. controls (292.31,151) and (280,138.69) .. (280,123.5) ; 
\draw  [fill={rgb, 255:red, 0; green, 0; blue, 0 }  ,fill opacity=1 ] (302,141) -- (314,147.5) -- (302,154) -- (308,147.5) -- cycle ; 
\draw  [fill={rgb, 255:red, 0; green, 0; blue, 0 }  ,fill opacity=1 ] (312,103) -- (300,96.5) -- (312,90) -- (306,96.5) -- cycle ;
\end{tikzpicture}}}$, \textit{i.e.}, the ion 1PI at the RPA level, which can be obtained
by simply replacing $n_{e}$ and $m_{e}$ with ion number density
$n_{N}$ and ion mass $m_{N}$ as follows, 
\begin{eqnarray}
\Pi_{N}\left(Q,\,\omega\right) & = & -\frac{n_{N}}{Q}\sqrt{\frac{m_{N}}{2\,T}}\left\{ \Phi\left[\sqrt{\frac{m_{N}}{2\,T}}\left(\frac{\omega}{Q}+\frac{Q}{2\,m_{N}}\right)\right]-\Phi\left[\sqrt{\frac{m_{N}}{2\,T}}\left(\frac{\omega}{Q}-\frac{Q}{2\,m_{N}}\right)\right]\right\} \nonumber \\
\nonumber \\
 &  & -i\,n_{N}\sqrt{\frac{2\pi}{m_{N}T}}\left(\frac{m_{N}}{Q}\right)\exp\left[-\left(\frac{m_{N}^{2}\,\omega^{2}}{Q^{2}}+\frac{Q^{2}}{4}\right)\frac{1}{2\,m_{N}T}\right]\sinh\left[\frac{\omega}{2\,T}\right].\label{eq:polarizability-1}
\end{eqnarray}
 Note that $\left(-Z\right)$ and $\left(-Z\right)^{2}$ describe
the electron-ion and ion-ion Coulomb interactions in the diagram,
respectively. Similar expression correspondent to $\Pi_{e}$ and $\Pi_{N}$
were derived in Ref.~\citep{An:2021qdl} using the kinetic theory,
where the solar plasma was treated as continuous fluid, and their
results are equivalent to the limit $Q\ll\sqrt{m_{e}T}$ and $Q\ll\sqrt{m_{N}T}$
in Eq.~(\ref{eq:polarizability}) and Eq.~(\ref{eq:polarizability-1}),
respectively. Thus in rare high-energy scattering processes where
$Q\apprge\sqrt{m_{e}T}$, a discrepancy is expected between the resulting
dielectric functions in this work and in Ref.~\citep{An:2021qdl}. 

To sum up, by plugging Eq.~(\ref{iron_goulomb}) into Eq.~(\ref{sum_iron_diagrams})
and generalizing the above discussion to include multiple atom species,
we have
\begin{eqnarray}
\mathrm{Im}\left(\chi_{\hat{\rho}\hat{\rho}}^{\mathrm{r}}\right) & = & \mathrm{Im}\left[\frac{\Pi_{e}\left(1-V_{e}\,{\displaystyle \sum\limits _{i}Z_{i}^{2}\,\Pi_{N_{i}}}\right)}{1-V_{e}\,\Pi_{e}-V_{e}\,{\displaystyle \sum\limits _{i}Z_{i}^{2}\,\Pi_{N_{i}}}}\right]\nonumber \\
 & = & \frac{\left|1-V_{e}\,{\displaystyle \sum\limits _{i}Z_{i}^{2}\,\Pi_{N_{i}}}\right|^{2}\mathrm{Im}\left(\Pi_{e}\right)}{\left|1-V_{e}\,\Pi_{e}-V_{e}\,{\displaystyle \sum\limits _{i}Z_{i}^{2}\,\Pi_{N_{i}}}\right|^{2}}+\frac{\sum\limits _{i}\,\left|Z_{i}V_{e}\,\Pi_{e}\right|^{2}\mathrm{Im}\left(\Pi_{N_{i}}\right)}{\left|1-V_{e}\,\Pi_{e}-V_{e}\,{\displaystyle \sum\limits _{i}Z_{i}^{2}\,\Pi_{N_{i}}}\right|^{2}},\label{eq:iron_chi}
\end{eqnarray}
where $\Pi_{N_{i}}$ denotes the bubble diagram of $i^{\mathrm{th}}$
ion species carrying a charge $Z_{i}e$. By inserting $\mathrm{Im}\left(\chi_{\hat{\rho}\hat{\rho}}^{\mathrm{r}}\right)$
in Eq.~(\ref{eq:iron_chi}) into Eq.~(\ref{eq:rate_NR}) we finally
arrive at the scattering rate including the ionic response in the
solar medium. The first and the second term on the right-hand side
of Eq.~(\ref{eq:iron_chi}) correspond to the DM-electron interaction,
and the induced DM-ion interaction, both screened jointly by electrons
and ions, respectively. While the first correspondence can be observed
from Eq.~(\ref{sum_iron_diagrams}) and Eq.~(\ref{iron_goulomb}),
the second can be seen from the induced interaction between DM particle
and the $i^{\mathrm{th}}$ ion species $V_{\chi N_{i}}$ in the position
space,
\begin{eqnarray}
V_{\chi N_{i}}\left(\mathbf{x},t\right) & = & \int\mathrm{d}^{3}x'\frac{\left(-Z_{i}\right)e^{2}}{4\pi\left|\mathbf{x}-\mathbf{x}'\right|}\rho_{\mathrm{ind}}\left(\mathbf{x}',t\right),
\end{eqnarray}
where $\rho_{\mathrm{ind}}$ is the polarized electron density triggered
by DM particles. In the momentum space it is written as~(see Eqs.~(\ref{eq:inverse_epsilon0},
\ref{electron_bubble} and \ref{eq:polarizability})),
\begin{eqnarray}
V_{\chi N_{i}}\left(Q,\omega\right) & = & \left(-Z_{i}\right)V_{e}\left(Q\right)\,\chi_{\hat{\rho}\hat{\rho}}^{\mathrm{r}}\left(Q,\,\omega\right)\,\frac{g_{\chi}g_{e}}{Q^{2}+m_{A'}^{2}}\nonumber \\
 & = & \left(-Z_{i}\right)V_{e}\left(Q\right)\,\frac{\Pi_{e}\left(Q,\,\omega\right)}{1-V_{e}\,\Pi_{e}\left(Q,\,\omega\right)}\,\frac{g_{\chi}g_{e}}{Q^{2}+m_{A'}^{2}},\label{eq:DM-ion interaction}
\end{eqnarray}
 and right matches the second term in Eq.~(\ref{eq:iron_chi}) except
for the term $V_{e}\,{\displaystyle \sum\limits _{i}Z_{i}^{2}\,\Pi_{N_{i}}}$
in the denominator, which should be included to describe the ionic
effect on a self-consistent basis. The presence of $\omega$ implies
that this interaction is time-retarded. 

\begin{figure}
\begin{centering}
\includegraphics[scale=0.7]{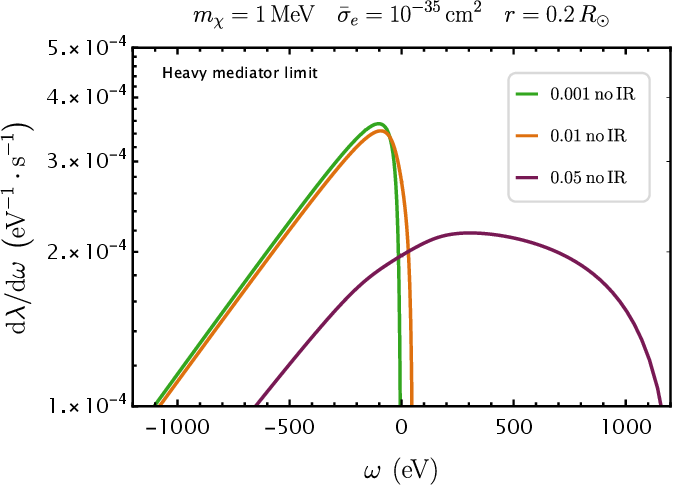}\hspace{1cm}\includegraphics[scale=0.7]{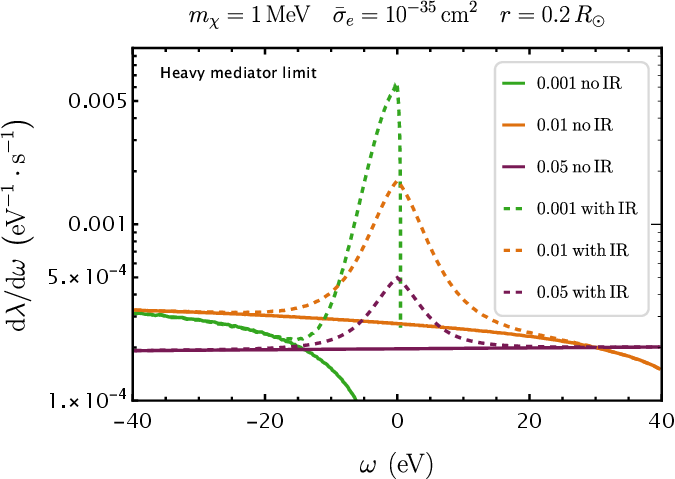}\vspace{0.5cm}
\par\end{centering}
\begin{centering}
\includegraphics[scale=0.7]{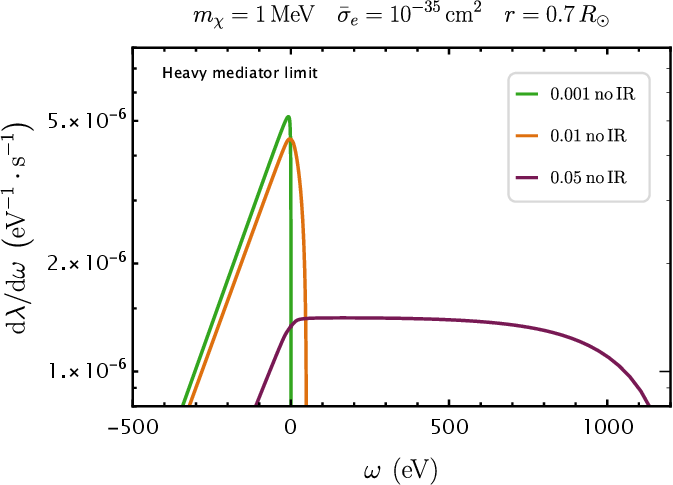}\hspace{1cm}\includegraphics[scale=0.7]{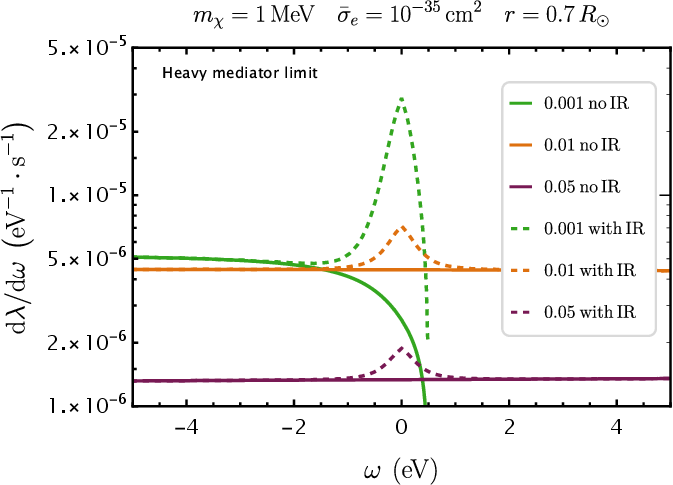}
\par\end{centering}
\caption{\label{fig:Heavy_Ionic}The differential scattering rates for DM particles
(with $m_{\chi}=1\,\mathrm{MeV}$ and $\bar{\sigma}_{e}=10^{-35}\,\mathrm{cm}^{2}$
in the heavy mediator limit) with benchmark velocities $v_{\chi}=1\times10^{-3}$~(\textbf{\textit{green}}),
$1\times10^{-2}$~(\textbf{\textit{orange}}), and $5\times10^{-2}$~(\textbf{\textit{purple}}),
with~(\textbf{\textit{right}} column) and without~(\textbf{\textit{left}}
column) the ionic response~(denoted as IR in the plot) at $r=0.2\,R_{\odot}$
(\textbf{\textit{top}}) and $r=0.7\,R_{\odot}$ (\textbf{\textit{bottom}})
inside the Sun, respectively. See text for details.}
\end{figure}

In the left column in Fig.~\ref{fig:Heavy_Ionic} we show the differential
event rates with only the electronic response effect for a $1\,\mathrm{MeV}$
DM particle with a coupling strength $\bar{\sigma}_{e}=10^{-35}\,\mathrm{cm}^{2}$
in the heavy mediator limit at velocities $v_{\chi}=1\times10^{-3}$,
$1\times10^{-2}$ and $5\times10^{-2}$, at $r=0.2\,R_{\odot}$ and
$r=0.7\,R_{\odot}$, respectively. For comparison, in the right column
we superimpose the differential rates including the ionic in-medium
effect~(in dashed lines). It is observed that the calculated rates
are remarkably enhanced within a width of $\mathcal{O}\left(10\right)\,\mathrm{eV}$
around $\omega=0\,\mathrm{eV}$ when the ion movements are involved.
In this regime, part of the DM energies are deposited into the thermal
ions via an effective DM-ion interaction induced by electrons. Since
ions are much heavier that DM particles, the directions of the DM
particles after scattering may be altered considerably. In the outer
region where temperature declines, the width of energy transfer narrows,
which is consistent with the expectation that slower ion movement
will accordingly reduce transferred energy to DM particles~(think
of a ping-pong ball reflected from a slowly moving wall). 
\begin{figure}
\begin{centering}
\includegraphics[scale=0.7]{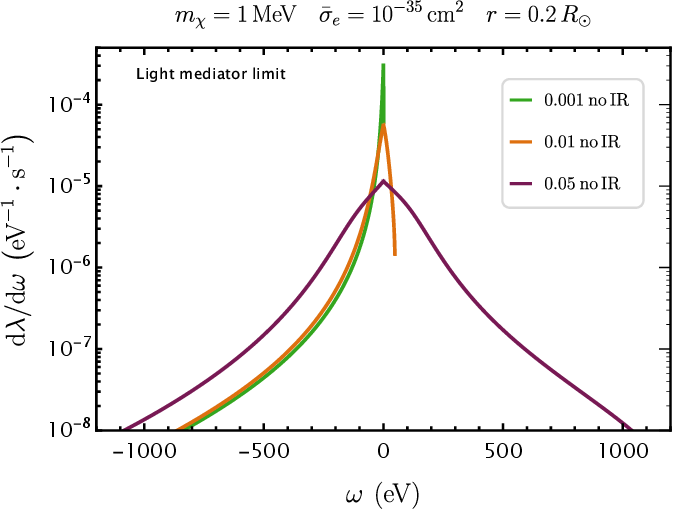}\hspace{1cm}\includegraphics[scale=0.7]{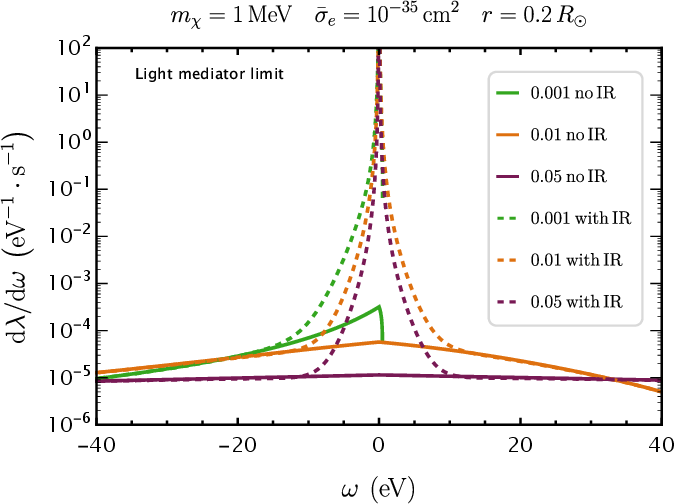}\vspace{0.5cm}
\includegraphics[scale=0.7]{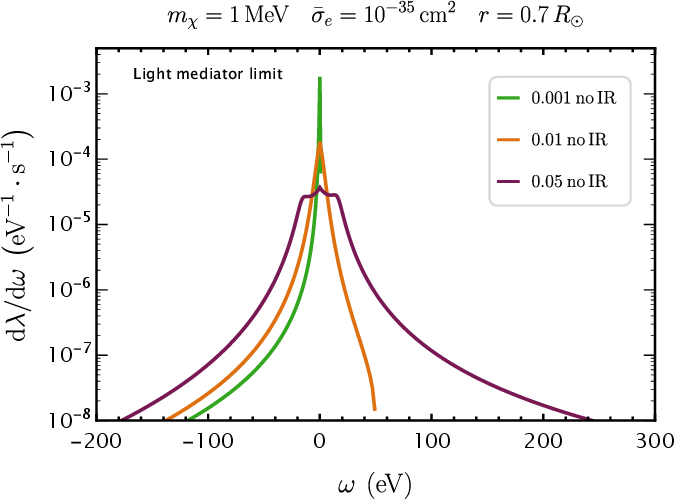}\hspace{1cm}\includegraphics[scale=0.7]{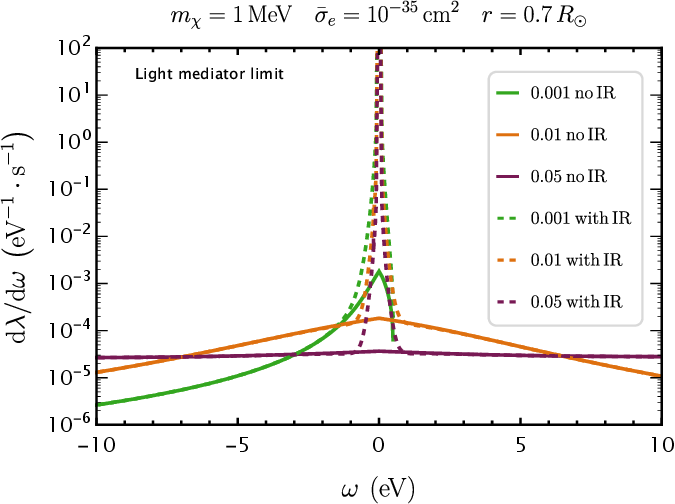}
\par\end{centering}
\caption{\label{fig:Light_Ionic}The differential scattering rates for DM particles
(with $m_{\chi}=1\,\mathrm{MeV}$ and $\bar{\sigma}_{e}=10^{-35}\,\mathrm{cm}^{2}$
in the light mediator limit) with various velocities $v_{\chi}=1\times10^{-3}$~(\textbf{\textit{green}}),
$1\times10^{-2}$~(\textbf{\textit{orange}}), and $5\times10^{-2}$~(\textbf{\textit{purple}})
with~(\textbf{\textit{right}} column) and without~(\textbf{\textit{left}}
column) the ionic response~(IR in the plot) at $r=0.2\,R_{\odot}$
(\textbf{\textit{top}}) and $r=0.7\,R_{\odot}$ (\textbf{\textit{bottom}})
inside the Sun, respectively.}
\end{figure}

While these behaviors qualitatively agree with expectation, it is
hard to give a simple estimate sketching all the features in the plots.
Here we try to get a sense of these numerical results. Particularly,
the spike in scattering rates around $\omega=0\,\mathrm{eV}$ can
be comprehended with the effective DM-ion interaction in Eq.~(\ref{eq:DM-ion interaction}),
\begin{eqnarray}
V_{\chi N}\left(Q,\omega\rightarrow0\right) & = & \frac{\left(-Z\right)4\pi\alpha}{Q^{2}}\,\left.\frac{\Pi_{e}\left(Q,\,\omega\right)}{1-V_{e}\,\Pi_{e}\left(Q,\,\omega\right)}\right|_{\omega\rightarrow0}\,\frac{g_{\chi}g_{e}}{Q^{2}+m_{A'}^{2}}\nonumber \\
\nonumber \\
 & = & \frac{\left(-Z\right)4\pi\alpha}{Q^{2}}\,\frac{Q^{4}\left(\Pi_{e}-V_{e}\left(Q\right)\left|\Pi_{e}\right|^{2}\right)}{\left[Q^{2}-4\pi\alpha\,\mathrm{Re}\left(\Pi_{e}\right)\right]^{2}+\left[4\pi\alpha\,\mathrm{Im}\left(\Pi_{e}\right)\right]^{2}}\,\frac{g_{\chi}g_{e}}{Q^{2}+m_{A'}^{2}},\label{eq:DM-ion-bare}
\end{eqnarray}
where $\Pi_{e}\simeq-\left(\kappa^{2}/4\pi\alpha\right)\left(1-x\Phi\left(x\right)\right)-i\left(\kappa^{2}/4\pi\alpha\right)\sqrt{\pi}\,x\,e^{-x^{2}}$
(see Appendix~\ref{sec:appendix1} for properties of function $\Phi$)
with $x=\sqrt{m_{e}/2\,T}\left(\omega/Q\right)$ for a small $Q$
($Q\ll\sqrt{2m_{e}T}$), and $\kappa=\sqrt{4\pi\alpha n_{e}/T}$ is
the Debye-Hückel scale for electrons, which is of order of $\mathrm{keV}$s
in the solar core. Considering the presence of the Coulomb interaction,
here we investigate the behavior of $V_{\chi N}$ in the limit $Q\rightarrow0$.
Note that the kinematical requirement $Q>\left|\omega\right|/v_{\chi}$
in the limit $Q\rightarrow0$ translates to an upper bound on $x$,
such that $\left|x\right|\simeq u_{0}^{-1}\left|\omega\right|/Q\apprle v_{\chi}/u_{0}\apprle\mathcal{O}\left(1\right)$.
As a consequence, $\left(1-x\Phi\left(x\right)\right)$ and $\sqrt{\pi}\,x\,e^{-x^{2}}$
in the denominator will not approach zero simultaneously. Thus in
the massive mediator scenario, these DM-ion interactions are well
regulated, making additional but finite contributions to the total
scattering rates around $\omega=0$. On the other hand, for larger
energy transfers~(or higher frequencies) $\left|\omega\right|>10\,\mathrm{eV}$,
considering $\left|x\right|\simeq u_{0}^{-1}\left|\omega\right|/Q$,
$V_{e}\,\Pi_{e}$ quickly drops to zero, which implies that the soft
electron medium cannot convey an energy larger roughly than $10\,\mathrm{eV}$,
so the induced ionic response becomes negligible. Here we note that
Eq.~(\ref{eq:DM-ion-bare}) is only the bare DM-ion interaction,
and a systematic treatment should also take into account the ionic
screening and other finite temperature effects (in Eq.~(\ref{eq:rate_NR})
and Eq.~(\ref{eq:iron_chi})).

In Fig.~\ref{fig:Light_Ionic}, in a parallel fashion we also plot
the differential scattering rates in the massless mediator scenario,
for the same set of parameters. In left column only the electrons
are taken into account for the in-medium effect. The differential
rates peak in the small energy regime, which corresponds to the forward
scattering. The electronic screening prevents the scattering rate
in Eq.~(\ref{eq:rate_NR}) from blowing up for a finite DM velocity
$v_{\chi}$. This can be deduced from the following observation for
a small $Q$:
\begin{eqnarray}
\left(-2\right)\mathrm{Im}\left(\chi_{\hat{\rho}\hat{\rho}}^{\mathrm{r}}\right) & = & \left.\frac{Q^{4}\,\kappa^{2}\left(2\sqrt{\pi}\right)^{-1}\,x\,e^{-x^{2}}}{\left[Q^{2}+\kappa^{2}\left(1-x\Phi\left(x\right)\right)\right]^{2}+\left[\kappa^{2}\sqrt{\pi}\,x\,e^{-x^{2}}\right]^{2}}\right|_{x=\sqrt{\frac{m_{e}}{2\,T}}\frac{\omega}{Q}},
\end{eqnarray}
where $Q^{4}$ in the numerator right cancels the characteristic $Q^{-4}$
factor in the massless mediator limit. In addition, since for an arbitrarily
small $\omega$ the variable $\left|x\right|\apprle\mathcal{O}\left(1\right)$
has an upper bound as $Q\rightarrow0$ in Eq.~(\ref{eq:polarizability}),
the denominator remains finite. Incidentally, when the DM velocity
$v_{\chi}$ is comparable to that of the electrons in the medium,
$x$ is not necessarily a small quantity, and the imaginary part of
Eq.~(\ref{eq:polarizability}) can also be comparable to the its
real counterpart. That's when the conventional argument in Debye static
screening no longer applies, which deals only with the situation where
the DM particles move much slower than the electron thermal velocity. 

Compared to the case in the heavy mediator limit in Fig.~\ref{fig:Heavy_Ionic},
the ionic response brings an even more significant boost to the scattering
rates in the low energy regime in the light mediator limit (in the
right panel of Fig.~\ref{fig:Light_Ionic}), which also takes effect
within only a few $\mathrm{eVs}$ around $\omega=0$. From Eq.~(\ref{eq:DM-ion-bare}),
we can see the interaction is proportional to $Q^{-2}$ in the limit
$Q\rightarrow0$, which implies that the induced electron density
can further induce a divergent DM-ion cross section at $\omega=0$,
and explains why the ionic response brings such significant enhancement
in scattering rates in the light mediator limit. 

Moreover, it should be noted that the quantity $\mathrm{Im}\left(\chi_{\hat{\rho}\hat{\rho}}^{\mathrm{r}}\right)$
is an odd function of $\omega$, and thus the scattering rate in Eq.~(\ref{eq:rate_NR})
is approximately symmetric about $\omega=0$ within a finite width,
so long as the kinetics allows depositing energy at the medium~$\left(\omega>0\right)$,
which is consistent with the observation from right panels of Fig.~\ref{fig:Heavy_Ionic}
and Fig.~\ref{fig:Light_Ionic} that the inclusion of the ironic
response has an equal probability accelerating and slowing the DM
particles by several $\mathrm{eV}$s for higher velocities~($v_{\chi}=1\times10^{-2}$
and $5\times10^{-2}$). For $v_{\chi}=1\times10^{-3}$, however, the
overall impact by the induced DM-ion interaction is to significantly
enhance the chance for DM particles to obtain energies $\sim10\,\mathrm{eV}$
from the medium. 
\begin{figure}
\begin{centering}
\includegraphics[scale=0.7]{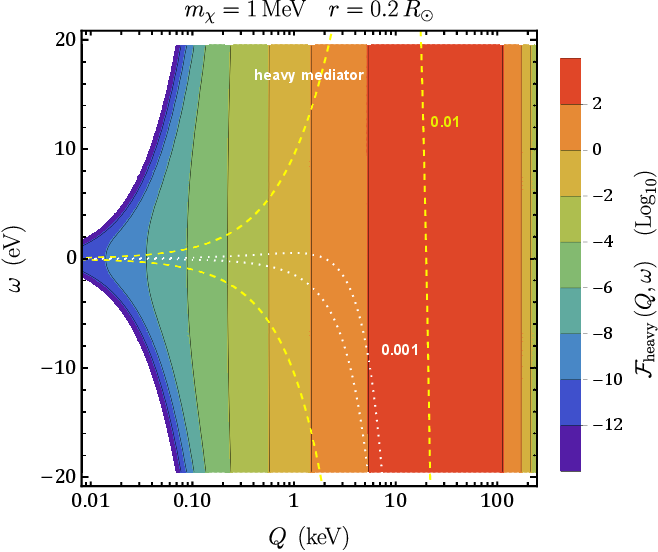}\hspace{1cm}\includegraphics[scale=0.7]{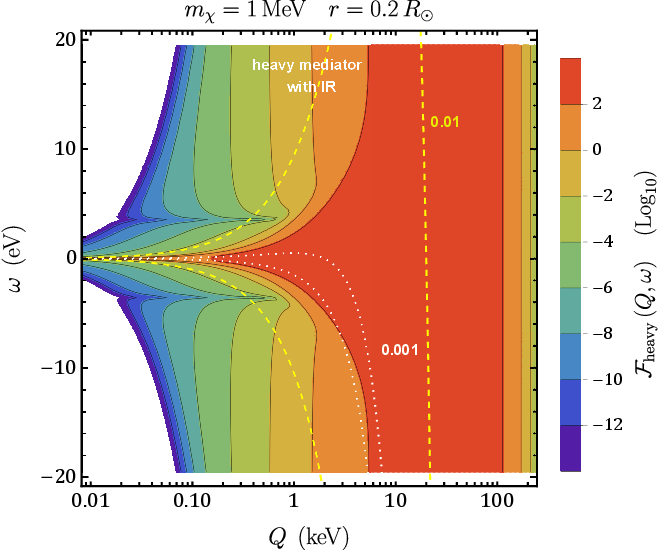}\vspace{0.5cm}
\par\end{centering}
\begin{centering}
\includegraphics[scale=0.7]{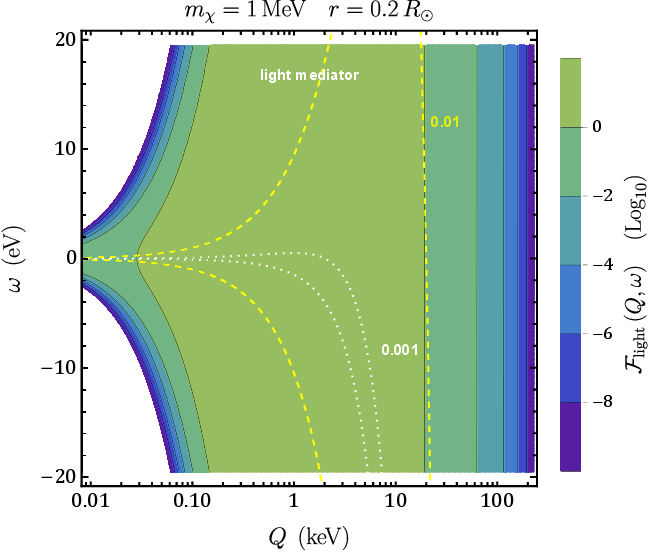}\hspace{1cm}\includegraphics[scale=0.7]{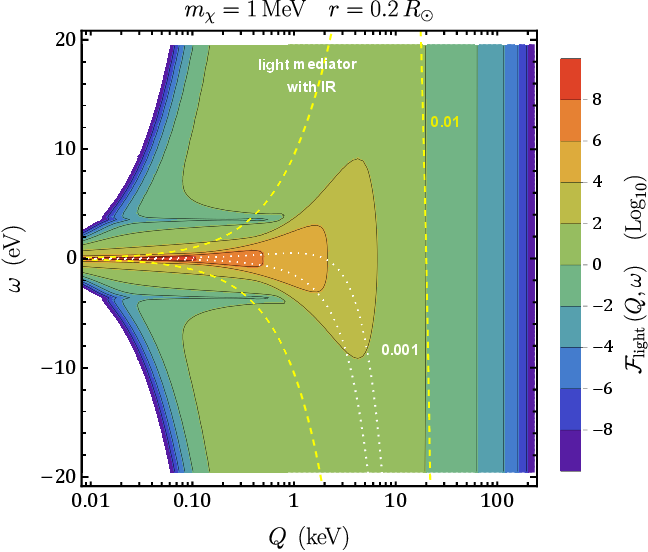}
\par\end{centering}
\caption{\label{fig:2d_plot}$\mathcal{F}_{\mathrm{heavy}}$~(\textbf{\textit{top}})
and $\mathcal{F}_{\mathrm{light}}$~(\textbf{\textit{bottom}}) for
a $1\,\mathrm{MeV}$ DM particle with velocities $v_{\chi}=1\times10^{-2}$
and $1\times10^{-3}$ at $0.2\,R_{\odot}$ with~(\textbf{\textit{right}})
and without~(\textbf{\textit{left}}) the ionic response~(denoted
as IR in the plot) in the heavy and the light mediator limit, respectively.
See text for details.}

\end{figure}

To better visualize the in-medium effect due to ions, in Fig.~\ref{fig:2d_plot}
we plot the dimensionless factors 
\begin{eqnarray}
\mathcal{F}_{\mathrm{heavy}}\left(Q,\,\omega\right) & = & \frac{Q^{2}}{\left(\alpha^{2}m_{e}^{2}\right)^{2}}\,\frac{\left(-2\right)\,\mathrm{Im}\left[\chi_{\hat{\rho}\hat{\rho}}^{\mathrm{r}}\left(Q,\,\omega\right)\right]}{1-e^{-\beta\omega}}
\end{eqnarray}
and 
\begin{eqnarray}
\mathcal{F}_{\mathrm{light}}\left(Q,\,\omega\right) & = & \frac{Q^{2}}{\left(\alpha^{2}m_{e}^{2}\right)^{2}}\left(\frac{\alpha^{2}m_{e}^{2}}{Q^{2}}\right)^{2}\,\frac{\left(-2\right)\,\mathrm{Im}\left[\chi_{\hat{\rho}\hat{\rho}}^{\mathrm{r}}\left(Q,\,\omega\right)\right]}{1-e^{-\beta\omega}},
\end{eqnarray}
on the $Q$-$\omega$ plane for the heavy and light mediator scenario
at $r=0.2\,R_{\odot}$, respectively, which are proportional to the
double differential scattering rate $\mathrm{d}^{2}\lambda/\left(\mathrm{d}\omega\,\mathrm{d}\log Q\right)$~(independent
of the coupling strength $\bar{\sigma}_{e}$, DM velocity $v_{\chi}$,
and DM mass $\mu_{\chi e}^{2}$). The yellow dashed contour~(white
dotted line) reflects the kinematical condition $\Theta\left[v_{\chi}-v_{\mathrm{min}}\left(Q,\,\omega\right)\right]\,\Theta\left[v_{\chi}+v_{\mathrm{min}}\left(Q,\,\omega\right)\right]$
in Eq.~(\ref{eq:rate_NR}) for a $1\,\mathrm{MeV}$ DM particle with
velocity $v_{\chi}=1\times10^{-2}$~($1\times10^{-3}$). It is observed
that both $\mathcal{F}_{\mathrm{heavy}}\left(Q,\,\omega\right)$ and
$\mathcal{F}_{\mathrm{light}}\left(Q,\,\omega\right)$ decrease rapidly
in a uniform manner around $Q\sim\mathcal{O}\left(10\right)\,\mathrm{keV}$,
where the factor $\exp\left(-\frac{Q^{2}}{8\,m_{e}T}\right)$ in the
imaginary component in Eq.~(\ref{eq:polarizability}) begins to decay.
Besides, while $\mathcal{F}_{\mathrm{heavy}}\left(Q,\,\omega\right)$
and $\mathcal{F}_{\mathrm{light}}\left(Q,\,\omega\right)$ are symmetric
about $\omega=0$, the differential rates for varied velocities are
merely determined by the kinetics, as we mentioned in the above discussion. 

It is also noticed from the upper right panel that the ionic response
can remarkably enhance the scattering rate at the region $Q\sim\mathcal{O}\left(1\right)\mathrm{keV}$
and $\omega\sim\mathcal{O}\left(1\right)\mathrm{eV}$. This area is
covered by the kinematical contour of a lower velocity $v_{\chi}=1\times10^{-3}$,
or a momentum $p_{\chi}=1\,\mathrm{keV}$, which means ions can deflect
trajectories of DM particles. Similar observation can be drawn from
the bottom right panel of Fig.~\ref{fig:2d_plot}, where the possibility
for the DM particle with a momentum $p_{\chi}=1\,\mathrm{keV}$ receiving
a momentum transfer $Q\sim\mathcal{O}\left(1\right)\mathrm{keV}$
is also significantly enhanced, while an even much larger enhancement
is observed in the region $\left(\omega,\,Q\right)\rightarrow0$.
As pointed out in Ref.~\citep{An:2021qdl}, a DM-ion interaction
may either shield DM particles from entering the solar core where
they can be effectively heated, or increase the chances of reentering
the solar core when they are heading outward. While ions do not exercise
an immediate impact on the energy spectrum of the accelerated DM particles,
the total effect should be evaluated by simulation methods. Here we
emphasize that the induced DM-ion interaction can exert a remarkable
influence on the DM propagation in the Sun.

\section{\label{sec:Conclusions}Summary and conclusion}

In this paper we have built up a theoretical framework for the description
of the DM-solar medium interaction in the context of the linear response
theory. In this framework, we no longer stick to a particle-particle
scattering picture, but instead we treat the entire solar medium as
a whole, with all the initial and final states of electrons and ions
being summed over and encapsulated in the dynamic structure factor.
Since we are only interested in the movement of DM particle in solar
environment, this approach is particularly neat and convenient to
handle. 

Taking this approach, we compare the differential scattering rates
computed with and without the solar in-medium effect. Our results
show that taking into account the in-medium effect can bring a remarkable
correction to the scattering rates in the heavy mediator scenario,
which can both enhance and suppress the scattering rate between DM
particle and solar medium, depending on the specific kinetics~(\textit{e.g.},
DM mass $m_{\chi}$ and velocity $v_{\chi}$). While the in-medium
effect mainly affects the calculation of scattering in the low-energy
regime~(\textit{i.e.}, $\left|\omega\right|\apprle1\,\mathrm{keV}$),
its ultimate impact on the spectrum of the reflected DM flux remains
to be determined through actual simulations.

Within this framework, we also discuss the implication of the electron-induced
ionic response on the DM scattering in the Sun. Our calculations show
that even for Lagrangian that contains only the DM-electron interaction,
the polarized electron density is able to trigger an effective DM-ion
interaction, which significantly prompts DM particles to absorb and
loose energy of several $\mathrm{eV}$s in every scattering event.
Such enhancement in scattering rate is particularly remarkable in
the light mediator scenario, due to the long range interaction, which
prefers a forward scattering. In both heavy and light mediator limits,
the induced DM-ion interaction can considerably change the direction
of the DM particles moving in the Sun, and may have a potential effect
on the energy spectrum of the up-scattered DM particles. 

The framework proposed in this paper is easily adapted for a wider
ranges of DM interactions. For example, in cases where the DM particle
field couples to the nucleus density operator $\hat{\psi}_{N}^{\dagger}\hat{\psi}_{N}$,
one can simply replace $\hat{\psi}_{e}$ with $\hat{\psi}_{N}$ to
obtain the ionic polarizability $\chi_{\hat{\rho}_{N}\hat{\rho}_{N}}^{\mathrm{r}}\left(Q,\,\omega\right)$,
along with other mixing terms $\chi_{\hat{\rho}_{N}\hat{\rho}}^{\mathrm{r}}\left(Q,\,\omega\right)$
and $\chi_{\hat{\rho}\hat{\rho}_{N}}^{\mathrm{r}}\left(Q,\,\omega\right)$,
and the total scattering rate consists of both the contributions from
the electrons and ions. Based on the formula Eq.~(\ref{eq:rate_NR}),
in the next step we will develop an efficient and robust tool for
simulating the scattering and propagation of the DM particle in the
Sun for the leptophilic interaction in Eq.~(\ref{eq:leptopihilic_interaction}),
so as to provide an accurate description of the reflection energy
spectrum  at the terrestrial detectors, where the in-medium effect
also plays an important role in boosting the DM signal rates in semiconductors
via the plasmon resonance.

\appendix

\renewcommand{\theequation}{A.\arabic{equation}}
\begin{acknowledgments}
We thank Fawei Zheng and Chongjie Mo for valuable discussions. This
work was partly supported by the Major Basic Program of Natural Science
Foundation of Shandong Province (Grant No. ZR2021ZD01), and by National
Natural Science Foundation of China under No.~11625415. 
\end{acknowledgments}

\section{\label{sec:appendix1}Evaluation of the polarizability of electron
gas in the Sun}

In this appendix~(further details and discussions can be found in
Ref.~\citep{fetter2012quantum}), we provide a detailed evaluation
of bubble diagram $\Pi_{e}$ for an electron gas in the Sun, where
the occupation number can be described with a Boltzmann form
\begin{eqnarray}
f_{\mathbf{p}} & = & n_{e}\left(\sqrt{\frac{2\pi}{m_{e}T}}\right)^{3}\exp\left[-\frac{p^{2}}{2m_{e}T}\right],
\end{eqnarray}
if the condition $n_{e}\left(\frac{2\pi}{m_{e}T}\right)^{3/2}\ll1$
is satisfied. Thus Eq.~(\ref{electron_bubble}) for the free electron
gas can be explicitly evaluated as follows, 

\begin{eqnarray}
\frac{1}{V}\sum_{\mathbf{p}}\frac{f_{\mathbf{p}}-f_{\mathbf{p}+\mathbf{Q}}}{\omega-\left(\varepsilon_{\mathbf{p}+\mathbf{Q}}-\varepsilon_{\mathbf{p}}\right)+i0^{+}} & = & \int\frac{\mathrm{d}^{3}p}{\left(2\pi\right)^{3}}f_{\mathbf{p}+\frac{\mathbf{Q}}{2}}\left(\frac{1}{\omega+\frac{\mathbf{p}\cdot\mathbf{Q}}{m_{e}}+i0^{+}}-\frac{1}{\omega-\frac{\mathbf{p}\cdot\mathbf{Q}}{m_{e}}+i0^{+}}\right)\nonumber \\
 & = & n_{e}\left(\sqrt{\frac{2\pi}{m_{e}T}}\right)^{3}\int\frac{\mathrm{d}^{3}p}{\left(2\pi\right)^{3}}\exp\left[-\frac{\left|\mathbf{p}+\frac{\mathbf{Q}}{2}\right|^{2}}{2m_{e}T}\right]\left(\cdots\right)\nonumber \\
 & = & n_{e}\left(\sqrt{\frac{2\pi}{m_{e}T}}\right)^{3}\int\frac{\mathrm{d}^{2}p_{\bot}}{\left(2\pi\right)^{2}}\exp\left[-\frac{p_{\bot}^{2}}{2m_{e}T}\right]\int_{-\infty}^{+\infty}\frac{\mathrm{d}p_{\Vert}}{2\pi}\exp\left[-\frac{\left(p_{\Vert}+\frac{Q}{2}\right)^{2}}{2m_{e}T}\right]\left(\cdots\right)\nonumber \\
 & = & n_{e}\sqrt{\frac{2\pi}{m_{e}T}}\int_{-\infty}^{+\infty}\frac{\mathrm{d}p_{\Vert}}{2\pi}\exp\left[-\frac{\left(p_{\Vert}+\frac{Q}{2}\right)^{2}}{2m_{e}T}\right]\left(\frac{1}{\omega+\frac{p_{\Vert}\cdot q}{m_{e}}+i0^{+}}-\frac{1}{\omega-\frac{p_{\Vert}\cdot q}{m_{e}}+i0^{+}}\right)\nonumber \\
 & = & n_{e}\sqrt{\frac{2\pi}{m_{e}T}}\,\mathcal{P}\int_{-\infty}^{+\infty}\frac{\mathrm{d}p_{\Vert}}{2\pi}\exp\left[-\frac{\left(p_{\Vert}+\frac{Q}{2}\right)^{2}}{2m_{e}T}\right]\left(\frac{1}{\omega+\frac{p_{\Vert}\cdot Q}{m_{e}}}-\frac{1}{\omega-\frac{p_{\Vert}\cdot Q}{m_{e}}}\right)\nonumber \\
 &  & -i\,\pi n_{e}\sqrt{\frac{2\pi}{m_{e}T}}\int_{-\infty}^{+\infty}\frac{\mathrm{d}p_{\Vert}}{2\pi}\exp\left[-\frac{\left(p_{\Vert}+\frac{Q}{2}\right)^{2}}{2m_{e}T}\right]\left[\delta\left(\omega+\frac{p_{\Vert}\cdot Q}{m_{e}}\right)-\delta\left(\omega-\frac{p_{\Vert}\cdot Q}{m_{e}}\right)\right],\nonumber \\
\nonumber \\
\label{eq:intermedia1}
\end{eqnarray}
where $\varepsilon_{\mathbf{p}}=p^{2}/\left(2m_{e}\right)$ is the
energy of a free electron of momentum $\mathbf{p}$, and we take the
substitution $\sum_{\mathbf{p}}\sim\frac{V}{\left(2\pi\right)^{3}}\int\mathrm{d^{3}}p$
$\left(\cdots\right)$ in the second and third line summarizes the
terms in parenthesis in the first line. In the third line we separate
the variable $\mathbf{p}$ into components perpendicular and parallel
to the momentum transfer $\mathbf{Q}$, such that $\mathbf{p}=\mathbf{p}_{\bot}+p_{\Vert}\hat{\mathbf{Q}}$$,$
with $\hat{\mathbf{Q}}$ being the unit vector of the $\mathbf{Q}$
direction. $\mathcal{P}$ denotes the Cauchy principal value of the
infinite integral. 
\begin{figure}
\begin{centering}
\includegraphics[scale=0.85]{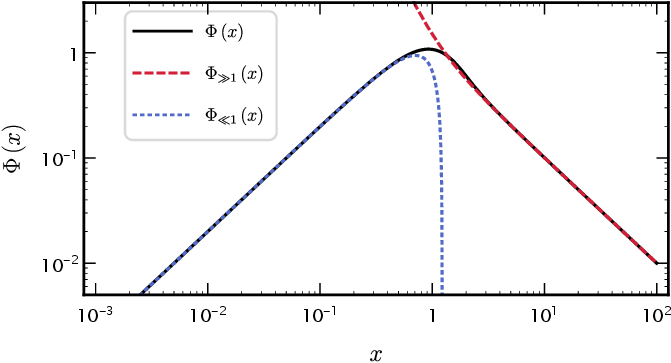}
\par\end{centering}
\caption{\label{fig:Phi} Numerical evaluation of function $\Phi\left(x\right)$,
also plotted are its asymptotic forms $\Phi_{\gg1}\left(x\right)$
and $\Phi_{\ll1}\left(x\right)$ for comparison. See text for details.}
\end{figure}

While further reducing the imaginary part of Eq.~(\ref{eq:intermedia1})~(the
last line) is straightforward, the real part (the fifth line) cannot
be expressed in terms of elementary functions. After a change of variable
$y=\left(p_{\Vert}+\frac{Q}{2}\right)/\sqrt{2m_{e}T}$, the real part
is rewritten as 
\begin{align}
 & n_{e}\sqrt{\frac{2\pi}{m_{e}T}}\,\mathcal{P}\int_{-\infty}^{+\infty}\frac{\mathrm{d}p_{\Vert}}{2\pi}\exp\left[-\frac{\left(p_{\Vert}+\frac{Q}{2}\right)^{2}}{2m_{e}T}\right]\left(\frac{1}{\omega+\frac{p_{\Vert}\cdot Q}{m_{e}}}-\frac{1}{\omega-\frac{p_{\Vert}\cdot Q}{m_{e}}}\right)\nonumber \\
= & -\frac{n_{e}}{Q}\sqrt{\frac{m_{e}}{2T}}\,\left\{ \Phi\left[\sqrt{\frac{m_{e}}{2\,T}}\left(\frac{\omega}{Q}+\frac{Q}{2\,m_{e}}\right)\right]-\Phi\left[\sqrt{\frac{m_{e}}{2\,T}}\left(\frac{\omega}{Q}-\frac{Q}{2\,m_{e}}\right)\right]\right\} 
\end{align}
with
\begin{eqnarray}
\Phi\left(x\right) & \equiv & \mathcal{P}\int_{-\infty}^{+\infty}\frac{\mathrm{d}y}{\sqrt{\pi}}\frac{e^{-y^{2}}}{x-y}\nonumber \\
 & = & \mathcal{P}\int_{-\infty}^{+\infty}\frac{\mathrm{d}y}{\sqrt{\pi}}\frac{e^{-y^{2}}\left(x+y\right)}{x^{2}-y^{2}}\nonumber \\
 & = & x\,\mathcal{P}\int_{-\infty}^{+\infty}\frac{\mathrm{d}y}{\sqrt{\pi}}\frac{e^{-y^{2}}}{x^{2}-y^{2}},\label{eq:Phi}
\end{eqnarray}
which approaches the asymptotic forms $\Phi_{\gg1}\left(x\right)=x^{-1}\left(1+\frac{1}{2}x^{-2}\right)$
and $\Phi_{\ll1}\left(x\right)=2x\left(1-\frac{2}{3}x^{2}\right)$
in the limits $x\gg1$ and $x\ll1$, respectively. The last line implies
that $\Phi\left(-x\right)=-\Phi\left(x\right)$. To get a sense, we
present $\Phi\left(x\right)$, $\Phi_{\gg1}\left(x\right)$ and $\Phi_{\ll1}\left(x\right)$
in Fig.~\ref{fig:Phi} for illustration. In addition, differentiating
Eq.~(\ref{eq:Phi}) with respect to $x$ and integrating by parts
yields a useful relation $\Phi'\left(x\right)=2-2x\Phi\left(x\right)$
in evaluating the polarizability of the solar medium.

\bibliographystyle{JHEP1}
\addcontentsline{toc}{section}{\refname}\bibliography{DMintheSun}

\providecommand{\href}[2]{#2}\begingroup\raggedright\begin{thebibliography}{10}

\bibitem{Barak:2020fql}
{\scshape SENSEI} collaboration, \emph{{SENSEI: Direct-Detection Results on
  sub-GeV Dark Matter from a New Skipper-CCD}},
  \href{https://doi.org/10.1103/PhysRevLett.125.171802}{\emph{Phys. Rev. Lett.}
  {\bfseries 125} (2020) 171802}
  [\href{https://arxiv.org/abs/2004.11378}{{\ttfamily 2004.11378}}].

\bibitem{Castello-Mor:2020jhd}
{\scshape DAMIC-M} collaboration, \emph{{DAMIC-M Experiment: Thick, Silicon
  CCDs to search for Light Dark Matter}},
  \href{https://doi.org/10.1016/j.nima.2019.162933}{\emph{Nucl. Instrum. Meth.
  A} {\bfseries 958} (2020) 162933}
  [\href{https://arxiv.org/abs/2001.01476}{{\ttfamily 2001.01476}}].

\bibitem{Amaral:2020ryn}
{\scshape SuperCDMS} collaboration, \emph{{Constraints on low-mass, relic dark
  matter candidates from a surface-operated SuperCDMS single-charge sensitive
  detector}}, \href{https://doi.org/10.1103/PhysRevD.102.091101}{\emph{Phys.
  Rev. D} {\bfseries 102} (2020) 091101}
  [\href{https://arxiv.org/abs/2005.14067}{{\ttfamily 2005.14067}}].

\bibitem{CDEX:2022kcd}
{\scshape CDEX} collaboration, \emph{{Constraints on Sub-GeV Dark
  Matter\textendash{}Electron Scattering from the CDEX-10 Experiment}},
  \href{https://doi.org/10.1103/PhysRevLett.129.221301}{\emph{Phys. Rev. Lett.}
  {\bfseries 129} (2022) 221301}
  [\href{https://arxiv.org/abs/2206.04128}{{\ttfamily 2206.04128}}].

\bibitem{Arnaud:2020svb}
{\scshape EDELWEISS} collaboration, \emph{{First germanium-based constraints on
  sub-MeV Dark Matter with the EDELWEISS experiment}},
  \href{https://doi.org/10.1103/PhysRevLett.125.141301}{\emph{Phys. Rev. Lett.}
  {\bfseries 125} (2020) 141301}
  [\href{https://arxiv.org/abs/2003.01046}{{\ttfamily 2003.01046}}].

\bibitem{Essig:2015cda}
R.~Essig, M.~Fernandez-Serra, J.~Mardon, A.~Soto, T.~Volansky and T.-T. Yu,
  \emph{{Direct Detection of sub-GeV Dark Matter with Semiconductor Targets}},
  \href{https://doi.org/10.1007/JHEP05(2016)046}{\emph{JHEP} {\bfseries 05}
  (2016) 046} [\href{https://arxiv.org/abs/1509.01598}{{\ttfamily
  1509.01598}}].

\bibitem{Hochberg:2015pha}
Y.~Hochberg, Y.~Zhao and K.~M. Zurek, \emph{{Superconducting Detectors for
  Superlight Dark Matter}},
  \href{https://doi.org/10.1103/PhysRevLett.116.011301}{\emph{Phys. Rev. Lett.}
  {\bfseries 116} (2016) 011301}
  [\href{https://arxiv.org/abs/1504.07237}{{\ttfamily 1504.07237}}].

\bibitem{Essig:2016crl}
R.~Essig, J.~Mardon, O.~Slone and T.~Volansky, \emph{{Detection of sub-GeV Dark
  Matter and Solar Neutrinos via Chemical-Bond Breaking}},
  \href{https://doi.org/10.1103/PhysRevD.95.056011}{\emph{Phys. Rev.}
  {\bfseries D95} (2017) 056011}
  [\href{https://arxiv.org/abs/1608.02940}{{\ttfamily 1608.02940}}].

\bibitem{Hochberg:2016ntt}
Y.~Hochberg, Y.~Kahn, M.~Lisanti, C.~G. Tully and K.~M. Zurek,
  \emph{{Directional detection of dark matter with two-dimensional targets}},
  \href{https://doi.org/10.1016/j.physletb.2017.06.051}{\emph{Phys. Lett.}
  {\bfseries B772} (2017) 239}
  [\href{https://arxiv.org/abs/1606.08849}{{\ttfamily 1606.08849}}].

\bibitem{Hochberg:2016sqx}
Y.~Hochberg, T.~Lin and K.~M. Zurek, \emph{{Absorption of light dark matter in
  semiconductors}},
  \href{https://doi.org/10.1103/PhysRevD.95.023013}{\emph{Phys. Rev.}
  {\bfseries D95} (2017) 023013}
  [\href{https://arxiv.org/abs/1608.01994}{{\ttfamily 1608.01994}}].

\bibitem{Derenzo:2016fse}
S.~Derenzo, R.~Essig, A.~Massari, A.~Soto and T.-T. Yu, \emph{{Direct Detection
  of sub-GeV Dark Matter with Scintillating Targets}},
  \href{https://doi.org/10.1103/PhysRevD.96.016026}{\emph{Phys. Rev.}
  {\bfseries D96} (2017) 016026}
  [\href{https://arxiv.org/abs/1607.01009}{{\ttfamily 1607.01009}}].

\bibitem{Hochberg:2017wce}
Y.~Hochberg, Y.~Kahn, M.~Lisanti, K.~M. Zurek, A.~G. Grushin, R.~Ilan et~al.,
  \emph{{Detection of sub-MeV Dark Matter with Three-Dimensional Dirac
  Materials}}, \href{https://doi.org/10.1103/PhysRevD.97.015004}{\emph{Phys.
  Rev.} {\bfseries D97} (2018) 015004}
  [\href{https://arxiv.org/abs/1708.08929}{{\ttfamily 1708.08929}}].

\bibitem{Knapen2018}
S.~Knapen, T.~Lin, M.~Pyle and K.~M. Zurek, \emph{{Detection of Light Dark
  Matter With Optical Phonons in Polar Materials}},
  \href{https://doi.org/10.1016/j.physletb.2018.08.064}{\emph{Phys. Lett.}
  {\bfseries B785} (2018) 386}
  [\href{https://arxiv.org/abs/1712.06598}{{\ttfamily 1712.06598}}].

\bibitem{Griffin:2018bjn}
S.~Griffin, S.~Knapen, T.~Lin and K.~M. Zurek, \emph{{Directional Detection of
  Light Dark Matter with Polar Materials}},
  \href{https://doi.org/10.1103/PhysRevD.98.115034}{\emph{Phys. Rev.}
  {\bfseries D98} (2018) 115034}
  [\href{https://arxiv.org/abs/1807.10291}{{\ttfamily 1807.10291}}].

\bibitem{Griffin:2019mvc}
S.~M. Griffin, K.~Inzani, T.~Trickle, Z.~Zhang and K.~M. Zurek,
  \emph{{Multichannel direct detection of light dark matter: Target
  comparison}}, \href{https://doi.org/10.1103/PhysRevD.101.055004}{\emph{Phys.
  Rev. D} {\bfseries 101} (2020) 055004}
  [\href{https://arxiv.org/abs/1910.10716}{{\ttfamily 1910.10716}}].

\bibitem{Trickle:2019ovy}
T.~Trickle, Z.~Zhang and K.~M. Zurek, \emph{{Detecting Light Dark Matter with
  Magnons}}, \href{https://doi.org/10.1103/PhysRevLett.124.201801}{\emph{Phys.
  Rev. Lett.} {\bfseries 124} (2020) 201801}
  [\href{https://arxiv.org/abs/1905.13744}{{\ttfamily 1905.13744}}].

\bibitem{Kurinsky:2019pgb}
N.~A. Kurinsky, T.~C. Yu, Y.~Hochberg and B.~Cabrera, \emph{{Diamond Detectors
  for Direct Detection of Sub-GeV Dark Matter}},
  \href{https://doi.org/10.1103/PhysRevD.99.123005}{\emph{Phys. Rev.}
  {\bfseries D99} (2019) 123005}
  [\href{https://arxiv.org/abs/1901.07569}{{\ttfamily 1901.07569}}].

\bibitem{Trickle:2019nya}
T.~Trickle, Z.~Zhang, K.~M. Zurek, K.~Inzani and S.~Griffin,
  \emph{{Multi-Channel Direct Detection of Light Dark Matter: Theoretical
  Framework}}, \href{https://doi.org/10.1007/JHEP03(2020)036}{\emph{JHEP}
  {\bfseries 03} (2020) 036}
  [\href{https://arxiv.org/abs/1910.08092}{{\ttfamily 1910.08092}}].

\bibitem{Campbell-Deem:2019hdx}
B.~Campbell-Deem, P.~Cox, S.~Knapen, T.~Lin and T.~Melia, \emph{{Multiphonon
  excitations from dark matter scattering in crystals}},
  \href{https://doi.org/10.1103/PhysRevD.101.036006}{\emph{Phys. Rev. D}
  {\bfseries 101} (2020) 036006}
  [\href{https://arxiv.org/abs/1911.03482}{{\ttfamily 1911.03482}}].

\bibitem{Coskuner:2019odd}
A.~Coskuner, A.~Mitridate, A.~Olivares and K.~M. Zurek, \emph{{Directional Dark
  Matter Detection in Anisotropic Dirac Materials}},
  \href{https://doi.org/10.1103/PhysRevD.103.016006}{\emph{Phys. Rev. D}
  {\bfseries 103} (2021) 016006}
  [\href{https://arxiv.org/abs/1909.09170}{{\ttfamily 1909.09170}}].

\bibitem{Geilhufe:2019ndy}
R.~M. Geilhufe, F.~Kahlhoefer and M.~W. Winkler, \emph{{Dirac Materials for
  Sub-MeV Dark Matter Detection: New Targets and Improved Formalism}},
  \href{https://doi.org/10.1103/PhysRevD.101.055005}{\emph{Phys. Rev. D}
  {\bfseries 101} (2020) 055005}
  [\href{https://arxiv.org/abs/1910.02091}{{\ttfamily 1910.02091}}].

\bibitem{Griffin:2020lgd}
S.~M. Griffin, Y.~Hochberg, K.~Inzani, N.~Kurinsky, T.~Lin and T.~Chin,
  \emph{{Silicon carbide detectors for sub-GeV dark matter}},
  \href{https://doi.org/10.1103/PhysRevD.103.075002}{\emph{Phys. Rev. D}
  {\bfseries 103} (2021) 075002}
  [\href{https://arxiv.org/abs/2008.08560}{{\ttfamily 2008.08560}}].

\bibitem{Prabhu:2022dtm}
A.~Prabhu and C.~Blanco, \emph{{Constraints on Dark Matter-Electron Scattering
  from Molecular Cloud Ionization}},
  \href{https://arxiv.org/abs/2211.05787}{{\ttfamily 2211.05787}}.

\bibitem{Esposito:2022bnu}
A.~Esposito and S.~Pavaskar, \emph{{Optimal anti-ferromagnets for light dark
  matter detection}},  \href{https://arxiv.org/abs/2210.13516}{{\ttfamily
  2210.13516}}.

\bibitem{Liang:2018bdb}
Z.-L. Liang, L.~Zhang, P.~Zhang and F.~Zheng, \emph{{The wavefunction
  reconstruction effects in calculation of DM-induced electronic transition in
  semiconductor targets}},
  \href{https://doi.org/10.1007/JHEP01(2019)149}{\emph{JHEP} {\bfseries 01}
  (2019) 149} [\href{https://arxiv.org/abs/1810.13394}{{\ttfamily
  1810.13394}}].

\bibitem{Griffin:2021znd}
S.~M. Griffin, K.~Inzani, T.~Trickle, Z.~Zhang and K.~M. Zurek, \emph{{Extended
  calculation of dark matter-electron scattering in crystal targets}},
  \href{https://doi.org/10.1103/PhysRevD.104.095015}{\emph{Phys. Rev. D}
  {\bfseries 104} (2021) 095015}
  [\href{https://arxiv.org/abs/2105.05253}{{\ttfamily 2105.05253}}].

\bibitem{Kahn:2021ttr}
Y.~Kahn and T.~Lin, \emph{{Searches for light dark matter using condensed
  matter systems}}, \href{https://doi.org/10.1088/1361-6633/ac5f63}{\emph{Rept.
  Prog. Phys.} {\bfseries 85} (2022) 066901}
  [\href{https://arxiv.org/abs/2108.03239}{{\ttfamily 2108.03239}}].

\bibitem{Trickle:2022fwt}
T.~Trickle, \emph{{Extended calculation of electronic excitations for direct
  detection of dark matter}},
  \href{https://doi.org/10.1103/PhysRevD.107.035035}{\emph{Phys. Rev. D}
  {\bfseries 107} (2023) 035035}
  [\href{https://arxiv.org/abs/2210.14917}{{\ttfamily 2210.14917}}].

\bibitem{Dreyer:2023ovn}
C.~E. Dreyer, R.~Essig, M.~Fernandez-Serra, A.~Singal and C.~Zhen, \emph{{Fully
  ab-initio all-electron calculation of dark matter--electron scattering in
  crystals with evaluation of systematic uncertainties}},
  \href{https://arxiv.org/abs/2306.14944}{{\ttfamily 2306.14944}}.

\bibitem{Emken:2019tni}
T.~Emken, R.~Essig, C.~Kouvaris and M.~Sholapurkar, \emph{{Direct Detection of
  Strongly Interacting Sub-GeV Dark Matter via Electron Recoils}},
  \href{https://doi.org/10.1088/1475-7516/2019/09/070}{\emph{JCAP} {\bfseries
  1909} (2019) 070} [\href{https://arxiv.org/abs/1905.06348}{{\ttfamily
  1905.06348}}].

\bibitem{Essig:2019xkx}
R.~Essig, J.~Pradler, M.~Sholapurkar and T.-T. Yu, \emph{{Relation between the
  Migdal Effect and Dark Matter-Electron Scattering in Isolated Atoms and
  Semiconductors}},
  \href{https://doi.org/10.1103/PhysRevLett.124.021801}{\emph{Phys. Rev. Lett.}
  {\bfseries 124} (2020) 021801}
  [\href{https://arxiv.org/abs/1908.10881}{{\ttfamily 1908.10881}}].

\bibitem{Trickle:2020oki}
T.~Trickle, Z.~Zhang and K.~M. Zurek, \emph{{Effective field theory of dark
  matter direct detection with collective excitations}},
  \href{https://doi.org/10.1103/PhysRevD.105.015001}{\emph{Phys. Rev. D}
  {\bfseries 105} (2022) 015001}
  [\href{https://arxiv.org/abs/2009.13534}{{\ttfamily 2009.13534}}].

\bibitem{Andersson:2020uwc}
E.~Andersson, A.~B\"okmark, R.~Catena, T.~Emken, H.~K. Moberg and
  E.~\r{A}strand, \emph{{Projected sensitivity to sub-GeV dark matter of
  next-generation semiconductor detectors}},
  \href{https://doi.org/10.1088/1475-7516/2020/05/036}{\emph{JCAP} {\bfseries
  05} (2020) 036} [\href{https://arxiv.org/abs/2001.08910}{{\ttfamily
  2001.08910}}].

\bibitem{Su:2020zny}
L.~Su, W.~Wang, L.~Wu, J.~M. Yang and B.~Zhu, \emph{{Atmospheric Dark Matter
  and Xenon1T Excess}},
  \href{https://doi.org/10.1103/PhysRevD.102.115028}{\emph{Phys. Rev. D}
  {\bfseries 102} (2020) 115028}
  [\href{https://arxiv.org/abs/2006.11837}{{\ttfamily 2006.11837}}].

\bibitem{Mitridate:2021ctr}
A.~Mitridate, T.~Trickle, Z.~Zhang and K.~M. Zurek, \emph{{Dark matter
  absorption via electronic excitations}},
  \href{https://doi.org/10.1007/JHEP09(2021)123}{\emph{JHEP} {\bfseries 09}
  (2021) 123} [\href{https://arxiv.org/abs/2106.12586}{{\ttfamily
  2106.12586}}].

\bibitem{Vahsen:2021gnb}
S.~E. Vahsen, C.~A.~J. O'Hare and D.~Loomba, \emph{{Directional Recoil
  Detection}},
  \href{https://doi.org/10.1146/annurev-nucl-020821-035016}{\emph{Ann. Rev.
  Nucl. Part. Sci.} {\bfseries 71} (2021) 189}
  [\href{https://arxiv.org/abs/2102.04596}{{\ttfamily 2102.04596}}].

\bibitem{Catena:2021qsr}
R.~Catena, T.~Emken, M.~Matas, N.~A. Spaldin and E.~Urdshals, \emph{{Crystal
  responses to general dark matter-electron interactions}},
  \href{https://arxiv.org/abs/2105.02233}{{\ttfamily 2105.02233}}.

\bibitem{Chen:2022pyd}
H.-Y. Chen, A.~Mitridate, T.~Trickle, Z.~Zhang, M.~Bernardi and K.~M. Zurek,
  \emph{{Dark matter direct detection in materials with spin-orbit coupling}},
  \href{https://doi.org/10.1103/PhysRevD.106.015024}{\emph{Phys. Rev. D}
  {\bfseries 106} (2022) 015024}
  [\href{https://arxiv.org/abs/2202.11716}{{\ttfamily 2202.11716}}].

\bibitem{Catena:2022fnk}
R.~Catena, D.~Cole, T.~Emken, M.~Matas, N.~Spaldin, W.~Tarantino et~al.,
  \emph{{Dark matter-electron interactions in materials beyond the dark photon
  model}}, \href{https://doi.org/10.1088/1475-7516/2023/03/052}{\emph{JCAP}
  {\bfseries 03} (2023) 052}
  [\href{https://arxiv.org/abs/2210.07305}{{\ttfamily 2210.07305}}].

\bibitem{Li:2022acp}
J.~Li, L.~Su, L.~Wu and B.~Zhu, \emph{{Spin-dependent sub-GeV inelastic dark
  matter-electron scattering and Migdal effect. Part I. Velocity independent
  operator}}, \href{https://doi.org/10.1088/1475-7516/2023/04/020}{\emph{JCAP}
  {\bfseries 04} (2023) 020}
  [\href{https://arxiv.org/abs/2210.15474}{{\ttfamily 2210.15474}}].

\bibitem{Wang:2023xgm}
W.~Wang, W.-L. Xu and J.~M. Yang, \emph{{Direct detection of finite-size dark
  matter via electron recoil}},
  \href{https://arxiv.org/abs/2304.13243}{{\ttfamily 2304.13243}}.

\bibitem{Kurinsky:2020dpb}
N.~Kurinsky, D.~Baxter, Y.~Kahn and G.~Krnjaic, \emph{{Dark matter
  interpretation of excesses in multiple direct detection experiments}},
  \href{https://doi.org/10.1103/PhysRevD.102.015017}{\emph{Phys. Rev. D}
  {\bfseries 102} (2020) 015017}
  [\href{https://arxiv.org/abs/2002.06937}{{\ttfamily 2002.06937}}].

\bibitem{Gelmini_2020}
G.~B. Gelmini, V.~Takhistov and E.~Vitagliano, \emph{Scalar direct detection:
  In-medium effects},
  \href{https://doi.org/10.1016/j.physletb.2020.135779}{\emph{Physics Letters
  B} {\bfseries 809} (2020) 135779}.

\bibitem{Kozaczuk:2020uzb}
J.~Kozaczuk and T.~Lin, \emph{{Plasmon production from dark matter
  scattering}}, \href{https://doi.org/10.1103/PhysRevD.101.123012}{\emph{Phys.
  Rev. D} {\bfseries 101} (2020) 123012}
  [\href{https://arxiv.org/abs/2003.12077}{{\ttfamily 2003.12077}}].

\bibitem{Knapen:2021run}
S.~Knapen, J.~Kozaczuk and T.~Lin, \emph{{Dark matter-electron scattering in
  dielectrics}}, \href{https://doi.org/10.1103/PhysRevD.104.015031}{\emph{Phys.
  Rev. D} {\bfseries 104} (2021) 015031}
  [\href{https://arxiv.org/abs/2101.08275}{{\ttfamily 2101.08275}}].

\bibitem{Hochberg:2021pkt}
Y.~Hochberg, Y.~Kahn, N.~Kurinsky, B.~V. Lehmann, T.~C. Yu and K.~K. Berggren,
  \emph{{Determining Dark-Matter\textendash{}Electron Scattering Rates from the
  Dielectric Function}},
  \href{https://doi.org/10.1103/PhysRevLett.127.151802}{\emph{Phys. Rev. Lett.}
  {\bfseries 127} (2021) 151802}
  [\href{https://arxiv.org/abs/2101.08263}{{\ttfamily 2101.08263}}].

\bibitem{Knapen:2021bwg}
S.~Knapen, J.~Kozaczuk and T.~Lin, \emph{{python package for dark matter
  scattering in dielectric targets}},
  \href{https://doi.org/10.1103/PhysRevD.105.015014}{\emph{Phys. Rev. D}
  {\bfseries 105} (2022) 015014}
  [\href{https://arxiv.org/abs/2104.12786}{{\ttfamily 2104.12786}}].

\bibitem{Liang:2021zkg}
Z.-L. Liang, C.~Mo and P.~Zhang, \emph{{In-medium screening effects for the
  Galactic halo and solar-reflected dark matter detection in semiconductor
  targets}}, \href{https://doi.org/10.1103/PhysRevD.104.096001}{\emph{Phys.
  Rev. D} {\bfseries 104} (2021) 096001}
  [\href{https://arxiv.org/abs/2107.01209}{{\ttfamily 2107.01209}}].

\bibitem{Chen:2022xzi}
M.~Chen, G.~B. Gelmini and V.~Takhistov, \emph{{Halo-independent dark matter
  electron scattering analysis with in-medium effects}},
  \href{https://doi.org/10.1016/j.physletb.2023.137922}{\emph{Phys. Lett. B}
  {\bfseries 841} (2023) 137922}
  [\href{https://arxiv.org/abs/2209.10902}{{\ttfamily 2209.10902}}].

\bibitem{An:2021qdl}
H.~An, H.~Nie, M.~Pospelov, J.~Pradler and A.~Ritz, \emph{{Solar reflection of
  dark matter}}, \href{https://doi.org/10.1103/PhysRevD.104.103026}{\emph{Phys.
  Rev. D} {\bfseries 104} (2021) 103026}
  [\href{https://arxiv.org/abs/2108.10332}{{\ttfamily 2108.10332}}].

\bibitem{DeRocco:2022rze}
W.~DeRocco, M.~Galanis and R.~Lasenby, \emph{{Dark matter scattering in
  astrophysical media: collective effects}},
  \href{https://doi.org/10.1088/1475-7516/2022/05/015}{\emph{JCAP} {\bfseries
  05} (2022) 015} [\href{https://arxiv.org/abs/2201.05167}{{\ttfamily
  2201.05167}}].

\bibitem{An:2017ojc}
H.~An, M.~Pospelov, J.~Pradler and A.~Ritz, \emph{{Directly Detecting MeV-scale
  Dark Matter via Solar Reflection}},
  \href{https://doi.org/10.1103/PhysRevLett.120.141801}{\emph{Phys. Rev. Lett.}
  {\bfseries 120} (2018) 141801}
  [\href{https://arxiv.org/abs/1708.03642}{{\ttfamily 1708.03642}}].

\bibitem{Garani:2017jcj}
R.~Garani and S.~Palomares-Ruiz, \emph{{Dark matter in the Sun: scattering off
  electrons vs nucleons}},
  \href{https://doi.org/10.1088/1475-7516/2017/05/007}{\emph{JCAP} {\bfseries
  1705} (2017) 007} [\href{https://arxiv.org/abs/1702.02768}{{\ttfamily
  1702.02768}}].

\bibitem{Liang:2018cjn}
Z.-L. Liang, Y.-L. Tang and Z.-Q. Yang, \emph{{The leptophilic dark matter in
  the Sun: the minimum testable mass}},
  \href{https://doi.org/10.1088/1475-7516/2018/10/035}{\emph{JCAP} {\bfseries
  10} (2018) 035} [\href{https://arxiv.org/abs/1802.01005}{{\ttfamily
  1802.01005}}].

\bibitem{Emken_2018}
T.~Emken, C.~Kouvaris and N.~G. Nielsen, \emph{The sun as a sub-gev dark matter
  accelerator},
  \href{https://doi.org/10.1103/physrevd.97.063007}{\emph{Physical Review D}
  {\bfseries 97} (2018) }.

\bibitem{Gaidau:2021vyr}
C.~Gaidau and J.~Shelton, \emph{{Singularities in the gravitational capture of
  dark matter through long-range interactions}},
  \href{https://doi.org/10.1088/1475-7516/2022/01/016}{\emph{JCAP} {\bfseries
  01} (2022) 016} [\href{https://arxiv.org/abs/2110.02234}{{\ttfamily
  2110.02234}}].

\bibitem{Emken:2021lgc}
T.~Emken, \emph{{Solar reflection of light dark matter with heavy mediators}},
  \href{https://doi.org/10.1103/PhysRevD.105.063020}{\emph{Phys. Rev. D}
  {\bfseries 105} (2022) 063020}
  [\href{https://arxiv.org/abs/2102.12483}{{\ttfamily 2102.12483}}].

\bibitem{Bose:2021cou}
D.~Bose, T.~N. Maity and T.~S. Ray, \emph{{Solar constraints on captured
  electrophilic dark matter}},
  \href{https://doi.org/10.1103/PhysRevD.105.123013}{\emph{Phys. Rev. D}
  {\bfseries 105} (2022) 123013}
  [\href{https://arxiv.org/abs/2112.08286}{{\ttfamily 2112.08286}}].

\bibitem{Acevedo:2023owd}
J.~F. Acevedo, R.~K. Leane and J.~Smirnov, \emph{{Evaporation Barrier for Dark
  Matter in Celestial Bodies}},
  \href{https://arxiv.org/abs/2303.01516}{{\ttfamily 2303.01516}}.

\bibitem{Boyd:2022tcn}
C.~Boyd, Y.~Hochberg, Y.~Kahn, E.~D. Kramer, N.~Kurinsky, B.~V. Lehmann et~al.,
  \emph{{Directional detection of dark matter with anisotropic response
  functions}},  \href{https://arxiv.org/abs/2212.04505}{{\ttfamily
  2212.04505}}.

\bibitem{bruus2004many}
H.~Bruus and K.~Flensberg, \emph{Many-Body Quantum Theory in Condensed Matter
  Physics: An Introduction}, Oxford Graduate Texts. OUP Oxford, 2004.

\bibitem{fetter2012quantum}
A.~Fetter and J.~Walecka, \emph{Quantum Theory of Many-Particle Systems}, Dover
  Books on Physics. Dover Publications, 2012.

\bibitem{Serenelli:2009yc}
A.~Serenelli, S.~Basu, J.~W. Ferguson and M.~Asplund, \emph{{New Solar
  Composition: The Problem With Solar Models Revisited}},
  \href{https://doi.org/10.1088/0004-637X/705/2/L123}{\emph{Astrophys.J.}
  {\bfseries 705} (2009) L123}
  [\href{https://arxiv.org/abs/0909.2668}{{\ttfamily 0909.2668}}].

\end{thebibliography}\endgroup

\end{document}